\documentclass[aps,nofootinbib,longbibliography,superscriptaddress,
tightenlines,notitlepage]{revtex4-1}

\usepackage{amsmath,amsthm,latexsym,amssymb,amsfonts}
\usepackage{graphicx,color,natbib}
\usepackage[colorlinks=true,citecolor=blue]{hyperref}
\usepackage{tikz}
\usetikzlibrary{patterns}

\def\be{\begin{equation}}
\def\ee{\end{equation}}
\def\ba{\begin{eqnarray}}
\def\ea{\end{eqnarray}}

\def\SU{\text{SU}}

\begin{document}

\title{Coarse graining flow of spin foam intertwiners}

\author{Bianca Dittrich} 
\email{bdittrich@perimeterinstitute.ca}
\affiliation{Perimeter Institute for Theoretical Physics,\\ 31 Caroline Street North, Waterloo, Ontario, Canada N2L 2Y5}
\author{Erik Schnetter}
\email{eschnetter@perimeterinstitute.ca}
\affiliation{Perimeter Institute for Theoretical Physics,\\ 31 Caroline Street North, Waterloo, Ontario, Canada N2L 2Y5}
\affiliation{Department of Physics, University of Guelph,\\ 50 Stone Road East, Guelph, Ontario, Canada N1G 2W1}
\affiliation{Center for Computation and Technology, Louisiana State University,\\ Baton Rouge, Louisiana, USA 70803}
\author{Cameron J. Seth}
\email{cjmpseth@uwaterloo.ca} 
\affiliation{Department of Applied Mathematics, University of Waterloo,\\ Waterloo, Ontario, Canada N2L 3G1}
\author{Sebastian Steinhaus}
\email{sebastian.steinhaus@desy.de}
\affiliation{II. Institute for Theoretical Physics, University of Hamburg,\\
Luruper Chaussee 149, D-22761 Hamburg, Germany}

\begin{abstract}

Simplicity constraints play a crucial role in the construction of spin foam models, yet their effective behaviour on larger scales is scarcely explored. In this article we introduce intertwiner and spin net models for the quantum group $\text{SU}(2)_k \times \text{SU}(2)_k$, which implement the simplicity constraints analogous to 4D Euclidean spin foam models, namely the Barrett-Crane (BC) and the Engle-Pereira- Rovelli-Livine/Freidel-Krasnov (EPRL/FK) model. These models are numerically coarse grained via tensor network renormalization, allowing us to trace the flow of simplicity constraints to larger scales. In order to perform these simulations we have substantially adapted tensor network algorithms, which we discuss in detail as they can be of use in other contexts.

The BC and the EPRL/FK model behave very differently under coarse graining: While the unique BC intertwiner model is a fixed point and therefore constitutes a  2D topological phase, BC spin net models flow away from the initial simplicity constraints and converge to several different topological phases.  Most of these phases correspond to decoupling spin foam vertices, however we find also a new phase in which this is not the case, and in which a non-trivial version of the simplicity constraints holds.  The coarse graining flow of the BC spin net models indicates furthermore that the  transitions between these phases are not of second order.  The EPRL/FK model by contrast reveals a far more intricate and complex dynamics. We observe an immediate flow away from the original simplicity constraints, however, with the truncation employed here, the models generically do not converge to a fixed point. 

The results show that the imposition of simplicity constraints can indeed lead to interesting, and also very complex dynamics. Thus we will need to further develop coarse graining tools to efficiently study the large scale behaviour of spin foam models, in particular for the EPRL/FK model.

\end{abstract}

\maketitle

\section{Introduction}

Spin foams provide a non--perturbative and background independent path integral quantization for general relativity \cite{reisenberger-rovelli,rovellibook,perezreview}. The construction of spin foam models involves an auxiliary discretization as a regulator for the path integral. A key outstanding task is the removal of this regulator, a process we refer to as continuum limit.  One also needs to establish whether spin foam models can reproduce in this limit familiar low energy physics, in particular a geometric phase in which the models resemble a smooth manifold. Related is the question whether diffeomorphism symmetry, which is deeply rooted into the dynamics of general relativity, can be restored \cite{dittrich08, bahrdittrich-broken,dittrich12a}.

There are two main paths to remove dependence of the auxiliary discretizations: a refinement limit of the underlying discretization, see e.g.  \cite{dittrich14review}, or summing over the discretizations \cite{oriti-gft,carrozza-review,warsaw-summing}. We will here consider the first approach, based on the refinement limit for the following reasons: a number of works \cite{improved,harmosci,regge-measure,q-spinnet,bahr-birenorm} have shown that discretization independent models, which at the same time restore a notion of diffeomorphism invariance, can be constructed via a refinement limit -- implemented in practice via a coarse graining flow. (For a review, and an explanation of the interplay between refinement and coarse graining, see \cite{dittcyl,timeevol,dittrich14review}.)  Secondly we are in particular interested in the fate of diffeomorphism symmetry, which in the canonical framework is implemented via constraints. The spin foam path integral is supposed to provide a projector onto wave functions satisfying the corresponding quantized constraints \cite{hartle,rovelli-proj}. However, summing over discretizations does in general not lead to a projector \cite{freidel-gft,zipfel}. In contrast one can show that with the restoration of diffeomorphism symmetry in the discrete path integral one also obtains a projector, implementing the constraints, including an anomaly free constraint algebra \cite{improved,harmosci,hoehn1,bonzom-dittrich-dirac}.

The main challenge for the investigation of spin foam models is their overwhelming algebraic complexity. This comes together with an incomplete understanding of possible infinities (possibly related to diffeomorphism symmetry), a question on which there has been recent progress however \cite{perini,aldo,bonzom-dittrich-bubble,lin-qing-1,lin-qing-2}. Furthermore a framework has been developed that clarifies a number of conceptual questions in the context of `background independent' renormalization \cite{dittcyl,timeevol,bahr-birenorm,dittrich14review}.   To condense this framework to what is important for the current work:  the initial models, which are constructed via an auxiliary discretization, are subjected to a coarse graining flow.  The models will typically flow to an attractive fixed point, defining a phase of the model. Such phases correspond to topological models (with local amplitudes), which are triangulation invariant (and also restore diffeomorphism symmetry), but  do not feature propagating degrees of freedom.  Fine tuning of some parameters in the initial models, which correspond e.g. to ambiguities in choosing the path integral measure, might allow to find phase transitions. In particular, second order phase transitions are characterized by unstable fixed points, which in the background dependent context describe conformal theories.  In the spin foam context, we are interested in the fact that these are fixed points, i.e. that the fixed point model is invariant under at least some subset of discretization changes, defined by the coarse graining flow. On such a fixed point we can construct a meaningful refinement limit, e.g. via an inductive limit as outlined in \cite{dittcyl,timeevol,dittrich14review}.  The corresponding model will feature non--local amplitudes, but its dynamics is accessible via a system of so--called dynamical embedding maps, that allow to extract the large scale dynamics in terms of coarse grained observables, which also capture the `most relevant degrees of freedom' \cite{dittcyl}.   This framework is particularly adapted to so--called tensor network renormalization schemes. Such schemes implement   real space renormalization,  based on (a) identifying the `most relevant' (or contributing)  degrees of freedom in the path integral using the dynamics of the system and (b) explicitly integrating out these (relevant) degrees of freedom \cite{levin,gu-wen,dec-TNW,vidal-evenbly}. In particular (b) is opposed to Monte--Carlo simulations, in which the integral is accessed via a sampling process. As spin foams are real time path integrals, that is are expected to have complex and highly oscillating amplitudes,  this latter method is however (in general) not applicable to spin foams. This yields another motivation for the use of tensor network renormalization. 

Coming back to the overwhelming algebraic complexity of the models, different lines of attack have been taken. First of all, spin foams can be understood as generalized lattice gauge theory models \cite{holonomy1}, allowing also a notion of these models for finite groups \cite{sffinite}. This latter approach inspired also so--called spin net models, which will be the main focus of this work. In short, spin net models share the same dynamical ingredients as lattice gauge theories, namely group elements and weights, which are associated to lower dimensional objects, namely vertices and edges respectively (instead of edges and faces). Instead of a local gauge symmetry, spin nets have a global symmetry. The Ising model is a typical example for a $\mathbb{Z}_2$ spin net. Remarkably, 2D spin net models of the same gauge group share statistical properties with the 4D lattice gauge theory \cite{Kogut}. Hence we study spin net models as dimensionally reduced analogues for spin foams, which capture a key ingredient of spin foam dynamics, namely the so--called simplicity constraints. Furthermore spin nets are equivalent to so--called melon spin foams\footnote{A melon spin foam consists only of two vertices, which are connected by many dual edges. From this melon one obtains a spin net by cutting through the dual edges, by mapping the projectors on the edges of the foam to the vertices of the spin net. The spin net vertices then have the same valency as the spin foam edges.}, which are spin foams defined on a discretization involving only two (spin foam) vertices but an arbitrary number of edges connecting these vertices. The coarse graining collects a number of spin foam edges to new `thicker' edges. The hope is that the coarse graining of spin net models would allow to study and understand the behaviour of the simplicity constraints under coarse graining, and that a similar coarse graining flow could hold in spin foams. In fact this avenue allowed the investigation of more and more complicated models \cite{eckert1,s3-spinnet,q-spinnet} via tensor network renormalization, with \cite{q-spinnet} studying spin nets based on the quantum group $\text{SU}(2)_k$, and revealing a rich phase diagram for these models.  Furthermore, tensor network renormalization has been also applied to 3D spin foam models, so far based on finite groups, where the results confirm the phase diagram obtained for the corresponding spin nets \cite{dec-TNW,clementtoappear}.  

The main aim of this work is to study spin nets with the full algebraic complexity of the full (Euclidean) spin foam models, in particular the Barrett Crane (BC) and the so--called Engle-Pereira-Rovelli-Livine / Freidel-Krasnov (EPRL/FK) models \cite{Barrett-Crane,eprl1,eprl2,fk,warsaw-sf}.  Implementing a (positive) cosmological constant, which at the same time provides a convenient cut--off on the summation range for the variables, we therefore need to consider spin nets with a structure group $\text{SU}(2)_k \times \text{SU(2)}_k$.  We will see that this requires a range of techniques to allow for the numerical implementation of the tensor network coarse graining. We will detail these techniques as we believe that these will be also helpful in more general contexts. 

As mentioned the tensor network coarse graining is based on explicit summation of the models. Furthermore, the space of models, in which the coarse graining flow can take place, is very large, in general given by all possible tensors of a predefined rank and index range.  The coarse graining process proceeds iteratively, the effective amplitudes of each coarse graining steps are encoded in a tensor (of very high dimension) which is updated in each step based on the previous tensor.  This allows in principle to keep track of many observables of the models, but is of course also a challenge for the numerical implementation.  

Most importantly this algorithms allows to track the coarse graining flow of the simplicity constraints, which are crucial for the spin foam dynamics. The simplicity constraints determine which spin values are allowed and the various models do differ in these sets. Under coarse graining one expects that the allowed set of spins changes: this is do to the coupling of `finer' spins to `coarser' spins, which does not need to respect the simplicity constraints. To understand the large scale behaviour of spin foams it is crucial to study how the simplicity constraints change under coarse graining. 

The recent work \cite{sf-cuboid,sf-cuboid-renorm} takes in some sense an opposite approach to the one taken here: one works with the full models (more precisely EPRL/FK  using coherent Livine-Speziale intertwiners \cite{liv-spez-int}), but implements a drastic, geometrically motivated, truncation, that for instance suppresses all curvature degrees of freedom, but keeps some torsion degrees of freedom. 
 \cite{sf-cuboid,sf-cuboid-renorm} employ furthermore a saddle point approximation for the spin foam amplitudes, such that the focus is on large spins. (In contrast, using quantum group models here, we rather concentrate on small spins.) As  curvature is suppressed the associated Regge action vanishes, such that Monte--Carlo methods can be readily employed. These allow the approximate computation of expectation values for observables arising in one coarse graining step. Such an expectation value is then also used as a criterion to truncate the amplitude for the coarse building block back to the initial one--parameter family of models.    This one parameter encodes a certain freedom in the choice of path integral measure.  From this procedure one can deduce a coarse graining flow which tracks only the parameter describing the path integral measure. Thus compared to the tensor network method, where the flow is computed in a very high dimensional parameter space, here one truncates the flow to a one parameter space. Despite these drastic truncations very interesting results were found:  this (truncated) flow shows indications for a phase transition, at which a  notion of residual diffeomorphism invariance is recovered.

Another approach relies even more heavily on analytical techniques \cite{jeffetal}. \footnote{So far particular (simplifying) features of the model studied in \cite{jeffetal} seem to be important in order to allow for analytical treatment, see also \cite{lin-qing-2}.}   The works \cite{jeffetal,lin-qing-1} consider Pachner moves in a general triangulation, which makes it however difficult to come up with an iterative (regular) coarse graining scheme. Thus one can compute the amplitudes for a coarser complex, the details of the truncation scheme and a full implementation of the flow still need to be explored. These methods do however allow for a general understanding of the divergence structure of the models \cite{lin-qing-1}.

Let us also mention the older works \cite{bc-monte} which studied the BC model with Monte Carlo simulations.
 Here one uses a property specific to the BC model, namely that it admits a representation in which the amplitudes are positive \cite{positivity-bc}. This does not hold for the EPRL model, prohibiting so far Monte Carlo simulations for the action contribution to the path integral. The work \cite{bc-monte} considered the $\SU(2)\times \SU(2)$ BC model on a very simple  4D triangulation, given by the 5--simplex. It also implemented a cut--off in the spins ($j=5/2$ and $j=25/2$). Three different choices for measure factors were tested: one which lead to a fast "divergence" of the model, i.e. a phase were large spins dominated (with respect to the cut--off).  One phase that led to a fast convergence and a partition function dominated by  $j=0$  spins. This motivated the introduction of a third choice, on the border between these two behaviours.  Also a quantum group BC model has been considered in \cite{Khavkine}. The most interesting point here is that the expectations values in the quantum group values do not converge to the classical group case, indicating a discontinuity. 
 
In these works \cite{bc-monte,Khavkine} one has measured e.g. the relative frequency of spin values in the probability distribution defined by the BC model. Although
 such observables give some insight -- in this case tested the suitability of measure factors -- we believe that we need a more systematic development of  order parameters admitting a diffeomorphism invariant meaning.  A main point of concern in \cite{bc-monte} are divergences and so--called bubbles that are actually a sign for a restoration of diffeomorphism invariance. 
 
The tensor network employed here has the advantage to test the model iteratively over a large range of scales (defined by the number of coarse graining steps). Divergences can in principle be dealt with (although these do not arise here due to using  quantum groups) by normalizing the partition function in each coarse graining step.  Also a key point is the ability to track how the simplicity constraints behave under coarse graining, as this understanding is crucial for the understanding of the (effective) dynamics in spin foams models. 

In agreement with results in \cite{bc-monte,sf-cuboid-renorm}, we find that the measure is a relevant factor which can drive phase transitions. This is also intuitively understandable: choosing a measure factor that suppresses larger spins drives the system to the Ashtekar-Lewandowski phase, in which all spins $j>0$ are not allowed. The similar question concerning the (dual) BF phase, which is characterized by allowing all representations weighted by their respective dimension, can be answered negatively: For both BC and EPRL analogue models we do not observe a flow to a BF phase for any choice of measure discussed here.


~\\

This article is organized into two main parts. One focusses on the computational methods used in this work, while the other one addresses the construction and results of the quantum gravity related models. We have designed these parts such that they can be read independently of one another:

The first part is aimed at researchers outside quantum gravity also using tensor network techniques. While avoiding technical details of the models under discussion we focus on the scope of the problem and the implemented improvements to the algorithm. In section \ref{sec:scope} we will discuss the scope of the models we intend to coarse grain with tensor network techniques, which has motivated the improvements to the algorithm we present in section \ref{sec:optimization}. 

The second part is aimed at people familiar with quantum gravity and spin foam models and can be read without reference to the computational / numerical details. In section \ref{sec:models} we briefly recall and motivate the general class of models under discussion, including their relation to lattice gauge theories and spin foam models. In section \ref{sec:qg-models} we construct models analogue to modern 4D spin foam models and discuss their behaviour under coarse graining.

We conclude with a discussion of the methods and results in section \ref{sec:discussion}.

Several of the methods used in this article have already been developed and used in previous articles. Thus we will not introduce them in full detail, but concisely introduce their main features in the appendices.

~\\
\begin{center}{\Large  PART I: IMPLEMENTATION}\end{center}


\section{Coarse graining of  spin net models: numerical challenges} \label{sec:scope}

\subsection{A brief introduction to tensor network renormalization}


Before introducing the models under discussion in this work let us briefly touch upon the numerical algorithms used to coarse grain said models, which are broadly summarized under the term tensor network renormalization. 

Before applying tensor network renormalization \cite{levin,gu-wen,vidal-evenbly} the partition function of the model is rewritten as a contraction of a tensor network. A tensor network is a collection of multi--dimensional arrays, i.e. tensors, at the vertices of a lattice, where each tensor has as many indices as the vertex has legs. Then the tensors are contracted according to the combinatorics of the network, that is a shared leg implies that the respective indices get contracted. 

There exist different ways to obtain such a tensor network, but they are usually straightforward. In the cases we are considering the partition function is already of tensor network form. Crucially even though the tensor network might coincide with the lattice the underlying model is defined on, the network is independent as it merely represents a rewriting of the model. This is particularly beneficial in the context of background independent approaches to quantum gravity as tensor networks do not refer to a background structure. In most cases one studies systems on regular lattices resulting also in a regular network with identical tensors at all vertices. Thus the coarse graining process can be straightforwardly iterated. 

The fundamental idea of tensor network renormalization is to locally manipulate the network, given by tensors $T$, such that the same partition function is (approximately) described by a coarser network of effective tensors $T'$:
\begin{equation}
Z = \text{Ttr} \; T \dots T \approx \text{Ttr} \; T' \dots T' \quad ,
\end{equation}
where $\text{Ttr}$ denotes the tensor trace, i.e. the contraction of the tensors according to the network. Hence one studies a flow of tensors $T \rightarrow T' \rightarrow \dots$ capturing the dynamics of the system, which lead to the original name of tensor renormalization group (TRG) \cite{levin}. In this article we use a method closely related to TRG. Nevertheless we would like to point out that a more advanced algorithm has been invented by Evenbly and Vidal \cite{vidal-evenbly}, named tensor network renormalization (TNR), which filters out short-range entanglement resulting in a proper renormalization group flow. This method is also closely related to the Multi-Scale entanglement renormalization ansatz (MERA) \cite{mera} used to construct ground states of condensed matter systems.

More concretely let us discuss the algorithm introduced in \cite{levin} as an example\footnote{In general many schemes to coarse grain tensor networks exist, e.g. one which more closely resembles block spin transformations: By contracting the edges connecting four tensors on the corner of a square one obtains a new coarse tensor. This step is exact, yet the new tensor has `double' edges with a bond dimension $\chi^2$. Due to this exponential growth of data and to relate the new tensor to the original one, approximations are necessary, which are usually implemented via variable transformations and truncations, such that the error is minimized. Graphically this is shown as a 3-valent tensor mapping the two edges into an effective one. We usually refer to these maps as embedding maps.}, as we are going to modify it in the rest of the article. Consider a 2D square tensor network of identical tensors. Let the indices of the tensor run from $1$ to $\chi$. This index range $\chi$ is frequently referred to as the (initial) bond dimension\footnote{As the models under discussion here are already of tensor network form, the tensor actually inherits the variables of the original model as labels on its edges. We refer to these spaces on the edges also as edge Hilbert spaces $\mathcal{H}_e$.}.

\begin{figure}[h]
\includegraphics[scale=0.5]{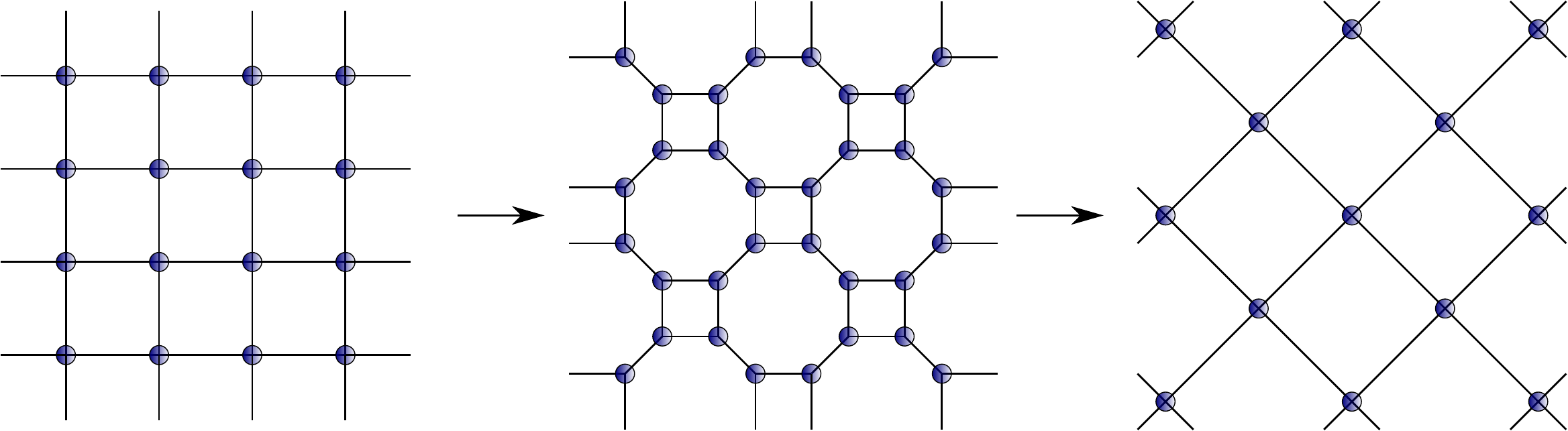}
\caption{\label{fig:algo-gen}
Scheme of 4-valent algorithm: Each 4-valent tensor is split along a new edge into two 3-valent tensors. Then four these 3-valent tensors are contracted along the original edges to give a new effective 4-valent tensor, whose coarser edges are those introduced by the splitting of the original tensors.}
\end{figure}

The general scheme of the algorithm is illustrated in fig. \ref{fig:algo-gen}. To coarse grain this network each 4-valent tensor is split first into two 3-valent ones, that is the 4-valent tensor is written as a contraction of two 3-valent tensors along a new edge. This new edge will be the effective edge of the coarse grained tensor network. A priori a 4-valent tensor can be split in many different ways, but as approximations during the numerical algorithm will be necessary this splitting should allow for error control. Thus the tensor is rewritten into a matrix by a pairwise grouping of its indices according to the intended splitting. This matrix is then split into two by a singular value decomposition (SVD) as follows:
\begin{equation} \label{eq:SVD}
T_{(ab);(cd)} =: M_{(ab);(cd)} = \sum_{i=1}^{\chi^2} U_{(ab);i} \; \lambda_i \; (V_{(cd);i})^\dagger \quad ,
\end{equation}
where $U$ and $V$ are the unitary matrices of singular vectors, $\lambda_i$ denote the singular values with $\lambda_1 \geq \lambda_2 \geq \dots \geq \lambda_{\chi^2} \geq 0$.

From $U$ and $V$ one then constructs 3-valent tensors, e.g. $S_{a b; i} = U_{(a b); i} \sqrt{\lambda_i}$. Four of these tensors $S$ are then contracted along the links of the original network to give the new effective tensors $T'$:
\begin{equation} \label{eq:effective-tensor}
T'_{i j k l} = \sum_{a b c d} S_{a b;i} S_{b c;j} S_{c d;k} S_{d a;l} \quad .
\end{equation}
For simplicity we have avoided to enumerate the tensors $S$; in principle they can be different, but this is not important to illustrate the scheme.

The coarse edges of the new network of tensors $T'$ are those obtained from splitting the initial tensors $T$. Thus the new tensors are actually labelled by the singular values. Note that the SVD \eqref{eq:SVD} is exact such that the coarse network is an exact rewriting of the original partition function. From this we can conclude the physical interpretation underlying tensor network renormalization:
\begin{itemize}
\item The SVD serves as a variable transformation, reshuffling the original degrees of freedom into effective degrees of freedom on a coarser scale. Since \eqref{eq:SVD} is exact no degrees of freedom are lost, while the relation to the original interpretation is encoded in the maps $U$ and $V$. Moreover the SVD arranges the degrees of freedom according to their significance, which is indicated by the relative size of the associated singular values.
\item In general the tensor $T'$ as obtained from \eqref{eq:SVD} and \eqref{eq:effective-tensor} has a bond dimension of $\chi^2$ compared to $\chi$ of the original tensor $T$. Without approximations this bond dimension grows exponentially with each iteration of the algorithm quickly rendering the scheme inefficient. Thus approximations must be implemented to keep the algorithm feasible. Due to the features of the SVD the quality of the approximations can be readily evaluated: The degrees of freedom are ordered in significance indicated by the size of the singular values. Hence it is straightforward to truncate less important degrees of freedom in \eqref{eq:SVD} by dropping e.g. all $\lambda_i$ with $i > \chi$. This approximation is actually the best one of $M_{(ab);(cd)}$ by a matrix of rank $\chi$ (with respect to least square error).
\end{itemize}
Of course the more singular values are taken over the better the approximation is, e.g. the position of a phase transition is more accurately determined.

Usually one iterates the algorithm for a fixed bond dimension until the system has converged to a fixed point tensor $T^*$. This tensor is then used to identify different phases of the model.

In the next section we introduce the models we will coarse grain in this article via tensor network renormalization.

\subsection{A brief introduction to quantum group spin nets}

In this article we successfully apply tensor network renormalization  to so--called spin net models \cite{eckert1,s3-spinnet,q-spinnet} based on the quantum group $\text{SU}(2)_k \times \text{SU}(2)_k$. The goal of this section is to give the reader an impression why this is a remarkable achievement made possible by several improvements of tensor network algorithms. After giving a very short introduction to spin net models we explain why the main challenge is the size of the tensors, encoding the models, leading to a memory consumption that is too large to handle even with HPC (high performance computing) resources. We then introduce several techniques which allow us to reduce memory usage enormously. We expect that these methods and ideas can also be facilitated by other researchers, in particular in high accuracy calculations.

Spin net models can be defined on lattices of arbitrary dimension, but we will restrict ourselves to a 2D square lattice in this article. The models are characterized  by a global  symmetry group, e.g. a non--Abelian finite or (compact) Lie group. The simplest non-trivial example is the $\mathbb{Z}_2$ Ising model, which is invariant under flipping all Ising spins.  This model can be represented in either the group picture, e.g. with the group $\mathbb{Z}_2$ defining the fundamental variables (the Ising spins), or in the dual picture, where the variables are given by the irreducible representation labels of the group. This latter picture defines also a tensor network representation for the spin net models. For non--Abelian groups, the representations are higher than one--dimensional and the representation labels are amended by vector space labels. 

The second representation, involving representations of the symmetry group, is also called `spin representation'. The name is due to the $\text{SU}(2)$ representations $j$, which are referred to as spins.  It is this representation which can be generalized also to quantum groups, in particular $\text{SU}(2)_k$ \cite{biedenharn,yellowbook}, which is thoroughly explained in \cite{q-spinnet}.  In this section we restrict the discussion to the most basic features of representation theory for $\text{SU}(2)_k$ in order to discuss the index range of the initial tensor.

\begin{itemize}
 \item $\text{SU}(2)_k$ is a Hopf algebra, the $q$-deformation of the universal enveloping algebra $\mathcal{U}(\text{SU}(2))$ for $q = \exp(\frac{i \pi}{k+2})$ at root of unity \cite{biedenharn,yellowbook}. $k \in \mathbb{N}$ is called the level of the quantum group. Very similar to $\text{SU}(2)$ these quantum groups have irreducible representations labelled by spins $j \in \frac{1}{2} \mathbb{N}$. These range from $0$ to $j_{\text{max}}=\frac{k}{2}$, the maximal spin of the quantum group. As for $\text{SU}(2)$ the representation vector spaces $V_j$ are $(2j+1)$-dimensional. In this work we will restrict ourselves to the integer representations of $\text{SU}(2)_k$\footnote{This can be understood as the $q$-deformation of the algebra of $\text{SO}(3)$. If $k$ is odd we take $\frac{k-1}{2}$ as the maximal spin.}.
 \item Each edge of $\text{SU}(2)_k \times \text{SU}(2)_k$ spin nets carries the edge Hilbert space $\mathcal{H}_e = \bigoplus_{j^+,j^- =0}^{j_{\text{max}}} V_{j^+} \otimes V_{{j^+}^*} \otimes V_{j^-} \otimes V_{{j^-}^*}$, where $j^+$ and $j^-$ are $\text{SU}(2)_k$ representations and $j^*$ denotes their dual representation. Thus expressed naively as a tensor network each leg of a tensor carries the indices $\{j^+,j^-,m^+,n^+,m^-,n^-\}$, where the so--called magnetic indices $m,n$ range from $-j$ to $j$, in integer steps.
 \item The tensor itself encodes the `quantum group symmetries', i.e. it is a projector onto the invariant subspace in the product space of all representation spaces meeting at the vertex, $\text{Inv}(\bigotimes_{e \supset v} V_{j_e})$. The projector onto the full invariant subspace is called the Haar projector, see \cite{q-spinnet} or appendix \ref{app:graph} for its definition.
\end{itemize}
Due to the symmetry of the model and the finite edge Hilbert spaces, it is in principle possible to directly turn the model into a tensor network. Thus one obtains the tensor:
\begin{equation}
T(\{j^+\},\{{j^+}'\},\{j^-\},\{{j^-}'\},\{m^+\},\{n^+\},\{m^-\},\{n^-\}) \quad .
\end{equation}
To not overburden the notation, we suppress the indices $i$ of the edges, which range from $1$ to $4$ in the case of square network. 

However this naive approach is not very feasible for neither non-Abelian groups \cite{s3-spinnet} nor quantum groups: A quick estimate of the dimension of the edge Hilbert space for a small quantum group, e.g. $k=4$ such that the spins range over  $j=0,1,2$, shows that the index range is roughly $6000$ if ${j^\pm}' \neq {j^\pm}^*$, due to the sheer amount of magnetic indices.

Fortunately due to the symmetries of the model, the dependence of the tensor $T$ on the magnetic indices  denoted is not arbitrary and given by the projector / intertwiner structure. This projector structure actually survives under tensor network renormalization  and can be exploited to significantly reduce the index range of the initial tensor. To do so, two measures were introduced in \cite{s3-spinnet,q-spinnet}.

\begin{figure}
\includegraphics[scale=0.4]{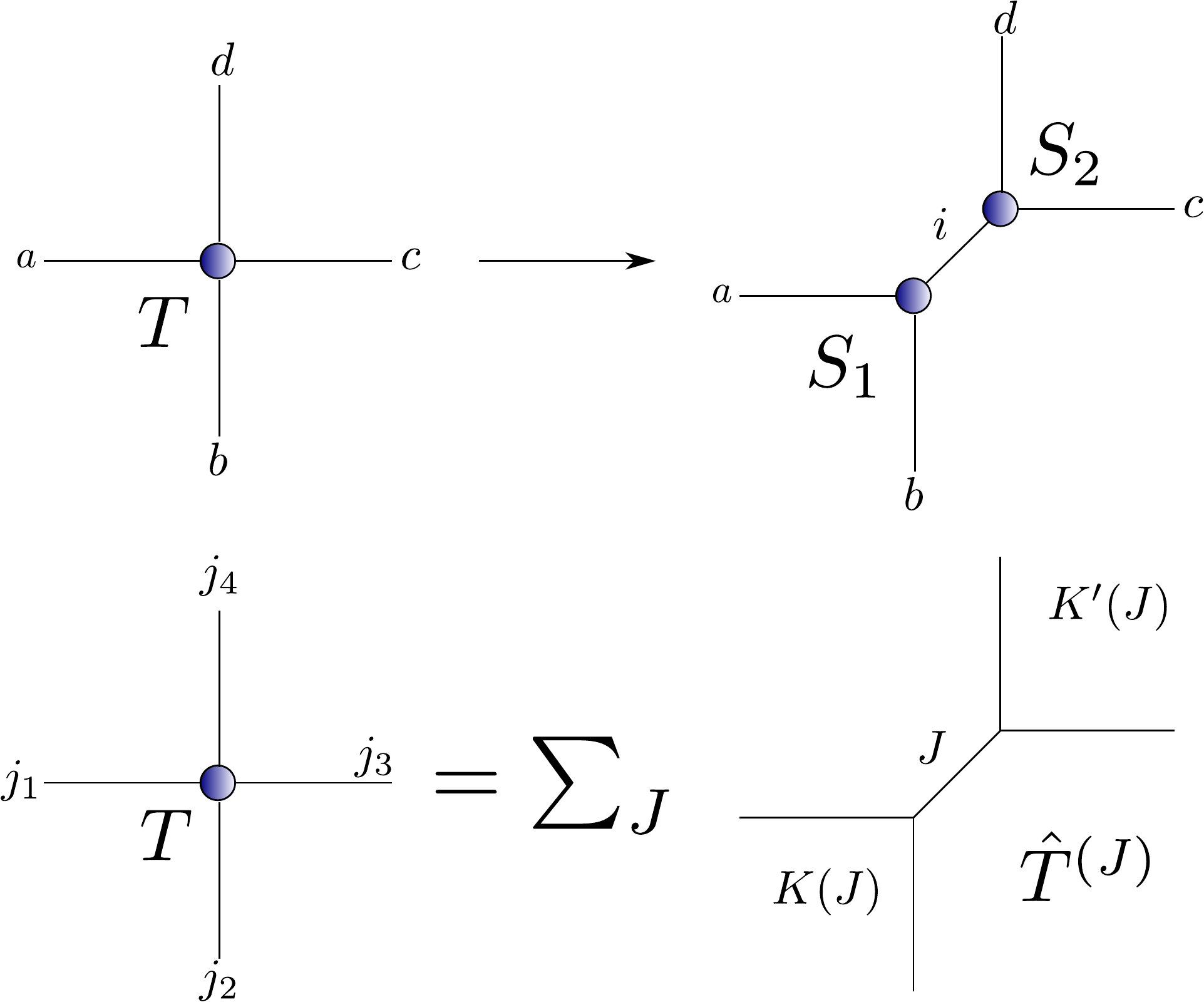}
\caption{\label{fig:recoupling}
Visualization of recoupling basis: Above we show the intended splitting of the coarse graining algorithm. The recoupling basis is chosen such that the variables that are to be split couple to the same intermediate spin $J$. Instead of storing all boundary data $\{j_i\}$, $i=1,\dots,4$ we only store the ones leading to non-vanishing tensors, stored in super-indices $K(J)$ and $K'(J)$ respectively. We have suppressed additional representations $j^\pm$ and magnetic indices.}
\end{figure}

\begin{figure}
\includegraphics[scale=0.75]{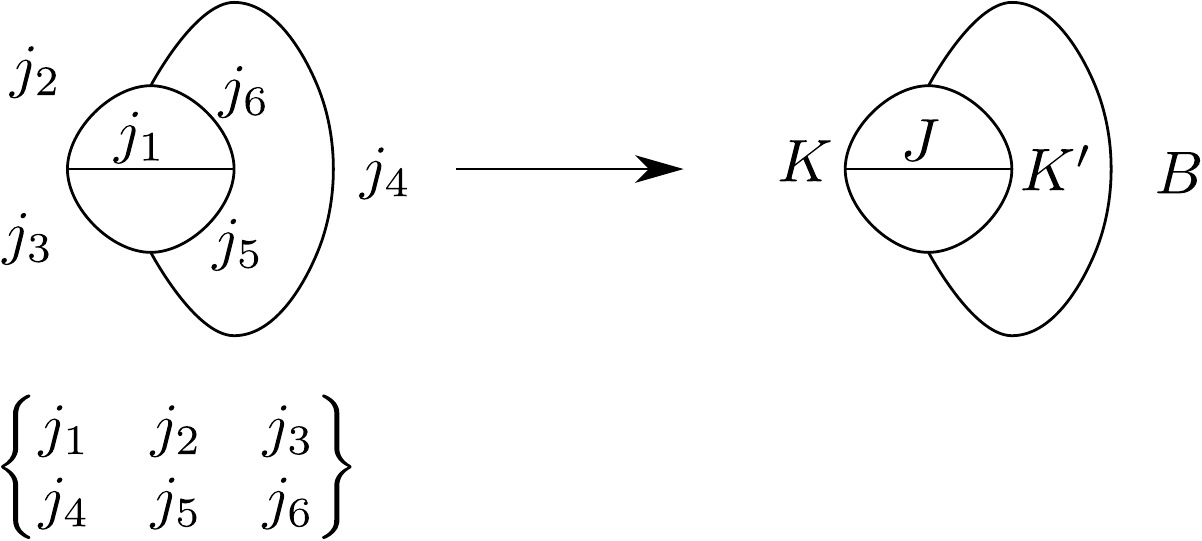}
\caption{\label{fig:6j-indices} 
Indices for the $\{ 6j \}$ symbol: We choose $J=j_1$ as the reference spin, then all non-vanishing choices for $j_2$, $j_3$ and $j_5$, $j_6$ are encoded in indices $K(J)$ and $K'(J)$ respectively. The non-vanishing choices for $j_4$ then depend on $J$, $K$ and $K'$ and are summarized in the super-index $B(J,K,K')$.
}
\end{figure}

\begin{itemize}
 \item The initial tensor was rewritten into a so-called recoupling (or intertwiner) basis, in which the tensor is expanded into a sum over  4-valent invariant tensors, which are labelled by (intermediate) spins $\{J^\pm,{J^\pm}'\}$. This basis is adapted to the intended splitting of the four--valent tensors into three--valent ones.  As the spins $ \{j^\pm{j^\pm}'\}$ associated to the original edges have to couple to the intermediate spins $\{J^\pm,{J^\pm}'\}$,  the tensor $T$ can be expressed in a block-diagonal form. Thus the crucial information on the tensor is encoded in an amplitude only depending on the spins $\{j^{\pm}\}$ on its edges and the spins $\{J^\pm\}$ labelling the basis. The projector structure, and with it the dependence of the tensor $T$ on the magnetic indices, is explicitly preserved under coarse graining. This allows to pre--contract the magnetic indices of the projective part into so--called recoupling symbols. Thus  during the coarse graining cycles itself we will only have to deal with the spin indices.
 \item The intertwiner structure introduced above can be exploited further by considering the coupling rules of $\text{SU}(2)_k$. In fact the intertwiner basis is written as the sum over two Clebsch-Gordan coefficients which are only non-vanishing if triangle inequalities are satisfied. Thus we introduce one superindex $K(J)$ for each intermediate label $J$ counting the allowed possibilities of the pair $(j_1,j_2)$ coupling to $J$, dismissing all vanishing entries due to coupling rules. Conversely one can translate $K(J)$ back into representations as $j_i(K(J))$ for $i=1,2$. Fig. \ref{fig:recoupling} illustrates the idea with reference to the splitting in the algorithm.
 \item A similar super index $B(J,K(J),K'(J))$ is also defined for the $\text{SU}(2)_k$ $\{6j\}$ recoupling symbol, which is a particular contraction of four Clebsch-Gordan coefficients and appears in the renormalization equations due to the treatment of the magnetic indices. Essentially one picks out one representation $J$, to which two pairs of two representations couple directly encoded in two superindices $K(J)$ and $K'(J)$. The last remaining representation has to satisfy several conditions and one stores the allowed choices in the index $B$, which depends on $J$, $K(J)$ and $K'(J)$. Thus one only stores (and sums over) the non-vanishing $\{6j\}$ symbols. 
 As before we can decode these indices back to the original spin values via functions $j_\alpha(J,K,K',B),\alpha=1,\ldots 6$. This index is explained in fig. \ref{fig:6j-indices}.
\end{itemize}
The first measure already drastically improves the initial index range of the tensors:
\begin{equation}
 T(\{j^\pm\},\{{j^\pm}'\},\{m^\pm\},\{n^\pm\}) = \sum_{J^{\pm},{J^{\pm}}'} \hat{T}^{(J^\pm,{J^\pm}')} (\{j_i^\pm\},\{{j_i^\pm}'\}) \times \left(P^{J^+}_{\{j_i^+\}} \otimes P^{{J^+}'}_{\{{j_i^+}'\}} \otimes P^{J^-}_{\{j_i^-\}} \otimes P^{{J^-}'}_{\{{j_i^-}'\}} \right) \quad ,
\end{equation}
where the $P$ represent a basis of 4-valent intertwiners for one copy of irreducible representations.  Their explicit form is not relevant here and can be found in section \ref{sec:models} and also \cite{q-spinnet}.
 Again the pivotal insight is that this structure is analytically dealt with and preserved under coarse graining, such that the information of the tensor is stored in $\hat{T}$ instead of $T$. Thus the tensor is specified by four $\text{SU}(2)_k$ representations on each edge, which for $k=4$ is an index range  of $\chi=3^4=81$. For the entire tensor we obtain a size of $81^5$ for the four edges and four intermediate spins labelling the basis.

This expression can be further simplified by using superindices $K(J)$. The superindices always combine two representations coupling to the spin $J$. In the 4-valent case we group two representations together according to the intended split, which we denote by sets $\{j_i\}_a$ and $\{j_i\}_b$, such that we also have superindices $K_a(J)$ and $K_b(J)$. Note that from here on we suppress the additional superscripts ${}^\pm$ and ${}'$ in order to simplify the notation. Unless specified otherwise $\{j\}$ is supposed to stand for $\{j^\pm,{j^\pm}'\}$, similar for $J$ and $K(J)$:
\begin{equation}
\hat{T}^{\{J\}} (\{j_i\}_a,\{j_i\}_b) = \left\{ \begin{matrix}
\hat{T}^{\{J\}} (\{j_i(K_a(J))\}_a,\{j_i(K_b(J)\}_b) & \quad \text{if all pairs } \{j_i\}_{a,b} = \{j_i(K_{a,b}(J))\}_{a,b} \\
0 & \quad \text{ else }
\end{matrix}  
\right .
\end{equation}
Since the projector structure is explicitly preserved it is sufficient to just consider the entries of $\hat{T}$ encoded in the superindices $K(J)$. Since these $K(J)$ combine the representations on two fine edges, one cannot talk about individual index ranges, but we can still give the size of the total tensor. Again for $k=4$ this is given by $43^4$, which is roughly $0.1\%$ of the data without using superindices.

To sum up the measures briefly described in this section provide an interpretational and a computational advantage: By introducing the intertwiner basis we isolate the relevant data of $T$ from the magnetic indices and encode it into a much smaller tensor $\hat{T}$. Moreover the tensor $\hat{T}^{\{J\}}$ is already in a block diagonal form, where each block is labelled by four $\text{SU}(2)_k$ representations $\{J\}$. Expressed in superindices $\{K_a(J)\},\{K_b(J)\}$ it can be readily rearranged into a matrix $M_{ab}$ to which SVD is applied in order to split $\hat{T}$. The new edges created in this SVD carry over the labels $\{J\}$ of the block, which are the same type of variables as in the original model. Thus one eventually obtains a new coarse tensor $\hat{T}'$ of the same form as the initial one. This procedure (with smaller index ranges) described here has already been used in \cite{q-spinnet}.

However these significant improvements quickly turn out to be insufficient as soon as one goes to higher levels $k$ of the quantum group $\text{SU}(2)_k$. In fact one easily reaches the limits of modern high performance computers, in particular in terms of memory usage. This issue is the subject of the next subsection.

\subsection{Memory costs of $\text{SU}(2)_k \times \text{SU}(2)_k$ spin nets} 

In this section we discuss the main numerical obstruction that has to be overcome if one intends to coarse grain $\text{SU}(2)_k \times \text{SU}(2)_k$ spin nets, which is the immense cost in memory. In order to give an idea of the order of magnitude it is sufficient to just consider the size of the initial tensor $\hat{T}$ (using superindices $K(J)$).

As one increases the level $k$ of the quantum group the size of the initial tensor increases due to two effects: $j_{\text{max}}$ and thus the number of irreducible representations increases and the range of the superindex $K(J)$ increases as more pairs of representations $(j_1,j_2)$ can couple to $J$. Because of the first effect, the number of intertwiners grows exponentially as they are labelled by four $\text{SU}(2)_k$ representations $\{J\}$. Moreover each of these grows in size as well due to the larger range of the superindices. We summarize this in table \ref{tab:tensor-size}.

\begin{table}
\begin{tabular}{| c | c | c | c | c | c | c | c |}
\hline
Level $k$ & $j_{\text{max}}$ & Maximal $K(J)$ & Number of blocks & Size of largest block & Size of block in GB & Size of $\hat{T}$ & Size of $\hat{T}$ in GB \\
\hline
4 & 2 & $K(1)=5$ & 81 & $25^4$ & $\sim 0.0058$ & $43^4$ & $\sim 0.051$ \\
\hline 
5 & 2 & $K(1)=6$ & 81 & $36^4$ & $\sim 0.025$ & $70^4$ & $\sim 0.36$ \\
\hline
6 & 3 & $K(1)=K(2)=8$ & 256 & $64^4$ & $\sim 0.25$ & $160^4$ & $\sim 9.77$ \\
\hline
7 & 3 & $K(2)=10$ & 256 & $100^4$ & $\sim 1.5$ & $246^4$ & $\sim 54.6$ \\
\hline
8 & 4 & $K(2)=13$ & 625 & $169^4$ & $\sim 12.2$ & $461^4$ & $\sim 673.1$ \\
\hline
9 & 4 & $K(2)=15$ & 625 & $225^4$ & $\sim 38.2$ & $671^4$ & $\sim 3021$ \\
\hline
10 & 5 & $K(2) = K(3) = 18$ & 1296 & $324^4 $ & $\sim 165$ & $1112^4$ & $\sim 23000$ \\
\hline
\end{tabular}
\caption{\label{tab:tensor-size}
Characteristic numbers for the initial tensor $\hat{T}$ of $\text{SU}(2)_k \times \text{SU}(2)_k$ spin nets for different values of $k$. We assume $16$ Byte per entry.}
\end{table}

From the data we can clearly observe an exponential increase in the size of the initial $\hat{T}$ for growing level $k$ of the quantum group. The memory used to store it roughly increases by an order of magnitude for each increase of $k$. While the models $k=4$ and $k=5$ appear to be small enough to still run on modern notebooks (at least concerning the memory usage), one has to move to high performance machines for $k \geq 6$. However already for $k=8$ the memory to only define the initial tensor exceeds the memory available on many modern machines. Note also that the memory cost for the initial $\hat{T}$ alone can only serve as a lower bound, in particular if one goes to higher bond dimension after several iterations of the algorithm.

It is apparent that the standard 4-valent algorithm \cite{levin,gu-wen,q-spinnet} is limited to smaller levels $k$ due to the sheer size of the tensors, which encode the dynamics of the theory. Thus in order to go to larger quantum groups one has to find a way of encoding the same information in smaller building blocks carrying less data. In a way the tensor network algorithm \cite{levin,gu-wen,q-spinnet} itself already holds the key to the solution of this problem. During the algorithm the 4-valent tensor $\hat{T}$ is split via a singular value decomposition (SVD) into two 3-valent tensors $\hat{S}_1$ and $\hat{S}_2$. This transformation is exact as long as no singular values get truncated such that one can encode the same information into 3-valent tensors. This allows to redesign the algorithm \cite{levin,gu-wen} based on four--valent tensors to an algorithm (first introduced in \cite{dec-TNW}) which is equivalent in precision, but  only involves 3-valent tensors  and requires moreover less computational time (scaling with $\chi^{3}$ instead of $\chi^{4}$). Fortunately in the intertwiner basis it is straightforward to define the initial 3-valent tensor without a SVD, such that one can readily work with 3-valent tensors $\hat{S}$ instead of $\hat{T}$. The sizes of 3-valent and 4-valent tensors are compared in table \ref{tab:tensor-size2}.

\begin{table}
\begin{tabular}{| c | c | c | c | c | c | c |}
\hline
Level $k$ & $j_{\text{max}}$ & Maximal $K(j)$ & Size of $\hat{S}$ & Size of $\hat{S}$ in GB & Size of $\hat{T}$ & Size of $\hat{T}$ in GB \\
\hline
4 & 2 & $K(1)=5$ &  $11^4$ & $\sim 0.00022$ & $43^4$ & $\sim 0.051$ \\
\hline 
5 & 2 & $K(1)=6$ &  $14^4$ & $\sim 0.0006$ & $70^4$ & $\sim 0.36$ \\
\hline
6 & 3 & $K(1)=K(2)=8$ &  $24^4$ & $\sim 0.005$ & $160^4$ & $\sim 9.77$ \\
\hline
7 & 3 & $K(2)=10$ & $30^4$ & $\sim 0.013$ & $246^4$ & $\sim 54.6$ \\
\hline
8 & 4 & $K(2)=13$ & $45^4$ & $\sim 0.062$ & $461^4$ & $\sim 673.1$ \\
\hline
9 & 4 & $K(2)=15$ & $55^4$ & $\sim 0.14$ & $671^4$ & $\sim 3021$ \\
\hline
10 & 5 & $K(2) = K(3) = 18$ & $76^4 $ & $\sim 0.5$ & $1112^4$ & $\sim 23000$ \\
\hline
\end{tabular}
\caption{\label{tab:tensor-size2}
Comparison of the size of 3-valent tensors $\hat{S}$ and 4-valent tensors $\hat{T}$ for various levels $k$ of the quantum group.}
\end{table}

Unsurprisingly the 3-valent tensors are far more economical in terms of memory usage compared to 4-valent ones, essentially because they are parametrized by only one superindex $K(J)$ per intertwiner label $J$ instead of two for 4-valent ones. However as beneficial as this fact may be one still requires a tensor network algorithm suited for 3-valent tensor which avoids higher valent intermediate tensors as much as possible. In the next section we present such an algorithm, called triangular algorithm, show that it uses significantly less memory than the original 4-valent one and discuss further numerical optimizations.

\section{Optimizing tensor network algorithms} \label{sec:optimization}

In this section we discuss the tensor network algorithm used to coarse grain $\text{SU}(2)_k \times \text{SU}(2)_k$ spin net models. Particular attention is given to the so--called triangular algorithm which can be understood as a modification of the familiar algorithms \cite{levin,gu-wen} using 3-valent tensors, denoted by $\hat{S}$, as its basic building block instead of 4-valent ones, called $\hat{T}$. Originally it was invented in \cite{dec-TNW} and already applied in \cite{matter-toy} yet it turns out to be indispensable for the models under discussion as motivated in the previous section. Furthermore we will also discuss improvements to the code itself which are recommended if one is dealing with tensors and matrices of the size mentioned before.

\subsection{Triangular algorithm}

As already discussed the step from the 4-valent algorithm \cite{levin,gu-wen} to a 3-valent one is almost directly built into the 4-valent algorithm. At an intermediate step each 4-valent tensor is split into two 3-valent ones by a SVD. Then four of the latter are glued together to form a new 4-valent tensor, rotated by 45 degrees. However in the next iteration of the algorithm this new coarse tensor is again split into two coarse 3-valent tensors that one in principle could directly construct from two fine 3-valent ones. Thus the question arises whether one can instead work just with 3-valent tensors as the basic building blocks. To make this idea more clear, let us consider the lattice dual to the tensor network.

The lattice dual to a 4-valent (square) tensor network is again a square lattice. By splitting the 4-valent tensor into 3-valent ones each square is cut along its diagonal into two triangles, each dual to a 3-valent tensor. As such one obtains a regular triangulation of the square lattice. In the next step of the algorithm four of these triangles are glued to form a coarser square. During the next iteration this square is again split into two triangles, where the (coarse) cut is precisely along the lines along which the finer triangles were glued. In fact it appears that the new coarse triangles are made up out of two fine triangles plus a variable redefinition, mapping a `subdivided edge into a coarse edge'. These  mappings\footnote{The maps can be seen as coarse graining maps or if one considers the inverse maps, as embedding maps, which are crucial for the continuum limit \cite{dittcyl}.} are instrumental for defining the truncation scheme underlying the tensor network algorithm \cite{dittcyl}. This motivates the triangular algorithm \cite{dec-TNW,matter-toy}.

\begin{figure}
\includegraphics[scale=0.6]{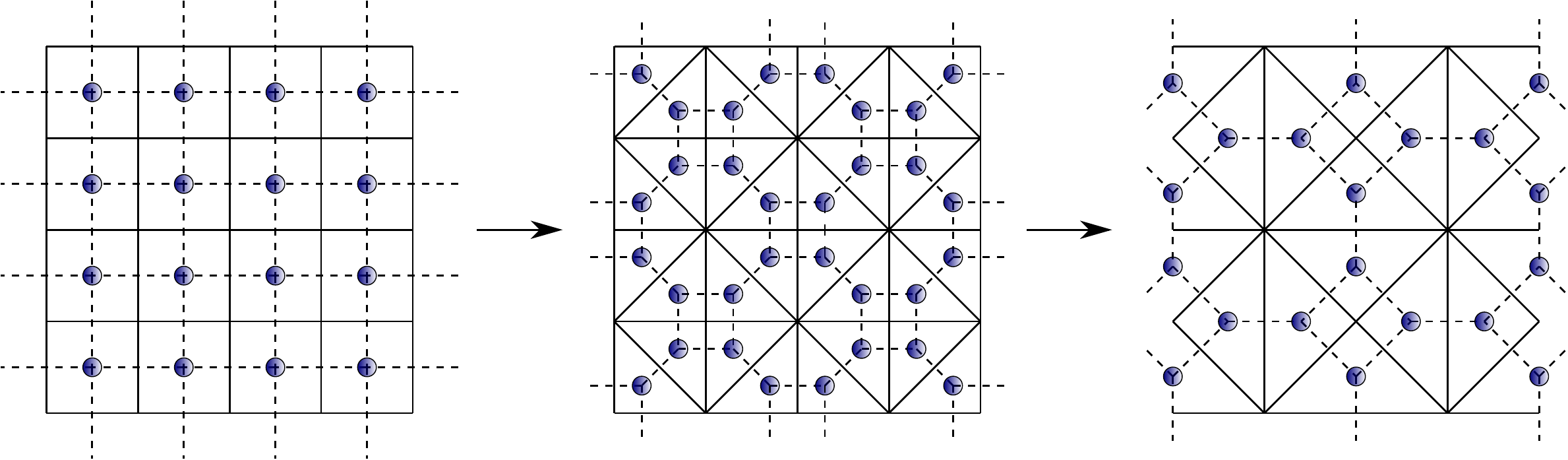}
\caption{\label{fig:tria-algo-gen}
Triangular algorithm for an extended network: First the square lattice is split into a regular triangulation as in the original algorithm \cite{levin,gu-wen}. Instead of defining new coarse squares / 4-valent tensors one combines two triangles into a coarse triangle, thus working with 3-valent tensors instead.
}
\end{figure}

Before we discuss this algorithm in detail note that it is defined for a square lattice split into a triangulation as described above. This is necessary in order to straightforwardly iterate the algorithm.

The triangular algorithm is demonstrated for a larger network in fig. \ref{fig:tria-algo-gen}. The first step is therefore to turn the 4-valent tensor network into a 3-valent one. In principle this can be done by performing the first step of the usual formalism, but it should rather be avoided if one has to deal with tensors of the size illustrated in section \ref{sec:scope}. Fortunately it is possible to analytically define the 3-valent tensors for many models including the spin nets under discussion; the Ising model is another example that is straightforward to split into 3-valent tensors.

\begin{figure}
\includegraphics[scale=0.6]{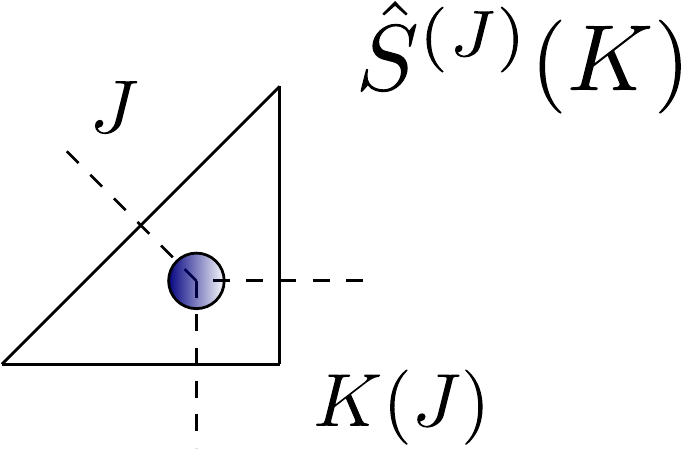}
\caption{\label{fig:tria-tensor}
Labels of triangular tensor: The triangular tensor essentially inherits its block diagonal form from the 4-valent one. The new coarse edge carries the reference spin $J$, the fine spins are combined into a super-index $K(J)$.}
\end{figure}

So assume the triangulation of the square lattice, or its dual tensor network of 3-valent tensors $\hat{S}$, is given, where $\hat{S}$ depends on the following variables:
\begin{equation}
\hat{S}^{\{J\}}(\{j_a(K_a(J))\}) \quad .
\end{equation}
The notation is purposely similar to the 4-valent tensor $\hat{T}$. On the one hand the spins $\{J\}$ label again the projector basis and also serve as the label for the superindices $\{K_a(J)\}$ summarizing the (non-vanishing) configurations of fine spin $\{j_a\}$. On the other hand $\{J\}$ actually represent a variable in the model attached to one edge of the 3-valent tensor / triangle. In fact this edge is distinguished both in the definition of $\hat{S}$ and in the lattice / network, as it is a coarser (that is by a factor of $\sqrt{2}$ longer) edge. This is also illustrated in fig. \ref{fig:tria-tensor}.

Given this triangulation it is straightforward to identify a coarse triangle made up out of two fine triangles glued along a fine edge. From the tensor network perspective one edge between two 3-valent tensor gets contracted resulting in a new effective 4-valent tensor describing a coarse triangle with a subdivided edge. Of course this immediately raises the question whether one runs into the same issue as the original algorithm of having to store an entire 4-valent tensor. Fortunately this problem can be nicely circumvented.

\begin{figure}
\includegraphics[scale=0.6]{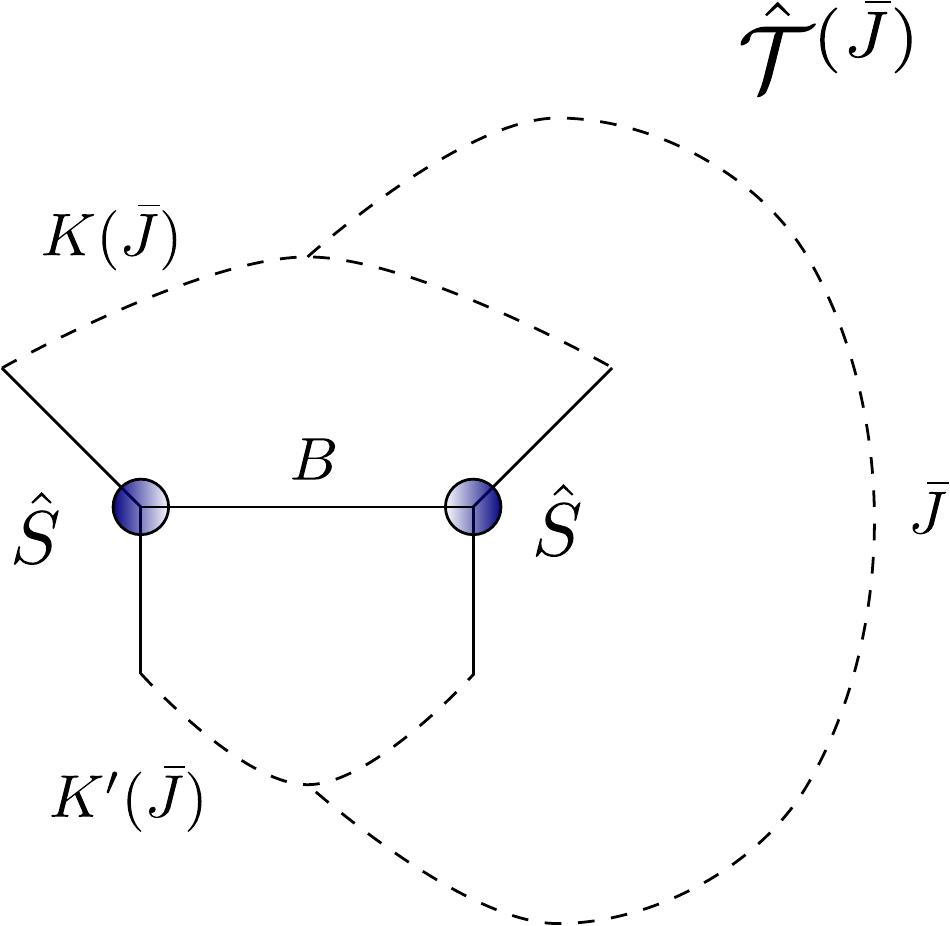}
\caption{\label{fig:int-tensor}
Intermediate 4-valent tensor $\hat{\mathcal{T}}$: Two 3-valent tensor are contracted to form an intermediate 4-valent tensor. It is directly defined in block-diagonal form labelled by $\bar{J}$ and indicated by the dashed recoupling lines. The combinatorics of this diagram exactly match the $\{ 6j \}$ symbol, such that one only sums over non-vanishing configurations labelled by the super-index $B$. 
}
\end{figure}

Recall that both the 3-valent $\hat{S}$ and the 4-valent tensor $\hat{T}$ were expressed in a so-called intertwiner basis, such that they are labelled by spins $\{J\}$ (denoting the basis and superindices $\{K(J)\}$). The same expansion can be applied to the new 4-valent tensor $\hat{\mathcal{T}}$ obtained from two 3-valent ones, see also fig. \ref{fig:int-tensor}:
\begin{align} \label{eq:int-4}
& \hat{\mathcal{T}}^{\{\bar{J}\}}(\{J_a(K_a(\bar{J}))\},\{j_b(K_b(\bar{J})\}) \nonumber \\
 = & \sum_{\{B(\bar{J},K_a,K_b)\}} S^{\{J_a(K_a(\bar{J})\})_1} (\{j_b(K_b(\bar{J})\}_1,\{j(B)\}) \; S^{\{J_a(K_a(\bar{J})\})_2} (\{j(B)\},\{j_b(K_b(\bar{J})\}_2) \times f(J,K_a,K_b,B) \quad .
\end{align} 
For the sake of simplicity we combine all $\text{SU}(2)_k$ specific expressions, that is recoupling symbols, quantum dimension factors, etc., in the function $f$, which is also expressed entirely as a function of $\{\bar{J},K_a,K_b,B\}$. See appendix \ref{app:formula} for the complete formula. Crucially the tensor $\hat{\mathcal{T}}$ is directly expressed in its new intertwiner basis labelled by new indices $\{\bar{J}\}$. Consequently the old indices $J$ of $\hat{S}$ get expressed in terms of the superindex $K_a(\bar{J})$, the fine (uncontracted) $j$ in terms of $K_b(\bar{J})$ and the fine contracted $j$ in terms of $B(\bar{J},K_a,K_b)$. Again this insures that one only sums over representations allowed by the coupling rules.

As we have discussed in section \ref{sec:scope}, storing the entire tensor $\hat{T}$ is a costly endeavour. Fortunately this is not necessary in this case: Recall that our goal is to construct a new 3-valent tensor $\hat{S}'$ from $\hat{\mathcal{T}}$, which then again serves as the starting point for the next iteration of the algorithm. To do so we have to combine the two indices of the subdivided coarse edge, which are labelled by the superindices $\{K_b\}$, into one effective index via a variable transformation, which is computed with a SVD. The form of $\hat{\mathcal{T}}^{\{\bar{J}\}}$ is precisely chosen with this goal in mind, as we intend to preserve the coarse edges labelled by $J_a(K_a)$ and (variable) transform the fine ones labelled by $j_b(K_b)$.

Thus we apply a SVD to:
\begin{equation} \label{eq:svd}
\hat{\mathcal{T}}^{\{\bar{J}\}}_{\{K_a\};\{K_b\}} =: M^{\{\bar{J}\}}_{ab} = \sum_i U^{\{\bar{J}\}}_{ai} \lambda_i \left( V^{\{\bar{J}\}}_{bi} \right)^\dagger \quad ,
\end{equation}
where $U_{ai}$ and $V_{bi}$ denote the $i$-th left- and right singular vectors of $M_{ab}$ respectively, and $\lambda_1 \geq \lambda_2 \geq ... \geq 0$ are the respective singular values ordered in size. The sum over the index $i$ counting the singular values runs over the full range of superindices $\{K(J)\}$.

The last step to obtain the new coarse tensor $\hat{S}'$ is the contraction of the indices $\{j_b(K_b(\bar{J}))\}$ with the singular vectors $V^{\{\bar{J}\}}_{\{j_b\},i}$:
\begin{equation} \label{eq:simple-3}
(\hat{S}')^{\{\bar{J},i\}}(\{J_a(K_a(\bar{J}))\}) = \sum_{\{j_b(K_b)\}} \hat{\mathcal{T}}^{\{\bar{J}\}}_{\{K_a\};\{K_b\}} V^{\{\bar{J}\}}_{\{j_b\},i} = U^{\{\bar{J}\}}_{\{J_a(K_a)\},i} \lambda_i \quad .
\end{equation}
The last identity follows from the fact that the matrices of singular vectors $U$, $V$ are unitary. For equation \eqref{eq:simple-3} to be valid, one has to insert a resolution of the identity $V V^\dagger$ into the tensor network in order not to change the partition function. As a consequence the tensors opposite of $\hat{\mathcal{T}}$ get contracted with $V^\dagger$, however there the second identity in \eqref{eq:simple-3} does not apply in general\footnote{To improve the algorithm one should actually perform the SVD for both pairs of tensors and compare, which map minimizes the error.}.

Having described the first iteration of the triangular algorithm, it is time to address the elephant in the room: the size of the intermediate tensor $\hat{\mathcal{T}}$ and how one can avoid saving all of it in the triangular algorithm. Therefore we would like to draw the readers attention to equations \eqref{eq:int-4}, \eqref{eq:svd} and \eqref{eq:simple-3}: the intertwiner basis $\{\bar{J}\}$ puts both the tensors $\hat{\mathcal{T}}$ and $\hat{S}$ into the same block diagonal form, as the new coarsest edge of $\hat{S}$ inherits the labels from $\hat{\mathcal{T}}$. Therefore, in order to compute one block $\{\bar{J}\}$ of $\hat{S}$ one only has to know the block $\{\bar{J}\}$ of $\hat{\mathcal{T}}$. As a result one can compute $\hat{S}'$ block by block, for which one only has to compute and store the respective block of $\hat{\mathcal{T}}$. Thus, as we can see from tables \ref{tab:tensor-size} and \ref{tab:tensor-size2}, the main limiting factor in the triangular algorithm is actually  the size of the largest block $\hat{\mathcal{T}}^{\{\bar{J}\}}$, which is roughly two orders of magnitude smaller than the full 4-valent tensor for any level $k$ of the quantum group.

This concludes the principle discussion of the 3-valent algorithm, in which we paid particular attention to the (avoidance of the) memory problem. In order to iterate the code one necessarily has to implement a truncation scheme after the SVD, as in any other tensor network algorithm. This is the subject of the next section. Furthermore we explain and justify the simplifications made in the algorithm.

\subsection{Truncation scheme and simplifications}

Any numerical tensor network algorithm requires a truncation scheme after the SVD has been performed. If we consider equation \eqref{eq:simple-3} again, we realize that the coarse 3-valent tensor $\hat{S}'$ comes with an additional label $i$ attached to $\{\bar{J}\}$ enumerating the singular values from the previous iteration in this block. Interpretation wise it tells us that $\{\bar{J}\}$ appears with a certain multiplicity, which have to be stored again. In the following iterations these index ranges keep growing exponentially such that one eventually is forced to truncate. However this should be done such that the error is as small as possible.

The decomposition of a tensor (rearranged as a matrix) by a SVD is optimal for that, as truncating the singular values to the largest $\chi$ values gives the best approximation (in terms of the least square error) of this matrix by another matrix of (lower) rank $\chi$. For the full tensor, that is all blocks $\{J\}$, one should compute all of the singular values for each block, compare them and take the largest $\chi$ of those. The approximation is improved if $\chi$ is increased.

As straightforward as this idea is it is rather cumbersome to realize in the context of this work. Due to the size of 4-valent tensors in this model, one cannot simply compute the SVD of all blocks one after another and store them, as the full $U$ and $V$ each take up as much memory as the original $\hat{\mathcal{T}}$. Moreover computing the full SVD, i.e. all singular values and vectors, of matrices of the size shown in table \ref{tab:tensor-size} (s. column `Size of largest block') is very costly.
 Instead one would have to compute one block $\{J\}$ of $\hat{\mathcal{T}}$, compute  only its singular values and store them. Then one deletes the current block and continues with the next one. Once all the singular values have been computed they are compared and the largest $\chi$ singular values are taken over. Afterwards one computes $\hat{\mathcal{T}}$ (block by block) again and computes the SVD only for the amount of singular values taken over.

Furthermore note that increasing the number of singular values taken over directly affects the sizes of the 3-valent and 4-valent tensors (and matrices) in the next and next-to-next iteration due to the asymmetry of the 3-valent tensors. Thus one may yet again run into memory issues. It has been noted in \cite{eckert1} that it is best advised not to lower the number of singular values in one block in the following iterations.

Instead of this elaborate truncation scheme we use the same simple one as in \cite{q-spinnet}, namely we take over only one singular value per block $\{J\}$. Even though it may appear to be very low at first sight, it actually does take many singular values into account: If we consult table \ref{tab:tensor-size} again we realize that for $k=8$, which is the largest model we have studied, we already have 625 blocks. Experience from previous models \cite{q-spinnet} showed that for many models this scheme is already suitable for exploring the phase structure of the model. Nevertheless for one model discussed in this article this approximation scheme clearly breaks down.

Concerning the triangular algorithm a further comment is in order: In the original tensor network algorithm \cite{levin,gu-wen} one usually gets four different 3-valent tensors from splitting the 4-valent tensor in two ways. Therefore in full generality the triangular algorithm is formulated for four 3-valent tensors, where each of these four gets renormalized. Fortunately in the model under discussion, as well as the Ising model \cite{dec-TNW}, one finds out that all four 3-valent tensors are identical. This also extends to the fact that equation \eqref{eq:simple-3} applies to all 3-valent tensors $\hat{S}$. Thus it is sufficient to perform the 3-valent algorithm with just one tensor $\hat{S}$, which slightly reduces memory usage but more importantly saves a lot of computational time, as less SVDs and summations have to be performed.

This concludes the discussion of the triangular algorithm. In the next section we briefly discuss several improvements of the code and parallelisation.

\subsection{Code improvements and parallelisation}

The triangular algorithm describes the current state as a
block-structured matrix, i.e. as a matrix consisting of dense blocks
of varying sizes, where some blocks are known to be identically zero
due to symmetry, and certain blocks are mere transpositions of others.
It goes without saying that one needs to make use of this structure to
improve performance and reduce memory requirements.

The operations described above implementing the renormalization flow
define a new block-structured matrix in terms of an existing one via
tensor contractions. This creates intermediate objects that can be
significantly larger than both the initial and the final state. It is
thus crucial to perform the tensor contractions in an optimal order to
reduce the amount of memory required for intermediate states.

As modern workstations have multiple cores, it is necessary to find
parallelism for good performance. The triangular algorithm can be
parallelized in two ways. First, each tensor operation (i.e. matrix
multiplication or singular value decomposition) can be executed in as
parallel operation. This is worthwhile only for sufficiently large
blocks, with more than about $20^2$ elements on a modern workstation.
Second, the blocks making up the resulting tensor can all be evaluated
simultaneously. We found the latter to lead to the best performance,
because while there are some large blocks, most blocks are too small
to be parallelized by themselves. We implemented this both via a
shared memory OpenMP \cite{openmp} parallelization with dynamically
scheduled loop in a C++ code, as well as via a distributed memory
parallelization in a Julia code \cite{julia}.\footnote{Readers interested in the codes can request them by contacting the authors.}

%

\section{Summary of triangular algorithm}

Before we continue to discuss the models studied in this article in more detail, in particular from the perspective of spin foam quantum gravity, we would like to summarize and conclude this part of the article about the employed tensor network algorithm. In section \ref{sec:scope} we have illustrated in detail that for models equipped with a large symmetry group, like $\text{SU}(2)_k \times \text{SU}(2)_k$ in our case, the 4-valent algorithm \cite{levin,gu-wen} is quickly limited by memory on modern machines. This even holds when one exploits the symmetries of the model as in \cite{s3-spinnet,q-spinnet}. By a shift of perspective from 4-valent to 3-valent tensors we have remedied this issue and invented the triangular algorithm \cite{dec-TNW,matter-toy}, which is only limited by the size of the largest block (in terms of the symmetries) instead of the size of the entire tensor. Indeed in our context the triangular algorithm allowed us to go to much larger levels $k$ of the quantum group.

In addition to that the shift to smaller fundamental building blocks should have more potential for other practitioners of tensor network algorithms, also for `smaller' symmetry groups. Since the triangular algorithm is more economical than its 4-valent counterpart it should be possible to further increase the bond dimension of the model and thus improve the approximation or lower the computational costs at the same level of accuracy. Moreover it might be worthwhile to modify the entanglement filtering algorithm \cite{vidal-evenbly} to the triangular case. 

This concludes the part of this article focussing on the technical and numerical aspects of this work. In the next sections we go into more details of the models under discussion, in particular establishing the relation to modern spin foam models and presenting the results of applying the before mentioned triangular algorithm to these models.

\newpage
\begin{center}{\Large  PART II: RESULTS}\end{center}

\section{Intertwiner and Spin net models} \label{sec:models}

After focussing on the technical and numerical challenges one faces when renormalizing quantum group spin net models, the rest of this article focusses more on the technical details of said models and their relation to spin foam models. This entails a brief introduction to and motivation of the models, including a discussion of intertwiner models \cite{wojtek}, a detailed construction of spin net models mimicking modern spin foam models and presenting the results from coarse graining these models. 

Spin net models are defined on a lattice of arbitrary dimension\footnote{This lattice need not be regular for tensor network methods to work. However regular lattices result in regular tensor networks, whose numerical coarse graining can be straightforwardly iterated.}. To each vertex one assigns a group element $g_v \in G$, e.g. a finite or a (compact) Lie group, and weight functions $\omega_e : G \rightarrow \mathbb{C}$. For concreteness and similarity to the quantum group case we assume $G$ to be finite. The edges of the lattice come with an orientation and the associated weights are evaluated on the product of group elements at the `source' and `target' of the edge, i.e. $\omega_e(g_{s(e)} g_{t(e)}^{-1})$, where $s(e)$ / $t(e)$ denote source / target vertex of $e$ respectively. Crucially the edge weights are invariant under conjugation, i.e. $\omega_e(h g h^{-1}) = \omega_e(g) \, \forall h \in G$. Thus the system possesses a global symmetry, as it is invariant under (left and right) multiplying the same group element to all group elements on the vertices. To conclude, the partition function of this system is given by
\begin{equation} \label{eq:partition}
Z = \frac{1}{|G|^E} \sum_{\{ g_v \}} \prod_e \omega_e( g_{s(e)} g_{t(e)}^{-1}) \quad .
\end{equation}
By $|G|$ we denote the number of group elements of $G$ and $E$ is the number of edges in the lattice. The simplest non-trivial model is the $\mathbb{Z}_2$ Ising model (for vanishing external magnetic field).

A `dual' description of spin net models can be derived from \eqref{eq:partition} by a group Fourier transform, as it can be found in \cite{savit,sffinite}. To do so one expands the class functions $\omega_e(g)$ into characters $\chi_\rho(g)$ of irreducible representations $\rho$ of the group $G$. As the characters factorise over group elements (into representations) so does \eqref{eq:partition}, such that the group summation can be performed for each vertex individually. Eventually one obtains a group theoretic object, the Haar projector $\mathcal{P}_v$\footnote{$\mathcal{P}_v : 
V_{\rho_1} \otimes ... \otimes V_{\rho_n}
\rightarrow \text{Inv}(V_{\rho_1} \otimes ... \otimes V_{\rho_n})$ is the projector onto the invariant subspace of the tensor product of representation vector spaces $V_\rho$. It can also be seen as a sum over an orthonormal basis of intertwiners $| \iota_d \rangle \in \text{Inv}(V_{\rho_1} \otimes ... \otimes V_{\rho_n})$.}, at each vertex of the lattice, such that \eqref{eq:partition} is rewritten as:
\begin{equation} \label{eq:partition2}
Z = \frac{1}{|G|^E} \sum_{\{\rho_e, m_e, n_e \}} \prod_e \tilde{\omega}_{\rho_e} \prod_v \mathcal{P}_v (\{\rho_e\}_{e \supset v}\})^{\{m_e\}}_{\{n_e\}} = \sum_{\{\rho_e, m_e, n_e \}} \prod_e \tilde{\omega}_{\rho_e} \prod_v \sum_{| \iota_d \rangle} {}^{\{m_e\}}| \iota_d \rangle \langle \iota_d |_{\{n_e\}}  \quad . 
\end{equation}
$\tilde{\omega}_{\rho_e}$ denotes the Fourier transformed edge weight. The indices of $\mathcal{P}_v$ are contracted with the indices of the other projectors according to the connectivity of the graph, such that each edge essentially carries the Hilbert space $\mathcal{H}_e = \bigoplus_{\rho} V_\rho \otimes V_{\rho^*}$, where $\rho^*$ denotes the dual representation to $\rho$. $\mathcal{P}$ itself is a projector satisfying $\mathcal{P} \cdot \mathcal{P}= \mathcal{P}$. More information and details on the derivation of these expressions can be found in \cite{sffinite}.

In order to apply tensor network algorithms, the partition function \eqref{eq:partition2} needs to be rewritten as a contraction of tensors, i.e. as a sum over tensor indices. This implies that we must assign all amplitudes to the vertices and variables to the edges\footnote{In general there exist many choices on how to rewrite a partition function as a tensor network. In fact the tensor network can be different from the underlying discretisation.}. As the projectors are already assigned to the vertices, it remains to split the edge weights $\omega_e$ by assigning $\sqrt{\omega_e}$ to source and target vertex of each edge $e$. Hence we define the following tensor $T$:
\begin{equation} \label{eq:org-tensor}
T(\{\rho\},\{m\},\{n\}) 
:= \frac{1}{|G|^2} \prod_{e \supset v} \sqrt{\tilde{\omega}_{\rho_e}} \, \mathcal{P}_v (\{\rho_e\}_{e \supset v}\})^{\{m_e\}}_{\{n_e\}} = \prod_{e \supset v} \sqrt{\tilde{\omega}_{\rho_e}}  \sum_{| \iota_d \rangle} {}^{\{m_e\}}| \iota_d \rangle \langle \iota_d |_{\{n_e\}}  \quad .
\end{equation}	
Given this $T$, the sums over irreducible representations and magnetic indices in \eqref{eq:partition2} are expressed as the contraction of tensor indices according to the combinatorics of the network, called the tensor trace $\text{Ttr}$:
\begin{equation} \label{eq:tensor-trace}
Z = \text{Ttr} \; T \dots T \quad .
\end{equation}

As already briefly discussed in section \ref{sec:scope} the representation \eqref{eq:partition} is not available for quantum groups, as these are not groups but Hopf algebras. Here we understand the quantum group $\text{SU}(2)_k$ as the $q$-deformation of the universal enveloping algebra $\mathcal{U}(\text{SU}(2))$ with $q = \exp(\frac{i \pi}{k+2})$ a root of unity \cite{biedenharn,yellowbook}. $k$ denotes the level of said quantum group. Since the representation theory of $\text{SU}(2)_k$ is well understood and actually very similar to the one of $\text{SU}(2)$, we take representation \eqref{eq:partition2} as our starting point. The most notable difference to the undeformed case is the cut-off in the spins $j_{\text{max}} = \frac{k}{2}$, which depends on the level $k$. Details on the derivation of the Haar projector for $\text{SU}(2)_k$ can be found in \cite{q-spinnet}.

At this stage we would like to briefly comment on the relation of spin net models and spin foams: the dynamical ingredients of spin foams are very similar to spin nets, however there are two major differences. First, group elements are assigned to edges instead of vertices and weights are assigned to faces. In representation \eqref{eq:partition2}, spin foams carry representations on the faces and intertwiners on the edges of the foam. The second difference is that spin foams possess a local gauge symmetry instead of a global one. As a result the partition functions of spin foams and spin nets are very similar in form. Another difference is the chosen dimension of the systems. Spin foam models are usually describing 4D spacetimes, whereas we study here 2D spin net models. There are several reasons for this choice. It is a known feature that 4D lattice gauge theories and 2D spin systems share certain statistical properties and have similar phase structures \cite{Kogut}. In addition to that 2D spin net models on a square lattice feature 4-valent projectors on their vertices as do 4D spin foam models defined on (the dual of a) triangulation.

Among these reasons, the similarity of dynamical ingredients was the main motivation of construction spin net models \cite{eckert1,s3-spinnet,q-spinnet} as it allows for capturing a key dynamical ingredient of spin foams, the so--called simplicity constraints. These simplicity constraints appear in the Plebansky formulation of general relativity \cite{plebanski} and break the symmetries of topological BF theory to obtain a theory with propagating degrees of freedom. Spin foam models take this Plebanski action as their starting point, but first discretise and quantize the topological theory. Discretized versions of the simplicity constraints are then imposed at the quantum level and are expected to result in propagating degrees of freedom. However this construction is not unique and the cause of different spin foam models, see again \cite{perezreview} for recent review. Due to their (dynamical) similarities we hope  that spin nets can serve as analogue models for spin foams and that the phases and possibly some key features of the coarse graining flow agree in these models. Due to their simpler structure spin nets allow the tracking of the simplicity constraints during coarse graining. Besides spin nets are also useful in developing coarse graining techniques that might be also applicable to spin foams.

In fact spin nets can even be identified with spin foams defined on a very special underlying discretization, known as melon spin foam \cite{q-spinnet}: Such a melon consists of two spin foam vertices glued together via many spin foam edges. The spin net coarse graining corresponds to a bundling of a number of spin foam edges into `thicker' edges. The simplicity constraints determine also how the two spin foam vertices are glued to each other. For instance later--on we will encounter so--called factorising spin net models, which translate to factorising, and hence unglued, spin foam vertices.


As mentioned simplicity constraints are of particular interest for the dynamics of spin foams. At the core of the construction of modern spin foam models is the insight of Plebanski that the Palatini action (a first order formulation of the Einstein-Hilbert action) can be written as a constrained topological field theory \cite{plebanski,perezreview}. Since it is understood how to discretize and quantize the unconstrained topological theory, known as BF theory, many modern spin foam models use it as the starting point \cite{Barrett-Crane,eprl2,fk,perezreview}. In order to obtain propagating degrees of freedom (a version of) the simplicity constraints are implemented at the discrete quantum level. As it is generically the case for discretizations this procedure is not unique and the root of differences among modern spin foam models. In more detail the simplicity constraints affect the projectors $\mathcal{P}$ onto the invariant subspace, e.g. in \eqref{eq:partition2}, by forbidding certain intertwiners $| \iota \rangle$ such that $\mathcal{P}$ projects only onto a subspace. In this regard spin foam models can be seen as extensions of lattice gauge theories \cite{holonomy1}.

Despite these difference in the construction of spin foam models they are remarkably in agreement in the semi-classical limit for one vertex amplitude, i.e. the amplitude assigned to a 4-simplex, which is given by the cosine of the Regge action \cite{regge}, a discretisation of general relativity on a triangulation, of this simplex \cite{baez10j,freidel-6j-10j,Conrady-Freidel,FrankEPRL}. While this is an encouraging result examinations for larger 2-complexes are scarce and the dynamics of spin foams is not well understood as well. Concerning the simplicity constraints this is a crucial issue to tackle as they play the crucial role of implementing the dynamics. Whether this dynamics is non-trivial, in the sense of describing propagating degrees of freedom, and furthermore compatible with general relativity in an appropriate limit is on open question. In turn a better understanding of the dynamics can lead to a improved construction of the models.

Thus progress towards this goal would already be achieved by examining the effect of the simplicity constraints if the foam consists of more than one building block. This is the question of how the subspaces the constraints project on change under coarse graining, i.e. how the constraints act effectively on a coarser scale. Due to their dynamical similarity to spin foams, yet for a simpler dynamics, we can address these questions in spin net models and subject them to a real space renormalization procedure. The purpose of this article is therefore to construct spin net models mimicking the properties of two 4D spin foam models, namely the Barrett-Crane (BC) \cite{Barrett-Crane} and the Engle-Pereira-Rovelli-Livine / Freidel-Krasnov (EPRL/FK) \cite{eprl2,fk} model, and coarse grain them via tensor network renormalization.

To this end the choice of the proper symmetry group is crucial, where studying spin nets for the quantum group $\text{SU}(2)_k \times \text{SU}(2)_k$ actually kills two birds with one stone. On the one hand one requires a cut-off on the (irreducible) representations in order to apply tensor network renormalization. On the other hand quantum groups also provide us with a physical motivation from quantum gravity as it is conjectured that spin foam models for quantum groups model gravity with a cosmological constant $\Lambda \neq 0$ \cite{qgroupmodels,qgroupmodels1,qgroupmodels2,qgroupmodels3,
catherine,muxin,qgroupmodels6,lqg-lambda2}. This insight stems from the Turaev-Viro model \cite{turaev-viro}, a spin foam model for discretised Euclidean quantum gravity in 3D with a positive cosmological constant. Its basic amplitude is precisely the $\text{SU}(2)_k$ $6j$-symbol describing a constantly (positively) curved quantum tetrahedron describing the case $\Lambda > 0$. In this setting the maximum spin $j_{\text{max}}$ can be interpreted as an infrared cut-off via the cosmological constant with $\Lambda \sim 1/j_{\text{max}}$. The small value of $\Lambda$ in current observations would thus suggest a large level $k$ of the quantum group. Extensions of these ideas to 4D spin foam models are a topical field of research \cite{catherine,muxin,Khavkine} and have recently uncovered an interesting connection to Chern-Simons theory \cite{cs-sf}. Moreover,  canonical loop quantum gravity frameworks, implementing a cosmological constant, can also be constructed, thus providing the boundary Hilbert spaces for these models \cite{lqg-lambda1,lqg-lambda2,ditt-ext-tqft,3D-4D-TQFT}.

After this motivation of quantum group spin nets let us return to the discussion of the model itself. As discussed above, the partition function is of the general form \eqref{eq:partition2} and can be written as a tensor network with \eqref{eq:org-tensor} and \eqref{eq:tensor-trace}. With $\text{SU}(2)_k \times \text{SU}(2)_k$ as the underlying quantum group irreducible representations are labelled by two irreducible $\text{SU}(2)_k$ representations $j^+$ and $j^-$, consequently the edge Hilbert space is $\mathcal{H}_e = \bigoplus_{j^+,j^-} V_{j^+} \otimes V_{j^-} \otimes V_{{j^+}^*} \otimes V_{{j^-}^*}$. Due to the quantum group symmetry encoded in the projectors the tensors $T$ are of a very specific form derived in \cite{q-spinnet} and alluded to in sections \ref{sec:scope} and \ref{sec:optimization}:
\begin{align} \label{eq:recoupling-basis}
& T(\{j^+\},\{j^-\},\{m^+\},\{m^-\},\{n^+\},\{n^-\}) = \nonumber \\
= & \sum_{J^\pm,(J^\pm)'} \hat{T}^{(J^\pm,(J^\pm)'}(\{j^+\},\{j^-\}) \; d_{J^+} d_{J^-} d_{(J^+)'} d_{(J^-)'}
\begin{tikzpicture}[baseline,scale=0.85]
\draw (-0.5,-1) -- (0,-0.5) -- (0,0.5) -- (-0.5,1)
      (0.5,-1) -- (0,-0.5)
      (0,0.5) -- (0.5,1)
      (-0.5,-1.25) node {$j^+_3$}
      (0.5,-1.25) node {$j^+_4$}
      (0.5,1.25) node {$j^+_1$}
      (-0.5,1.25) node {$j^+_2$}
      (0.4,0) node {$J^+$};
\end{tikzpicture}
\; \otimes \;
\begin{tikzpicture}[baseline,scale=0.85]
\draw (-0.5,-1) -- (0,-0.5) -- (0,0.5) -- (-0.5,1)
      (0.5,-1) -- (0,-0.5)
      (0,0.5) -- (0.5,1)
      (-0.5,-1.25) node {$j^-_3$}
      (0.5,-1.25) node {$j^-_4$}
      (0.5,1.25) node {$j^-_1$}
      (-0.5,1.25) node {$j^-_2$}
      (0.4,0) node {$J^-$};
\end{tikzpicture}
\; \otimes \;
\begin{tikzpicture}[baseline,scale=0.85]
\draw (-0.5,0.75) -- (-0.25,1) arc(180:0:0.5) -- (0.75,-1) arc(0:-180:0.5) -- (-0.5,-0.75)
      (-0.25,1) -- (0,0.75)
      (-0.25,-1) -- (0,-0.75)
      (-0,0.5) node {$j^+_1$}
      (-0.5,0.5) node {$j^+_2$}
      (-0,-0.5) node {$j^+_4$}
      (-0.5,-0.5) node {$j^+_3$}
      (1.25,0) node {$(J^+)'$};
\end{tikzpicture}
\; \otimes \;
\begin{tikzpicture}[baseline,scale=0.85]
\draw (-0.5,0.75) -- (-0.25,1) arc(0:180:0.5) -- (-1.25,-1) arc(-180:0:0.5) -- (-0.5,-0.75)
      (-0.25,1) -- (0,0.75)
      (-0.25,-1) -- (0,-0.75)
      (-0,0.5) node {$j^-_1$}
      (-0.5,0.5) node {$j^-_2$}
      (-0,-0.5) node {$j^-_4$}
      (-0.5,-0.5) node {$j^-_3$}
      (-1.75,0) node {$(J^-)'$};
\end{tikzpicture}
\quad .
\end{align}
This equation expresses the change of basis, namely to the recoupling basis, at the heart of the tensor network algorithm. $d_J$ denotes the quantum dimension $[2J+1]$ of the representation $J$, with $[n]$ being the quantum number of $n$ (see appendix \ref{app:q-basics}). $\hat{T}$, which is a function solely of representation labels, is the tensor expressed in the new, block-diagonal basis. The graphical expressions denote the Haar projector of $\text{SU}(2)_k \times \text{SU}(2)_k$ (modulo dimension factors and signs) and encode the dependence on the magnetic indices. Essentially each trivalent vertex is dual to a Clebsch-Gordan coefficient ${}_q \mathcal{C}^{\{j\}}_{\{m\}}$, where the first two diagrams encode the indices $\{m^\pm\}$ and the latter two encode $\{n^\pm\}$. A short explanation on this graphical calculus can be found in appendix \ref{app:graph}, see also \cite{q-spinnet} for more details.

The crucial point about identity \eqref{eq:recoupling-basis} is that any spin net model equipped with the quantum group symmetry can be written in this form. Thus the different methods and concepts of implementing simplicity constraints will result in different tensors $\hat{T}$. Moreover the symmetry structure of the model is precisely preserved under coarse graining, such that we can directly examine the renormalization group flow of $\hat{T}$, which one could then interpret as the `flow' of the simplicity constraints. Again we would also like to recall that the basis \eqref{eq:recoupling-basis} is crucial for developing and optimizing the algorithm such that it can be applied to larger levels $k$ of the quantum group, which is the subject of sections \ref{sec:scope} and \ref{sec:optimization}.

Before we go on and discuss in more detail the construction of BC- and EPRL- spin nets, we would like to also introduce $\text{SU}(2)_k \times \text{SU}(2)_k$ intertwiner models and give the reader a brief context to other models examined before.

\subsection{Intertwiner models}

Seen from spin net models, intertwiner models resemble simpler version with less degrees of freedom. Looking at the spin net for a arbitrary group \eqref{eq:partition2}, intertwiner models only have an edge Hilbert space $\mathcal{H}_e = \bigotimes_{\rho} V_\rho$, so the dual representations are missing. Thus the model is considerably simpler, in particular to numerically coarse grain, but of a similar form as \eqref{eq:recoupling-basis}:
\begin{equation}
t(\{j^+\},\{j^-\},\{m^+\},\{m^-\},
) = 
 \sum_{J^\pm,(J^\pm)'} \hat{t}^{(J^\pm)}(\{j^+\},\{j^-\}) \; d_{J^+} d_{J^-}
\begin{tikzpicture}[baseline,scale=0.85]
\draw (-0.5,-1) -- (0,-0.5) -- (0,0.5) -- (-0.5,1)
      (0.5,-1) -- (0,-0.5)
      (0,0.5) -- (0.5,1)
      (-0.5,-1.25) node {$j^+_3$}
      (0.5,-1.25) node {$j^+_4$}
      (0.5,1.25) node {$j^+_1$}
      (-0.5,1.25) node {$j^+_2$}
      (0.4,0) node {$J^+$};
\end{tikzpicture}
\; \otimes \;
\begin{tikzpicture}[baseline,scale=0.85]
\draw (-0.5,-1) -- (0,-0.5) -- (0,0.5) -- (-0.5,1)
      (0.5,-1) -- (0,-0.5)
      (0,0.5) -- (0.5,1)
      (-0.5,-1.25) node {$j^-_3$}
      (0.5,-1.25) node {$j^-_4$}
      (0.5,1.25) node {$j^-_1$}
      (-0.5,1.25) node {$j^-_2$}
      (0.4,0) node {$J^-$};
\end{tikzpicture}
\quad .
\end{equation}
Indeed these models can prove useful in better understanding the full spin net models, as some of the fixed points obtained via coarse graining actually factorise, that is the sets of representations decouple.

In previous work a similar behaviour has already been encountered: $\text{SU}(2)_k$ intertwiner models have been introduced in \cite{wojtek}, where several (families of) topological fixed points were derived by requiring triangulation independence. These topological fixed points were then used in \cite{q-spinnet} as initial data for $\text{SU}(2)_k$ spin nets and a rich phase structure with possibly second order phase transitions were found. Some of the fixed points describing the phases turned out to be factorising as the representation $j$ and its dual $j^*$ completely decoupled. Indeed taking the tensor product of fixed points of intertwiner models is also a fixed point of spin nets, as long as the representations $j$ and $j' \neq j^*$ are uncoupled. Note however, e.g. in \eqref{eq:recoupling-basis}, that in the initial spin net models $j'=j^*$. So one can interpret a spin net as two entangled, or rather interacting, intertwiner models\footnote{This is also the reason that taking topological fixed points of intertwiner models results in a flow of the spin net model under coarse graining.}. Pushing these analogies even further, one can interpret $\text{SU}(2)_k \times \text{SU}(2)_k$ spin nets as a tensor product of two interacting $\text{SU}(2)_k$ spin nets or as a tensor product of four interacting $\text{SU}(2)_k$ intertwiner models.

In the next section we will discuss the construction of two spin net models in detail, an analogue BC and an analogue EPRL model.

\section{BC and EPRL models} \label{sec:qg-models}

As we have discussed before most 4D spin foam models are built by imposing a version of the simplicity constraints on a discretised and quantised BF theory, with the goal to break the (too many) symmetries of the latter to obtain a theory with propagating degrees of freedom. In this article, we intend to construct analogue models by mimicking the 4D procedures. In order to keep this concise, we will focus on the two models which have been studied most thoroughly in the literature, namely the BC- and the EPRL-model, which differ significantly in their construction.

\subsection{The BC construction}

The Barrett-Crane (BC) model is one of the first 4D spin foam models that have been constructed, both for Euclidean \cite{Barrett-Crane} and Lorentzian signatures \cite{Barrett-Crane2}. As discussed before we will focus on the Euclidean version in this article. A BC spin foam model implementing a cosmological constant has been constructed in \cite{Khavkine}. This work also includes the Monte Carlo simulation of some observables, like relative frequencies of spin values.

Nowadays the BC model is disfavoured, for mainly two reasons \cite{eprl,perezreview,yasha-area}. On the one hand it suffers from metric discontinuities in the semi-classical limit due to non-matching shapes of tetrahedra along which 4-simplices are glued. On the other hand the BC model (intentionally) does not make contact with the (kinematical) Hilbert space of loop quantum gravity (LQG) (as a boundary Hilbert space), such that it cannot be used to define a physical inner product in LQG. In particular the latter point is overcome in the EPRL model. Indeed, a motivation of the EPRL model was the failure of the BC model to reproduce all properties of the continuum graviton propagator \cite{bc-prop}.

Despite these disadvantages the BC model is an interesting model to study due to its geometric construction and remarkable simplicity. Essentially it is constructed by describing a triangulated 4D Riemannian manifold by assigning bivectors to all triangles of the 4-simplices plus constraints. These are identified with Lie algebra elements and quantised by expressing them as group theoretic objects, namely by assigning $\text{SU}(2) \times \text{SU}(2)$ representations $(j^+,j^-)$ to the triangles and intertwining maps to the tetrahedra. Simplicity constraints are implemented by requiring the bivectors to be simple, which is translated to the representation labels $j^+ = j^-$\footnote{This condition is inferred from the classical condition on the bivectors stating that self-dual and anti-self-dual parts have the same norm.}. Representations $(j,j)$ are thus also called simple representations. 4-valent intertwiners assigned to tetrahedra are shown to be unique. They can be expanded into 3-valent ones, where one requires that the intermediate representations are simple again. This splitting is not unique, however all possible recoupling schemes are related to one another by fusion, where the fusion coefficients are given by $\{6j\}$ symbols. 
Following this general idea, we define the following spin net tensor implementing the BC conditions on the representations:
\begin{equation} \label{eq:BC-tensor}
T_{\text{BC}} (\{j\},\{m^\pm\},\{n^\pm\}) = c_{\{j_i\}} \; \;
\begin{tikzpicture}[baseline,scale=0.85]
\draw (-0.25,-1.5) arc(180:0:0.5);
\draw (-0.75,-1.5) arc(180:0:1);
\draw (0.75,1.5) arc(0:-180:0.5);
\draw (1.25,1.5) arc(0:-180:1);
\draw (0.25,0.25) node {$j_1$}
(0.25,1.5) node {$j_2$}
(0.25,-0.25) node {$j_4$}
(0.25,-1.5) node {$j_3$}
(-0.5,0) node {$m^-$}
(1.25,0) node {$m^+$};
\end{tikzpicture}
\; \otimes \;
\begin{tikzpicture}[baseline,scale=0.85]
\draw (-0.25,0.5) arc(180:0:0.5);
\draw (-0.75,0.5) arc(180:0:1);
\draw (0.75,-0.5) arc(0:-180:0.5);
\draw (1.25,-0.5) arc(0:-180:1);
\draw (0.25,0.5) node {$j_1$}
(0.25,1.75) node {$j_2$}
(0.25,-0.5) node {$j_4$}
(0.25,-1.75) node {$j_3$}
(-0.5,0) node {$n^+$}
(1.25,0) node {$n^-$};
\end{tikzpicture} \quad .
\end{equation}
The first diagram represents the representation $j_i$, the second one their duals $j_i^*$. The normalization constant $c_{\{j_i\}}$ will turn out to only depend on the quantum dimension of the representations $\{j_i\}$. In the graphical calculus the half-circle represents a Clebsch-Gordan coefficient for representations $(j_i^+,j_i^-)$ coupling to $j=0$, i.e. ${}_q \mathcal{C}^{j_i^+ \, j_i^- \, 0}_{m_i^+ \, m_i^- \, 0}$. This is only non-vanishing if $j_i^+ = j_i^- =: j_i$.

Note that the tensor $T_{\text{BC}}$ generically is not a projector onto the invariant subspace given by the basis in \eqref{eq:recoupling-basis}. In order to define the spin net one thus has to project it down onto the invariant subspace. To this end we contract both the sets of magnetic indices of $T_{\text{BC}}$ with the Haar projector $\mathcal{P}$ for $\text{SU}(2)_k \times \text{SU}(2)_k$:
\begin{align}
\mathcal{P} \, \circ \, T_{\text{BC}} \, \circ \, \mathcal{P} = & \sum_{\{J\}} c_{\{j_i\}} \; d_{J^+} d_{J^-} d_{(J^+)'} d_{(J^-)'} \quad
\underbrace{
\begin{tikzpicture}[baseline,scale=0.85]
\draw (-0.25,0.5) arc(180:0:0.5);
\draw (-0.75,0.5) arc(180:0:1);
\draw (0.75,-0.5) arc(0:-180:0.5);
\draw (1.25,-0.5) arc(0:-180:1);
\draw (-0.25,0.5) -- (-0.5,0.25) -- (-0.75,0.5);
\draw (-0.25,-0.5) -- (-0.5,-0.25) -- (-0.75,-0.5);
\draw (-0.5,-0.25) -- (-0.5,0.25);
\draw (0.75,-0.5) -- (1.,-0.25) -- (1.25,-0.5);
\draw (0.75,0.5) -- (1.,0.25) -- (1.25,0.5);
\draw (1.0,0.25) -- (1.0,-0.25);
\draw (0.25,0.5) node {$j_1$}
      (0.25,1.75) node {$j_2$}
      (0.25,-0.5) node {$j_4$}
      (0.25,-1.75) node {$j_3$}
      (-1.,0) node {$(J^+)'$}
      (1.5,0) node {$(J^-)'$};
\end{tikzpicture}
\quad
\begin{tikzpicture}[baseline,scale=0.85]
\draw (-0.25,-1.5) arc(180:0:0.5);
\draw (-0.75,-1.5) arc(180:0:1);
\draw (0.75,1.5) arc(0:-180:0.5);
\draw (1.25,1.5) arc(0:-180:1);
\draw (-0.25,-1.5) -- (-0.5,-1.75) -- (-0.75,-1.5);
\draw (1.25,1.5) -- (1.0,1.75) -- (0.75,1.5);
\draw (-0.25,1.5) -- (-0.5,1.75) -- (-0.75,1.5);
\draw (1.25,-1.5) -- (1.0,-1.75) -- (0.75,-1.5);
\draw (-0.5,-1.75) arc(0:-180:0.5) -- (-1.5,1.75) arc(180:0:0.5);
\draw (1.,-1.75) arc(-180:0:0.5) -- (2.,1.75) arc(0:180:0.5);
\draw (0.25,0.25) node {$j_1$}
      (0.25,1.5) node {$j_2$}
      (0.25,-0.25) node {$j_4$}
      (0.25,-1.5) node {$j_3$}
      (1.75,0) node {$J^+$}
      (-1.2,0) node {$J^-$};
\end{tikzpicture}
}_{= \frac{\delta_{J^+,J^-} \delta_{(J^+)',(J^-)'}}{d_J d_{J'}} \prod_{i=1}^4 \delta_{j_i^+,j_i^-}}
\; \nonumber \\
&
\begin{tikzpicture}[baseline,scale=0.85]
\draw (-0.5,-1) -- (0,-0.5) -- (0,0.5) -- (-0.5,1)
      (0.5,-1) -- (0,-0.5)
      (0,0.5) -- (0.5,1)
      (-0.5,-1.25) node {$j^+_3$}
      (0.5,-1.25) node {$j^+_4$}
      (0.5,1.25) node {$j^+_1$}
      (-0.5,1.25) node {$j^+_2$}
      (0.4,0) node {$J^+$};
\end{tikzpicture}
\; \otimes \;
\begin{tikzpicture}[baseline,scale=0.85]
\draw (-0.5,-1) -- (0,-0.5) -- (0,0.5) -- (-0.5,1)
      (0.5,-1) -- (0,-0.5)
      (0,0.5) -- (0.5,1)
      (-0.5,-1.25) node {$j^-_3$}
      (0.5,-1.25) node {$j^-_4$}
      (0.5,1.25) node {$j^-_1$}
      (-0.5,1.25) node {$j^-_2$}
      (0.4,0) node {$J^-$};
\end{tikzpicture}
\; \otimes \;
\begin{tikzpicture}[baseline,scale=0.85]
\draw (-0.5,0.75) -- (-0.25,1) arc(180:0:0.5) -- (0.75,-1) arc(0:-180:0.5) -- (-0.5,-0.75)
      (-0.25,1) -- (0,0.75)
      (-0.25,-1) -- (0,-0.75)
      (-0,0.5) node {$j^+_1$}
      (-0.5,0.5) node {$j^+_2$}
      (-0,-0.5) node {$j^+_4$}
      (-0.5,-0.5) node {$j^+_3$}
      (1.25,0) node {$(J^+)'$};
\end{tikzpicture}
\; \otimes \;
\begin{tikzpicture}[baseline,scale=0.85]
\draw (-0.5,0.75) -- (-0.25,1) arc(0:180:0.5) -- (-1.25,-1) arc(-180:0:0.5) -- (-0.5,-0.75)
      (-0.25,1) -- (0,0.75)
      (-0.25,-1) -- (0,-0.75)
      (-0,0.5) node {$j^-_1$}
      (-0.5,0.5) node {$j^-_2$}
      (-0,-0.5) node {$j^-_4$}
      (-0.5,-0.5) node {$j^-_3$}
      (-1.75,0) node {$(J^-)'$};
\end{tikzpicture} \quad .
\end{align}
The diagrams are straightforwardly calculated by using several identities and orthogonality relations of the Clebsch-Gordan coefficients \cite{biedenharn}. As it can be read off from the result $T_{\text{BC}}$ precisely implements the Barrett-Crane conditions on the representations.

The only component left to define $\hat{T}_{\text{BC}}$ is the normalization constant $c_{\{j_i\}}$.  This can be fixed by requiring the projector condition \cite{warsaw1}, see appendix \ref{app:BC-norm} for a derivation:
\begin{equation}
T_{\text{BC}} \circ T_{\text{BC}} \overset{!}{=} T_{\text{BC}} \quad  \implies c_{j_1,j_2,j_3,j_4} = (d_{j_1} d_{j_2} d_{j_3} d_{j_4})^{-1} .
\end{equation}
We will however introduce a free parameter in this normalization, allowing a power $\alpha$, instead of just $(-1)$. The reason is that this normalization determines the (path integral) measure in spin foams. Different principles have been suggested to fix this measure \cite{bahrknotting,warsaw1,bonzom-dittrich-bubble,marseille-face}, leading to different proposals.  There is however one very strong requirement which is expected to give a unique answer, namely to ensure a restoration of diffeomorphism invariance and triangulation independence in the continuum limit \cite{bojowald-measure,improved,harmosci}. This principle does in fact fix the measure in 3D Regge calculus uniquely \cite{regge-measure}. For 4D Regge calculus one can show that there is no local measure satisfying this requirement \cite{non-local-measure}, which again emphasizes the need to study which measures could lead to a diffeomorphism invariant model via coarse graining.  Thus it is important to allow for some freedom of choice in the initial measure, as this can also determine the phase the models are flowing to. (In \cite{sf-cuboid-renorm} the measure is the only free parameter and its tuning does indeed indicate a phase transition.)

To conclude the construction, we obtain the following initial tensor $\hat{T}_{\text{BC}}$ in block diagonal form:
\begin{equation}
\hat{T}^{\{J\}}_{\text{BC}} (\{j_i\}) = (d_{j_1} d_{j_2} d_{j_3} d_{j_4})^\alpha \; (d_J d_{J'})^{-1} \delta_{J^+,J^-} \delta_{(J^+)',(J^-)'} \prod_{i=1}^4 \delta_{j_i^+,j_i^-} \quad .
\end{equation}

For the triangular algorithm, which we are using in this work, one rather has to define a 3-valent tensor from the 4-valent one.
This is given by
\begin{equation} \label{eq:BC-3-valent}
\hat{S}^{\{J\}}_{\text{BC}} (\{j_i\}) = (d_{j_1} d_{j_2})^\alpha (d_J d_{J'})^{-1} \delta_{J^+,J^-} \delta_{(J^+)',(J^-)'} \prod_{i=1}^4 \delta_{j_i^+,j_i^-} \quad .
\end{equation}
Note that in the triangular algorithm, the indices $\{J\}$ do not have the interpretation of an intermediate label, but rather are the irreducible representations assigned to a coarser edge of the tensor (obtained from splitting a square along its diagonal).

This concludes the construction of the BC model. In the next section we will present the construction of the EPRL model, which is based on very different concepts and more elaborate.

\subsection{The EPRL construction}

Originally the EPRL model is motivated as a modification of the BC model, in particular in the imposition of the simplicity constraints onto discretised and quantised BF theory \cite{eprl1,eprl2,liv-spez-int}. These constraints do not form a closed algebra, more precisely the off-diagonal constraints are second class. Imposing them strongly, as it is done in the BC model, might therefore restrict the degrees of freedom of the model more than in the classical theory. It is frequently argued that the uniqueness of the BC intertwiner supports this reasoning, as the intertwiner degrees of freedom are completely constrained.

The EPRL model lifts this issue by imposing the constraints weakly, that is not as an operator equation but at the level of expectation values, e.g. by a Gupta-Bleuler criterion. In this article, we will not use the original derivation, but rather follow the more recent and straightforward method which was motivated by the closely related Freidel-Krasnov (FK) model \cite{fk}. Instead of imposing the quadratic simplicity constraints, which can be shown to reduce BF theory to general relativity, one imposes so-called linear simplicity constraints. In the classical and discrete setting imposing these linear constraints on each face of a triangulation is actually equivalent to imposing the quadratic simplicity constraints. Also the linear constraints are second class, thus they are also imposed weakly. One possible solution, in the Gupta-Bleuler condition, results in a condition on the $\text{SU}(2) \times \text{SU}(2)$ representations $(j^+,j^-)$. In the rest of this article we assume the Barbero-Immirzi parameter $\gamma < 1$:
\begin{equation} \label{eq:eprl-map}
(j^+ \, , \, j^-) := \left (\frac{1+\gamma}{2} l \, , \, \frac{1-\gamma}{2} l \right) \quad .
\end{equation}
$l$ denotes another $\text{SU}(2)$ representation. Given this relation of $\text{SU}(2)$ representations to $\text{SU}(2) \times \text{SU}(2)$ representations labelled by $\gamma$ one defines a map $Y_\gamma : \mathcal{H}_{(1+\gamma) l/2, (1-\gamma) l/2} \rightarrow \mathcal{H}_l$ relating the respective Hilbert spaces. Essentially this map restricts the representations $(j^+,j^-)$ to those compatible with the simplicity constraints.

Similarly one  also constructs $\text{SU}(2) \times \text{SU}(2)$ intertwiners from $\text{SU}(2)$ ones. First one maps all $\text{SU}(2)$ representations to $\text{SU}(2) \times \text{SU}(2)$, however the resulting vector is not necessarily an intertwiner, i.e. it does not lie in the invariant subspace. Thus one has to contract this object again with the Haar projectors from both sides in order to obtain an $\text{SU}(2) \times \text{SU}(2)$ intertwiner. Note that this map between invariant subspaces is not an isometry \cite{eprl-int}, that is the norm of the intertwiners are not preserved under this map. Nevertheless, due to this construction of the EPRL model its boundary Hilbert space is actually isomorphic to the (kinematical) Hilbert space of loop quantum gravity (for a fixed graph). Thus the EPRL model lifts the second shortcoming of the BC model as it can be used to define transition amplitudes for states of loop quantum gravity.

In the construction of the EPRL spin nets we essentially follow the same route outlined in the previous two paragraphs. First we consider a map from $\text{SU}(2)_k$ representations to $\text{SU}(2)_k \times \text{SU}(2)_k$ ones implementing simplicity constraints, where the maximum spin $j_{\text{max}}$ of the quantum group requires particular care. Then we lift the Haar projector of $\text{SU}(2)_k$ to a $\text{SU}(2)_k \times \text{SU}(2)_k$ representation theoretic object, which we denote as the EPRL tensor $T_{\text{EPRL}}$. As this generically is not a projector onto the $\text{SU}(2)_k \times \text{SU}(2)_k$ invariant subspace, it is then contracted by Haar projectors of $\text{SU}(2)_k \times \text{SU}(2)_k$. To put it in a nutshell we essentially restrict the model to projectors that can arise from the $\text{SU}(2)_k$ Haar projector given the map implementing the simplicity constraints.

The spirit of the construction is very similar to the corresponding spin foam model as defined by Meusburger and Fairbairn \cite{catherine}. We work here with the Euclidean version, Lorentzian spin foam vertex amplitudes have been also constructed by \cite{catherine} and \cite{muxin}.

As already mentioned above, the map \eqref{eq:eprl-map} from $l \rightarrow (j^+(l),j^-(l))$ requires some attention in the Euclidean theory: Both the representations $l$ and $j^+$, $j^-$ must be $\frac{1}{2} \mathbb{N}$. If this is not the case, the particular mapping is forbidden, i.e. will be assigned a vanishing weight. As we usually start from a representation $l$, this gives restrictions onto the Barbero-Immirzi parameter $\gamma$ in order to obtain a non-trivial map, i.e. beyond just mapping the trivial representations to one another. Thus one quickly realizes that $\gamma \in \mathbb{Q}$ is the necessary condition to do so. Again this is a particular condition on the Euclidean theory, a similar restriction does not exist for the Lorentzian one. In the case we are considering here, there is a further restriction, as we only consider integer representations. On a more technical level, one can understand this identification as a map from $V_l \rightarrow V_{j^+} \otimes V_{j^-}$, so essentially a Clebsch-Gordan coefficient. Thus the coupling rules of $\text{SU}(2)$ also influence whether a non-trivial map exists.

In the case of quantum groups further restrictions occur, as it has been already studied in \cite{catherine} (see also \cite{muxin} for an independent derivation of the Lorentzian model) in the case $\gamma < 1$. As discussed above $\text{SU}(2)_k$ (at root of unity) has a natural cut-off on the spins, $j_{\text{max}} = \frac{k}{2}$. Representations labelled by larger $\frac{1}{2} \mathbb{N}$ exist, but are referred to as having vanishing quantum dimension\footnote{For $j > j_{\text{max}}$ the quantum dimension $d_j = [2j+1]_q$ is no longer positive definite.}. The $\text{SU}(2)_k$ spin nets in \cite{q-spinnet} have been explicitly constructed to avoid these representations, thus we have to ensure that no allowed spin $l$ gets mapped to such a representation. Similar to \cite{catherine}, we achieve this by requiring:
\begin{equation}
 \left(j^+(j_{\text{max}}),j^-(j_{\text{max}})\right) = \left( \frac{1+\gamma}{2} j_{\text{max}} , \frac{1 - \gamma}{2} j_{\text{max}} \right) \in \left \{ (j,j') \in \left( \mathbb{N}, \mathbb{N} \right) : j,\, j' \leq j_{\text{max}} \right \} \quad .
\end{equation}
Again this puts many restrictions on the possible choices of $\gamma$, in particular for small levels $k$ of the quantum group. In many cases only the trivial map exists. The following non-trivial cases are possible (we omit the trivial identification):
\begin{itemize}
 \item $k=6$ ($j_{\text{max}} = 3$) for $\gamma = \frac{1}{3}$: $l=3 \mapsto (j^+ = 2, j^-=1)$.
 \item $k=10$ ($j_{\text{max}} = 5$) for $\gamma = \frac{3}{5}$: $l=5 \mapsto (j^+ = 4, j^-=1)$.
 \item $k=12$ ($j_{\text{max}} = 6$) for $\gamma = \frac{1}{3}$: $l=3 \mapsto (j^+ = 2, j^-=1)$ and $l=6 \rightarrow (j^+ = 4, j^-=2)$. \end{itemize}
As we will argue below, due to freedom in the normalization as in the BC case, the model for $k=12$ is the most interesting one, as it will actually be a whole one-parameter family of models. However, at least for spin nets, $k=12$ is currently beyond efficient simulation, despite the optimization efforts described in sections \ref{sec:scope} and \ref{sec:optimization}. Nevertheless, the associated intertwiner model can be studied without problems.

Concretely for spin nets, we first construct the EPRL tensor $T_{\text{EPRL}}$ from the $\text{SU}(2)_k$ Haar projector:
\begin{equation} \label{eq:EPRL-tensor}
T_{\text{EPRL}} (\{j\},\{m\},\{n\}) := \sum_l c_{\{l\}}
\begin{tikzpicture}[baseline,scale=0.85]
\draw (-0.5,-1) -- (0,-0.5) -- (0,0.5) -- (-0.5,1)
      (0.5,-1) -- (0,-0.5)
      (0,0.5) -- (0.5,1)
      (-0.45,-0.55) node {$l_3$}
      (0.45,-0.55) node {$l_4$}
      (0.45,0.55) node {$l_1$}
      (-0.45,0.55) node {$l_2$}
      (0.2,0) node {$l$};
\draw (-0.5,1) -- (-0.75,1.25);
\draw (-0.5,1) -- (-0.25,1.25);
\draw (-0.5,-1) -- (-0.75,-1.25);
\draw (-0.5,-1) -- (-0.25,-1.25);
\draw (0.5,1) -- (0.75,1.25);
\draw (0.5,1) -- (0.25,1.25);
\draw (0.5,-1) -- (0.25,-1.25);
\draw (0.5,-1) -- (0.75,-1.25);
\draw (0.25,-1.5) node {$j^-_4$};
\draw (0.75,-1.5) node {$j^+_4$};
\draw (-0.25,-1.5) node {$j^+_3$};
\draw (-0.75,-1.5) node {$j^-_3$};
\draw (0.25,1.5) node {$j^-_1$};
\draw (0.75,1.5) node {$j^+_1$};
\draw (-0.25,1.5) node {$j^+_2$};
\draw (-0.75,1.5) node {$j^-_2$};
\end{tikzpicture}
\; \otimes \;
\begin{tikzpicture}[baseline,scale=0.85]
\draw (-0.75,1.0) -- (-0.25,1.5) arc(180:0:0.75) -- (1.25,-1.5) arc(0:-180:0.75) -- (-0.75,-1.0)
      (-0.25,1.5) -- (0.25,1.)
      (-0.25,-1.5) -- (0.25,-1.0)
      (0.4,1.45) node {$l_1$}
      (-0.9,1.45) node {$l_2$}
      (0.4,-1.45) node {$l_4$}
      (-0.9,-1.45) node {$l_3$}
      (1.5,0.) node {$l$};
\draw (0.25,1.) -- (0.5,0.75);
\draw (0.25,1.0) -- (0.,0.75);
\draw (-0.75,1.0) -- (-1.0,0.75);
\draw (-0.75,1.0) -- (-0.5,0.75);
\draw (-0.75,-1.0) -- (-0.5,-0.75);
\draw (-0.75,-1.0) -- (-1.0,-0.75);
\draw (0.25,-1.0) -- (0.5,-0.75);
\draw (0.25,-1.0) -- (0.,-0.75);
\draw (-1.0,0.5) node {$j^-_2$};
\draw (-0.5,0.5) node {$j^+_2$};
\draw (0.5,0.5) node {$j^+_1$};
\draw (0,0.5) node {$j^-_1$};
\draw (-1.0,-0.5) node {$j^-_3$};
\draw (-0.5,-0.5) node {$j^+_3$};
\draw (0.5,-0.5) node {$j^+_4$};
\draw (0.,-0.5) node {$j^-_4$};
\end{tikzpicture} \quad .
\end{equation}
Note again that $j^\pm_i = (1 \pm \gamma) l_i/2$. $c_{\{l\}}$ is the normalization constant. Also it is important that both copies of $\text{SU}(2)_k \times \text{SU}(2)_k$ representations are generated from the same $\text{SU}(2)_k$ intertwiner (and thus identical, i.e. $(j^\pm_i)'=(j^\pm_i)^*$). If the latter were independent, i.e. replace the second $l$ by $l'$, this would result in a different, factorising model.

As for the BC case, $T_{\text{EPRL}}$ is not a projector onto the invariant subspace, therefore it has to be contracted with the Haar projector from both sides:

\begin{align} \label{eq:EPRL-formula}
\mathcal{P} \, \circ \, T_{\text{EPRL}} \, \circ \, \mathcal{P} = & \sum_{\{J\}} c_{\{j_i\}} \; d_{J^+} d_{J^-} d_{(J^+)'} d_{(J^-)'} \quad
\begin{tikzpicture}[baseline,scale=0.85]
\draw (-0.75,1.0) -- (-0.25,1.5) arc(180:0:0.75) -- (1.25,-1.5) arc(0:-180:0.75) -- (-0.75,-1.0)
      (-0.25,1.5) -- (0.25,1.)
      (-0.25,-1.5) -- (0.25,-1.0)
      (0.4,1.45) node {$l_1$}
      (-0.9,1.45) node {$l_2$}
      (0.4,-1.45) node {$l_4$}
      (-0.9,-1.45) node {$l_3$}
      (1.5,0.) node {$l$};
\draw (0.25,1.) -- (0.5,0.75);
\draw (0.25,1.0) -- (0.,0.75);
\draw (-0.75,1.0) -- (-1.0,0.75);
\draw (-0.75,1.0) -- (-0.5,0.75);
\draw (-0.75,-1.0) -- (-0.5,-0.75);
\draw (-0.75,-1.0) -- (-1.0,-0.75);
\draw (0.25,-1.0) -- (0.5,-0.75);
\draw (0.25,-1.0) -- (0.,-0.75);
\draw (-1.25,0.75) node {$j^-_2$};
\draw (0.75,0.75) node {$j^+_1$};
\draw (-1.25,-0.75) node {$j^-_3$};
\draw (0.75,-0.75) node {$j^+_4$};
\draw (-1.,0.) node {$(J^-)'$};
\draw (0.65,0.) node {$(J^+)'$};
\draw (-1.0,-0.75) -- (-0.5,-0.25)  -- (-0.35,-0.4);
\draw (-0.15,-0.6) -- (0.,-0.75);
\draw (-0.5,-0.75) -- (-0.,-0.25) -- (0.5,-0.75);
\draw (-1.0,0.75) -- (-0.5,0.25) -- (0.,0.75);
\draw (-0.5,0.75) -- (-0.4,0.65);
\draw (-0.15,0.4) -- (0.,0.25) -- (0.5,0.75);
\draw (0,0.25) -- (0,-0.25);
\draw (-0.5,0.25) -- (-0.5,-0.25);
\end{tikzpicture}
\quad
\begin{tikzpicture}[baseline,scale=0.85]
\draw (-0.5,-1) -- (0,-0.5) -- (0,0.5) -- (-0.5,1)
      (0.5,-1) -- (0,-0.5)
      (0,0.5) -- (0.5,1)
      (-0.45,-0.55) node {$l_3$}
      (0.45,-0.55) node {$l_4$}
      (0.45,0.55) node {$l_1$}
      (-0.45,0.55) node {$l_2$}
      (0.2,0) node {$l$};
\draw (-0.5,1) -- (-0.75,1.25);
\draw (-0.5,1) -- (-0.25,1.25);
\draw (-0.5,-1) -- (-0.75,-1.25);
\draw (-0.5,-1) -- (-0.25,-1.25);
\draw (0.5,1) -- (0.75,1.25);
\draw (0.5,1) -- (0.25,1.25);
\draw (0.5,-1) -- (0.25,-1.25);
\draw (0.5,-1) -- (0.75,-1.25);
\draw (0.75,-1.75) node {$j^+_4$};
\draw (-0.75,-1.75) node {$j^-_3$};
\draw (0.75,1.75) node {$j^+_1$};
\draw (-0.75,1.75) node {$j^-_2$};
\draw (1.,0.) node {$J^+$};
\draw (2.25,0.) node {$J^-$};
\draw (-0.25,1.25) -- (0.25,1.75) -- (0.75,1.25);
\draw (-0.75,1.25) -- (-0.25,1.75) -- (-0.10,1.6);
\draw (0.1,1.4) -- (0.25,1.25);
\draw (-0.25,-1.25) -- (-0.1,-1.4);
\draw (0.75,-1.25) -- (0.25,-1.75) -- (0.1,-1.6);
\draw (-0.75,-1.25) -- (-0.25,-1.75) -- (0.25,-1.25);
\draw (-0.25,-1.75) arc(-180:0:1.0) -- (1.75,1.75) arc(0:180:1.0);
\draw (0.25,-1.75) arc(-180:0:0.5) -- (1.25,1.75) arc(0:180:0.5);
\end{tikzpicture}
\; \nonumber \\
&
\begin{tikzpicture}[baseline,scale=0.85]
\draw (-0.5,-1) -- (0,-0.5) -- (0,0.5) -- (-0.5,1)
      (0.5,-1) -- (0,-0.5)
      (0,0.5) -- (0.5,1)
      (-0.5,-1.25) node {$j^+_3$}
      (0.5,-1.25) node {$j^+_4$}
      (0.5,1.25) node {$j^+_1$}
      (-0.5,1.25) node {$j^+_2$}
      (0.4,0) node {$J^+$};
\end{tikzpicture}
\; \otimes \;
\begin{tikzpicture}[baseline,scale=0.85]
\draw (-0.5,-1) -- (0,-0.5) -- (0,0.5) -- (-0.5,1)
      (0.5,-1) -- (0,-0.5)
      (0,0.5) -- (0.5,1)
      (-0.5,-1.25) node {$j^-_3$}
      (0.5,-1.25) node {$j^-_4$}
      (0.5,1.25) node {$j^-_1$}
      (-0.5,1.25) node {$j^-_2$}
      (0.4,0) node {$J^-$};
\end{tikzpicture}
\; \otimes \;
\begin{tikzpicture}[baseline,scale=0.85]
\draw (-0.5,0.75) -- (-0.25,1) arc(180:0:0.5) -- (0.75,-1) arc(0:-180:0.5) -- (-0.5,-0.75)
      (-0.25,1) -- (0,0.75)
      (-0.25,-1) -- (0,-0.75)
      (-0,0.5) node {$j^+_1$}
      (-0.5,0.5) node {$j^+_2$}
      (-0,-0.5) node {$j^+_4$}
      (-0.5,-0.5) node {$j^+_3$}
      (1.25,0) node {$(J^+)'$};
\end{tikzpicture}
\; \otimes \;
\begin{tikzpicture}[baseline,scale=0.85]
\draw (-0.5,0.75) -- (-0.25,1) arc(180:0:0.5) -- (0.75,-1) arc(0:-180:0.5) -- (-0.5,-0.75)
      (-0.25,1) -- (0,0.75)
      (-0.25,-1) -- (0,-0.75)
      (-0,0.5) node {$j^-_1$}
      (-0.5,0.5) node {$j^-_2$}
      (-0,-0.5) node {$j^-_4$}
      (-0.5,-0.5) node {$j^-_3$}
      (1.25,0) node {$(J^-)'$};
\end{tikzpicture} \quad .
\end{align}
A comment on the choice of Haar projector is in order. In contrast to the BC model, we have slightly changed the Haar projector\footnote{The $\text{SU}(2)_k \times \text{SU}(2)_k$ Haar projector used for the BC model is a  tensor product $\mathcal{P}_q \otimes \mathcal{P}_{\bar{q}}$, whereas the one for the EPRL is $\mathcal{P}_q \otimes \mathcal{P}_q$.}, which allows us to simplify the diagrams.

Another peculiarity of quantum groups are the over- and undercrossings of representations, which one has to keep track off since they do not commute. They can be transferred into one another employing the so-called $\mathcal{R}$  matrix \cite{biedenharn,wojtek}:
\begin{align}
\mathcal{R} = & 
\begin{tikzpicture}[baseline,scale=0.85]
\draw (-0.5,-0.5) -- (0.5,0.5);
\draw (0.5,-0.5) -- (0.1,-0.1);
\draw (-0.1,0.1) -- (-0.5,0.5);
\draw (-0.75,-0.5) node {$j_1$};
\draw (0.75,-0.5) node {$j_2$};
\end{tikzpicture}
\; = \; 
\sum_j \, d_j \; q^{-\frac{1}{2} ( j_1 (j_1 + 1) + j_2(j_2 + 1) - j(j+1)}
\begin{tikzpicture}[baseline,scale=0.85]
\draw (-0.5,-0.75) -- (0.0,-0.25) -- (0.0,0.25) -- (-0.5,0.75);
\draw (0.5,-0.75) -- (0.0,-0.25);
\draw (0.,0.25) -- (0.5,0.75);
\draw (-0.75,-0.5) node {$j_1$};
\draw (0.75,-0.5) node {$j_2$};
\draw (-0.75,0.5) node {$j_2$};
\draw (0.75,0.5) node {$j_1$};
\draw (0.25,0.0) node {$j$};
\end{tikzpicture}  \quad , \nonumber \\
\mathcal{R}^{-1} = &
\begin{tikzpicture}[baseline,scale=0.85]
\draw (-0.5,-0.5) -- (-0.1,-0.1);
\draw (0.5,-0.5) -- (-0.5,0.5);
\draw (0.1,0.1) -- (0.5,0.5);
\draw (-0.75,-0.5) node {$j_1$};
\draw (0.75,-0.5) node {$j_2$};
\end{tikzpicture}
\; = \; 
\sum_j \, d_j \; q^{\frac{1}{2} ( j_1 (j_1 + 1) + j_2(j_2 + 1) - j(j+1)}
\begin{tikzpicture}[baseline,scale=0.85]
\draw (-0.5,-0.75) -- (0.0,-0.25) -- (0.0,0.25) -- (-0.5,0.75);
\draw (0.5,-0.75) -- (0.0,-0.25);
\draw (0.,0.25) -- (0.5,0.75);
\draw (-0.75,-0.5) node {$j_1$};
\draw (0.75,-0.5) node {$j_2$};
\draw (-0.75,0.5) node {$j_2$};
\draw (0.75,0.5) node {$j_1$};
\draw (0.25,0.0) node {$j$};
\end{tikzpicture}  \quad .
\end{align}
Using these identities we can replace the crossings in the diagrams. Furthermore the diagrams can be manipulated further using identities derived in \cite{q-spinnet} (see also appendix \ref{app:graph}). Eventually we can define the block diagonal form of $T_{\text{EPRL}}$, namely $\hat{T}_{\text{EPRL}}$. Here we restrict ourselves to the triangular version $\hat{S}_{\text{EPRL}}$. See appendices \ref{app:EPRL-norm} and \ref{app:EPRL-diagram} for derivations of the normalisation and the diagrams respectively:
\begin{align} \label{eq:eprl-3-valent}
\hat{S}^{\{J\}}_{\text{EPRL}} (\{j_i\}) = & (-1)^{j^+_1 + j^-_1 + j^+_2 + j^-_2 - l} \; (d_{l_1} d_{l_2})^\alpha \; d_l \nonumber \\
& \times \left[ \sum_{j} d_j \, q^{-\frac{1}{2} ( j^+_2 (j^+_2 + 1) + j^-_1(j^-_1 + 1) - j(j+1)} 
\begin{tikzpicture}[baseline,scale=0.85]
\draw (-0.75,0.75) -- (-0.25,1.25) arc(180:0:1.0) -- (1.75,-1.25) arc(0:-180:1.0)
	  (0.25,0.75) -- (-0.25,1.25)
	  (0.25,0.75) -- (0.75,0.25) -- (0.75,-0.25) -- (0.25,-0.75)
	  (0.25,0.75) -- (-0.25,0.25) -- (-0.25,-0.25) -- (-0.75,-0.75)
	  (-0.25,-0.25) -- (0.25,-0.75)
	  (-0.75,0.75) -- (-0.25,0.25)
	  (-0.75,0.75) -- (-1.25,0.25) -- (-1.25,-0.25) -- (-0.75,-0.75)
	  (-0.75,-0.75) -- (-0.25,-1.25)
	  (0.25,-0.75) -- (-0.25,-1.25)
      (0.4,1.2) node {$l_1$}
      (-0.9,1.2) node {$l_2$}
      (2.0,0.) node {$l$}
      (-1.5,0.) node {$j^-_2$}
      (0.,0.) node {$j$}
      (1.2,0.) node {$j^+_1$}
      (-0.7,0.4) node {$j^+_2$}
      (0.2,0.4) node {$j^-_1$}
      (0.6,-1.2) node {$(J^+)'$}
      (-1.1,-1.2) node {$(J^-)'$}
      (-0.7,-0.4) node {$j^-_1$}
      (0.2,-0.4) node {$j^+_2$};
\end{tikzpicture}
\right] \nonumber \\
& \times
\left[ \sum_{j} d_j \, q^{-\frac{1}{2} ( j^+_1 (j^+_1 + 1) + j^-_2(j^-_2 + 1) - j(j+1)} 
\begin{tikzpicture}[baseline,scale=0.85]
\draw (-0.75,0.75) -- (-0.25,1.25) arc(180:0:1.0) -- (1.75,-1.25) arc(0:-180:1.0)
	  (0.25,0.75) -- (-0.25,1.25)
	  (0.25,0.75) -- (0.75,0.25) -- (0.75,-0.25) -- (0.25,-0.75)
	  (0.25,0.75) -- (-0.25,0.25) -- (-0.25,-0.25) -- (-0.75,-0.75)
	  (-0.25,-0.25) -- (0.25,-0.75)
	  (-0.75,0.75) -- (-0.25,0.25)
	  (-0.75,0.75) -- (-1.25,0.25) -- (-1.25,-0.25) -- (-0.75,-0.75)
	  (-0.75,-0.75) -- (-0.25,-1.25)
	  (0.25,-0.75) -- (-0.25,-1.25)
      (0.4,1.2) node {$J^-$}
      (-0.9,1.2) node {$J^+$}
      (2.0,0.) node {$l$}
      (-1.5,0.) node {$j^+_2$}
      (0.,0.) node {$j$}
      (1.2,0.) node {$j^-_1$}
      (-0.7,0.4) node {$j^+_1$}
      (0.2,0.4) node {$j^-_2$}
      (0.6,-1.2) node {$l_1$}
      (-1.1,-1.2) node {$l_2$}
      (-0.7,-0.4) node {$j^-_2$}
      (0.2,-0.4) node {$j^+_1$};
\end{tikzpicture}
\right] \quad .
\end{align}
The normalization constant $c_{\{l_i\}}$ is again computed  by contracting $T_{\text{EPRL}}$ with itself, and as in the BC case we introduce more freedom by allowing it to appear with a power $\alpha$. Note that the normalization only depends on the quantum dimensions of the $\text{SU}(2)_k$ representations $l_i$. Therefore $\alpha$ only plays a role if there exists a non-trivial map $l_i \mapsto (j^+,j^-)$ for $l_i \neq 0, j_{\text{max}}$, as $d_0 = d_{j_{\text{max}}} = 1$. The smallest $k$ for which this is possible (in the model discussed here) is $k=12$. Even though the diagrams turn out to be nicely symmetric, we have not found a simpler expression for them. Interestingly both diagrams turn out to give the same expression.

Before we continue with the discussion of the related intertwiner models, we would like to comment on the simplicity constraints. As discussed above, these are implemented at the level of representations, here explicitly in the maps $l_i \mapsto (j^+_i,j^-_i)$. In the diagrams we then observe that $j^\pm_1$ and $j^\pm_2$ couple to $J^\pm$, which are then again coupled to $l$. Note that the simplicity constraints are not explicitly implemented in the latter coupling, such that we expect a flow of the simplicity constraints. As one would interpret the theory at a coarser scale as an effective theory of the finer one, we a priori do not see a reason to enforce the constraints there, too.

In the next section we will discuss the respective intertwiner models and their behaviour under coarse graining.

\section{Intertwiner models} \label{sec:intertwiners}

As already discussed above intertwiner models can be motivated from spin nets as simpler versions of those. In fact one can interpret the latter as the tensor product of two interacting intertwiner models, thus one could also denote them as `entangled'\footnote{This is in full analogy to quantum information where two subsystems are entangled when they cannot be written as a product state, e.g. the Bell states.}. This insight in itself is already helpful in interpreting the fixed point of spin net models, as they often turn out to be factorising.

Nevertheless the study of intertwiner models themselves is already interesting in itself as their construction is analogous to spin nets, such that coarse graining them gives us first hints and insights into the behaviour of the simplicity constraints under coarse graining with much lower computational costs. The latter is crucial for studying the EPRL model, as the interesting case, i.e. $k=12$, for spin net models is currently out of reach.

Therefore in the next two subsections we will very briefly introduce the respective BC and EPRL intertwiner models and briefly discuss the results under coarse graining.

\subsection{BC intertwiner model}

For completeness, let us give the initial 3-valent tensor (in block diagonal form) for BC intertwiner models, which can be straightforwardly defined by omitting the dual representations (compare also with \eqref{eq:BC-3-valent}):
\begin{equation} \label{eq:BC-3-int}
\hat{S}^{\{J\}}_{\text{BC}} (\{j_i\}) = (d_{j_1} d_{j_2})^\alpha (d_J)^{-1} \delta_{J^+,J^-} \prod_{i=1}^4 \delta_{j_i^+,j_i^-} \quad .
\end{equation}
As before we keep the factor $\alpha$ as a means to study different models. It reflects the fact that the normalization is not uniquely defined, as it is also the case for edge and face amplitudes in spin foam models.

Under coarse graining we find a very simple pattern valid for all levels $k$ of quantum groups: the first important fact is that the BC intertwiner is a fixed point of the renormalization group flow for $\alpha_c=\frac{1}{2}$. This is not surprising as the original 4-valent BC intertwiner \cite{Barrett-Crane} is unique, i.e. it does not depend on the recoupling scheme chosen. On that specific fixed point the model thus is discretisation independent.

If we consider $\alpha \neq \alpha_c$ we do observe the following behaviour: As the model is not on the fixed point, we observe that channels other than the BC ones, i.e. $J^+ \neq J^-$, get excited, that is come associated with a non-vanishing singular value. This signifies a weakening of the simplicity constraints under the coarse graining flow. As we will see below the BC simplicity conditions are however restored at the fixed points. 

If one orders the singular values into a matrix $J^+ = J^-$ give the diagonal elements, $J^+ \neq J^-$ are off-diagonal elements. Crucially this matrix is symmetric, that is the model is invariant under exchanging $J^+$ and $J^-$. Eventually the models flow back to only BC channels and converge to one of two different fixed points. One of them is again the usual BC fixed point (for $\alpha = \frac{1}{2}$), that is all channels $J^+ = J^-$ are equally excited with $J^\pm \in \{0,1,\dots,j_{\text{max}}\}$. The model flows back to this fixed point for roughly all $\alpha > 0$.

For $\alpha < 0$ we observe a flow to a different fixed point. Again only channels $J^+ = J^-$ are allowed, however only $J^\pm \in \{0,j_{\text{max}}\}$ for $k$ even and $J^\pm = 0$ for $k$ odd. (For odd $k$ the maximal representation is half integer, which we have excluded.) This new fixed point, which is similar to the Ashtekar-Lewandowski vacuum in LQG, also exists in the initial model if $\alpha \rightarrow - \infty$. Then all representations $j \neq 0, j_{\text{max}}$ in \eqref{eq:BC-3-int} get suppressed as they possess a quantum dimension $d_j > 1$. Due to the coupling rules of $\text{SU}(2)_k$ any combinations of the trivial and the maximal representation can only couple to either the trivial or the maximal representation.

Thus, to sum up, we find that the BC intertwiner model has a simple phase structure valid for all levels $k$ of the quantum group. There exist two (attractive) fixed points of the renormalization group flow, both compatible with the BC condition $J^+ = J^-$. The BC intertwiner, given for $\alpha = \frac{1}{2}$, is the fixed point allowing all representations, whereas the other fixed point (for $\alpha \rightarrow -\infty$) is similar to the Ashtekar-Lewandowski vacuum of LQG, where only the trivial and the maximal representations are allowed. That the maximal representation remains is a peculiarity of the quantum group as it also has a quantum dimension of $1$ as the trivial one. Thus we would expect that only the trivial one remains  in the limit towards $\text{SU}(2)$. For any other value of $\alpha$ the system initially flows away from the BC condition before eventually converging to one of the two fixed points. It appears that this condition is very strictly implemented and deviations from it are possible, but are dynamically disfavoured. Therefore it is interesting to study the BC spin net model under similar aspects.

In the next subsection we will study the EPRL intertwiner model in detail.

\subsection{EPRL intertwiner model}

Before we discuss the behaviour of the EPRL model under coarse graining, let us present the initial 3-valent tensor (again compare with \eqref{eq:eprl-3-valent}):
\begin{align}
\hat{S}^{\{J\}}_{\text{EPRL}} (\{j_i\}) = & (-1)^{j^+_1 + j^-_1 + j^+_2 + j^-_2 - l} \; (d_{l_1} d_{l_2})^\alpha \; \sqrt{d_l} \nonumber \\
& \times \left[ \sum_{j} d_j \, q^{-\frac{1}{2} ( j^+_1 (j^+_1 + 1) + j^-_2(j^-_2 + 1) - j(j+1)} 
\begin{tikzpicture}[baseline,scale=0.85]
\draw (-0.75,0.75) -- (-0.25,1.25) arc(180:0:1.0) -- (1.75,-1.25) arc(0:-180:1.0)
	  (0.25,0.75) -- (-0.25,1.25)
	  (0.25,0.75) -- (0.75,0.25) -- (0.75,-0.25) -- (0.25,-0.75)
	  (0.25,0.75) -- (-0.25,0.25) -- (-0.25,-0.25) -- (-0.75,-0.75)
	  (-0.25,-0.25) -- (0.25,-0.75)
	  (-0.75,0.75) -- (-0.25,0.25)
	  (-0.75,0.75) -- (-1.25,0.25) -- (-1.25,-0.25) -- (-0.75,-0.75)
	  (-0.75,-0.75) -- (-0.25,-1.25)
	  (0.25,-0.75) -- (-0.25,-1.25)
      (0.4,1.2) node {$J^-$}
      (-0.9,1.2) node {$J^+$}
      (2.0,0.) node {$l$}
      (-1.5,0.) node {$j^+_2$}
      (0.,0.) node {$j$}
      (1.2,0.) node {$j^-_1$}
      (-0.7,0.4) node {$j^+_1$}
      (0.2,0.4) node {$j^-_2$}
      (0.6,-1.2) node {$l_1$}
      (-1.1,-1.2) node {$l_2$}
      (-0.7,-0.4) node {$j^-_2$}
      (0.2,-0.4) node {$j^+_1$};
\end{tikzpicture}
\right] \quad .
\end{align}
Compared to the BC model this expression is clearly more complicated and involved, which will be reflected in the flow under coarse graining. Again we keep the parameter $\alpha$ free in order to keep a freedom in the normalization choice. As discussed above this is actually only relevant when the map from $\text{SU}(2)_k$ representations to $\text{SU}(2)_k \times \text{SU}(2)_k$ ones is non-trivial, i.e. there exists an $l$ different from $j=0$ or $j=j_{\text{max}}$ that gets mapped to an $\text{SU}(2)_k \times \text{SU}(2)_k$ representation. This is the case for $k=12$, thus we will focus on this one in the following.

Before we start with the discussion of the flow under coarse graining it is instructive to focus first on the diagram in the definition of the model. We see that the two representations $J^+$ and $J^-$ couple to $l$, however this coupling does not need to fulfil the conditions of the simplicity constraints. Thus, if one considers the coupling rules of $\text{SU}(2)_k$, one quickly realizes that channels $(J^+,J^-)$ are allowed that are initially not part of the conditions given by the simplicity constraints. These include for example several diagonal channels (with $J^+=J^-$). Therefore it is clear that this EPRL intertwiner model will immediately flow away from the original one for any value of $\alpha$. Note that this is the first clear difference to the BC model, where the BC simplicity condition $J^+ = J^-$ automatically followed from recoupling theory and thus remained valid at least in the first iteration. In this regard the EPRL model appears to restrict the intertwiner degrees of freedom less. Moreover, the symmetry of the BC model under exchange of $J^+$ and $J^-$ does not exist in the EPRL one by construction.

Indeed under coarse graining we observe an almost orthogonal behaviour of the EPRL model compared to the BC model. The truncation scheme we have used in this article, see also section \ref{sec:optimization}, is taking over the most relevant (effective) degree of freedom per block $\{J^\pm\}$. While this scheme works very well for the BC model, as there usually one degree of freedom is clearly the most important one (given by the size of its singular value), this does not hold any more for the EPRL model. Already from the first iteration, one finds that the several effective degrees of freedom per block are too relevant to be truncated, that is are not negligible in reference to the most important one. In subsequent iterations this gets even more pronounced for more of the channels. Therefore in order to study the system in more detail one should employ a more elaborate truncation scheme. However we refrain from doing so for three reasons: first our main goal is to study the respective spin net models, for which simply increasing the number of degrees of freedom is out of reach (for the interesting models). The second reason concerns the increasing number of degrees of freedom that are too relevant to be truncated: previously we have observed this only close to (arguably) second order phase transitions, which are indicated by an (almost) scale invariance. We do not find such an indication here. Although this might be due to the truncation scheme, it hints at the possibility of a very intricate behaviour of the simplicity constraints. This brings us to the third reason, which is the need to have a geometric interpretation of the additional degrees of freedom appearing in each block. With the truncation of one singular value per block the models also preserved part of their initial form, and thus allowing us to retain the original geometrical interpretation. This would change with a more complicated truncation scheme, which also should be accompanied with a geometric understanding of the emerging effective degrees of freedom.  Nevertheless it would be more promising to study this model with an algorithm more suited for cases where the number of relevant degrees of freedom increase  \cite{vidal-evenbly}, in order to better understand the origin of these degrees of freedom.

Despite these shortcomings we would still like to report on some of the observations that we have made under renormalization. Of course these results should be taken with a grain of salt as they might change under a more accurate algorithm. One of the first observations is that phases are very difficult to identify as in many cases the model does not converge to a fixed point, but rather oscillates. This behaviour is most likely related to the low cut-off per block, such that it can happen that two degrees of freedom in the same block $\{J\}$ `change' their significance, leading to a different new tensor and subsequent flow. Nevertheless, for $\alpha < 0.3$ we observe a flow of the model towards a phase similar to the Ashtekar-Lewandowski vacuum of LQG, where only $(J^+,J^-) = (0,0)$ is allowed.

\begin{figure}
\includegraphics[width=0.95 \textwidth]{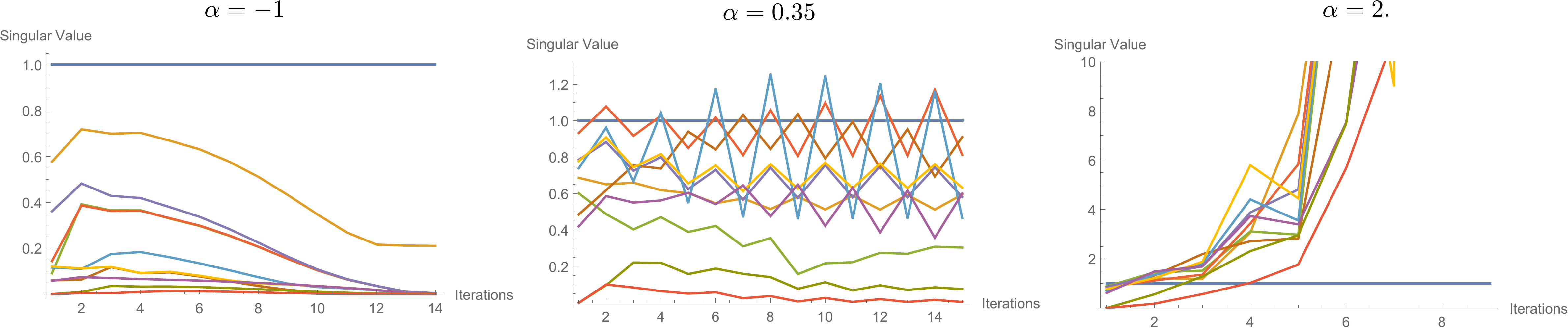}
\caption{ \label{fig:eprl-plots}
Plots of singular values for EPRL-intertwiner model for $k=12$: We plot the flow of singular values for three different initial values of $\alpha$ from the regions mentioned in the main text. For $\alpha=-1$, we observe a convergence to the Ashtekar-Lewandowski phase, where only $(J^+,J^-) = (0,0)$ is allowed. For the other values the behaviour is ambiguous, as we observe oscillations (for $\alpha = 0.35$ and $\alpha = 2$) and diverging singular values for $\alpha=2.$. The singular values are normalized with respect to channel $(J^+,J^-) = (0,0)$. We plot only a selection of values, namely all diagonal ones with $J^+=J^-$, as well as $(J^+,J^-) =(1,2),(2,1),(2,4),(4,2)$, in which  the EPRL model is explicitly asymmetric ($(2,1)$ and $(4,2)$ are initially allowed). As we cannot identify interesting phases, we refrain from labelling the singular values.
}
\end{figure}

For (roughly) $\alpha > 0.3$ we observe another peculiarity that we have never encountered in related models so far. Usually the trivial representation, here with the channel $(J^+,J^-) = (0,0)$ always appears as the most important one, i.e. with the largest singular value, such that it is suitable to normalize the tensor with respect to it. However for $\alpha > 0.3$ other representations become more relevant than the trivial one, i.e. their associated singular value is larger. That is unexpected as inevitably all degrees of freedom couple to the trivial representation again. Moreover, if we increase $\alpha$ further to roughly $\alpha > 1$ we also observe that some of the singular values associated to non-trivial representations increase over several iterations and appear to diverge, e.g. are several orders of magnitude larger than the one associated to the trivial representation. The analysis is made difficult also by the oscillation of the singular values as no clear pattern can be uncovered under coarse graining, such that changing the normalization is not straightforward as well. One example for this is again the behaviour for $\alpha > 1$, where we observe some singular values increasing rapidly iteration after iteration before suddenly dropping to values smaller than the trivial representation before converging to the Ashtekar-Lewandowski vacuum. We again attribute these features to the cut-off scheme. In fig. \ref{fig:eprl-plots} we plot the singular values for three different values of $\alpha$ to illustrate the behaviour of the model in the three different regimes described above.

One reason for the rather irregular behaviour of the EPRL model might be due to the few  spin values that are allowed in the initial model, which are moreover asymmetric under exchange of $j^+$ and $j^-$.  For the case $k=12$ with maximal spin $j^\pm_{\text{max}}=6$ (and the Barbero--Immirzi parameter chosen as $\gamma = \frac{1}{3} $) these include besides the trivial spin assignment only the two cases $(j^+,j^-)=(2,1)$ and $(j^+,j^-)=(4,2)$.  Compare this with the BC model (for finite $\alpha$) in which all configurations satisfying $j^+=j^-$ appear. We expect that this might actually hinder the Euclidean EPRL-$\Lambda$ models, as defined in their current form \cite{catherine}, to display a suitable continuum limit, at least for large cosmological constant (correspondingly small $k$). For larger $k$ this issue should be attenuated as more configurations become allowed and more choices for the Immirzi parameter lead to non-trivial configurations.   In the Lorentzian case one has a priori infinitely many representations appearing, however the imposition of the (EPRL) simplicity constraints does lead to a cut--off in the spins. That is also in this case, the combination of using a quantum group and implementing a Barbero--Immirzi parameter suppresses (infinitely) many spin combinations. 

Clearly one should not interpret too much into the results obtained from coarse graining the EPRL intertwiner model, due to the reasons mentioned above. Nevertheless we have reported on several qualitative features which are strikingly different from the BC model or other models studied so far, namely the quickly growing number of relevant degrees of freedom for any parameter of the model, that is without a sign of a nearby phase transition.
 Thus it is apparent that the EPRL construction allows for a far more intricate dynamics than the BC one, which is difficult to characterize yet. 
The dynamics of the model has to be studied in more detail, which requires new tools, possibly combining both analytical and numerical methods.

Let us further remark that the fact that the BC intertwiner model allows for a fixed point (for $\alpha=1/2$), that is defines a triangulation invariant 2D model, is crucial for the arguments that were invoked to show uniqueness of the model in \cite{reisenberger}, see also the discussion in \cite{q-spinnet}. In contrast we have not found a corresponding  topological model that would be triangulation invariant and originates from the EPRL intertwiner models by coarse graining. Again one reason seems to be the restrictions on the allowed spins imposed by the Barbero--Immirzi parameter.  Also our coarse graining method are not sufficient to capture the phase structure, that is find {\it local}, triangulation invariant models, to which the initial models flow under coarse graining. It might however be that the fixed points for the EPRL model feature non--local amplitudes, which in particular applies to second order phase transitions \cite{he,non-local-measure}.

This concludes the section on intertwiner models. In the next section we move on towards coarse graining spin net models.

\section{BC spin net models}

In this section we will present the results obtained by applying the tensor network renormalization algorithm introduced and thoroughly discussed in sections \ref{sec:scope} and \ref{sec:optimization} to BC spin nets constructed in section \ref{sec:qg-models}. Note that the details of the algorithm are not vital for understanding the results.

Before doing this however we would like to mention again why we do not consider the EPRL spin net model here as well. On the technical side this is due to the size of the tensors for $k=12$, the first non-trivial model we can study. With all the optimizations developed and implemented, the largest obstacle remaining is the size of the largest block $\{J\}$ of the tensor. For $k=12$ this block consists of $25^8$ entries, which equals roughly $2.3$ TByte of memory usage. Moreover as we have already observed for the intertwiner models, a more accurate cut-off scheme, which further increases the memory usage, as well as an algorithm better suited in dealing with increasing number of relevant degrees of freedom \cite{vidal-evenbly} are necessary. Therefore we do not expect reliable results and thus leave this question for future research.

Returning to the BC model, the initial tensor is given in \eqref{eq:BC-3-valent} and is of a very simple form. The thoughtful reader might wonder why we expect an interesting flow of this tensor under coarse graining when it essentially consists of two BC intertwiners, which have been previously identified (individually) as fixed points of the renormalization procedure. As we have already commented before the two intertwiners making up the projector sitting at the vertices of the spin net are not independent, but depend on the same labels. In particular the second copy of representations satisfy $(j^\pm_i)' = (j^\pm_i)^*$. Due to the particular conditions of the BC model this actually also translates to the new effective labels $J^\pm$. The two intertwiners would only be completely independent if all $(j^\pm_i)'$ are completely unrelated to the $j^\pm_i$, such that the model could be written exactly as a tensor product of two intertwiners. As the renormalization algorithm does not mix the two copies of the representations, i.e. $\{j\}$ and $\{j'\}$, they would stay independent during the entire coarse graining process. Thus tensor products of fixed points are indeed fixed points of the renormalization group flow. Using the same terminology, the intertwiner degrees of freedom in the BC spin net are interacting, one could say the intertwiners are entangled (as they do not factorize). As a result one observes a non-trivial renormalization group flow (at least initially) beyond factorising models.

In this article we do not consider only one BC spin net model, but a one-parameter family described by the parameter $\alpha$, which appears as the power of the normalization constant. Again this relates to the ambiguities in the choice of face and edge amplitudes in spin foam models. From the intertwiner models in section \ref{sec:intertwiners} we have already observed that $\alpha$ influences the model quite significantly: if $\alpha < 0$ it will suppress representations $j$ in the model with quantum dimension $d_j > 1$, which is all other than the trivial and maximal one. For $\alpha > 0$ it conversely emphasizes said representations. We will see that this also affects the BC spin net model.

A further comment on the simplicity constraints is in order: as discussed thoroughly above, the simplicity constraints in the BC model are essentially imposed by requiring that all $j^+_i = j^-_i$. This holds also for the new edge labels $J^+ = J^-$ introduced in the first coarse graining iteration, however, as we have already seen for the intertwiner models away from one of its two fixed points, this condition gets violated dynamically under coarse graining. Thus we also expect this for spin nets, as long as the initial model is not a fixed point.

This immediately leads us to one of the first observations: the BC spin net model is not a fixed point of the renormalization group flow for any finite value of $\alpha$; it is only a fixed point if one sends $\alpha \rightarrow - \infty$, such that only the trivial and the maximal representation are allowed. Similar to the intertwiner model we find at least two (extended) phases, i.e. fixed points the model converges two, for each level $k$ of the quantum group. Additionally we find up to two more phases for particular levels $k$.

The phases / fixed points of the model are characterised by the singular values assigned to the intertwiner channels $\{J\}$ and are best organized in a matrix, with $(J^+,J^-)$ and $((J^+)',(J^-)')$ denoting rows and columns respectively\footnote{The ordering of $(J^+,J^-)$ in the (rows and columns of the) matrix is as follows: $(J^+,J^-) = (0,0),$ $(0,1), \dots (0,j_{\text{max}}),$ $(1,0),$ $(1,1), \dots (j_{\text{max}},j_{\text{max}}-1),$ $(j_{\text{max}},j_{\text{max}})$.}. Actually this matrix is (and stays) symmetric with respect to both diagonals, i.e. under exchanging both sets of representations and also under exchanging $J^+$ and $J^-$.

\begin{figure}
\includegraphics[scale=0.35]{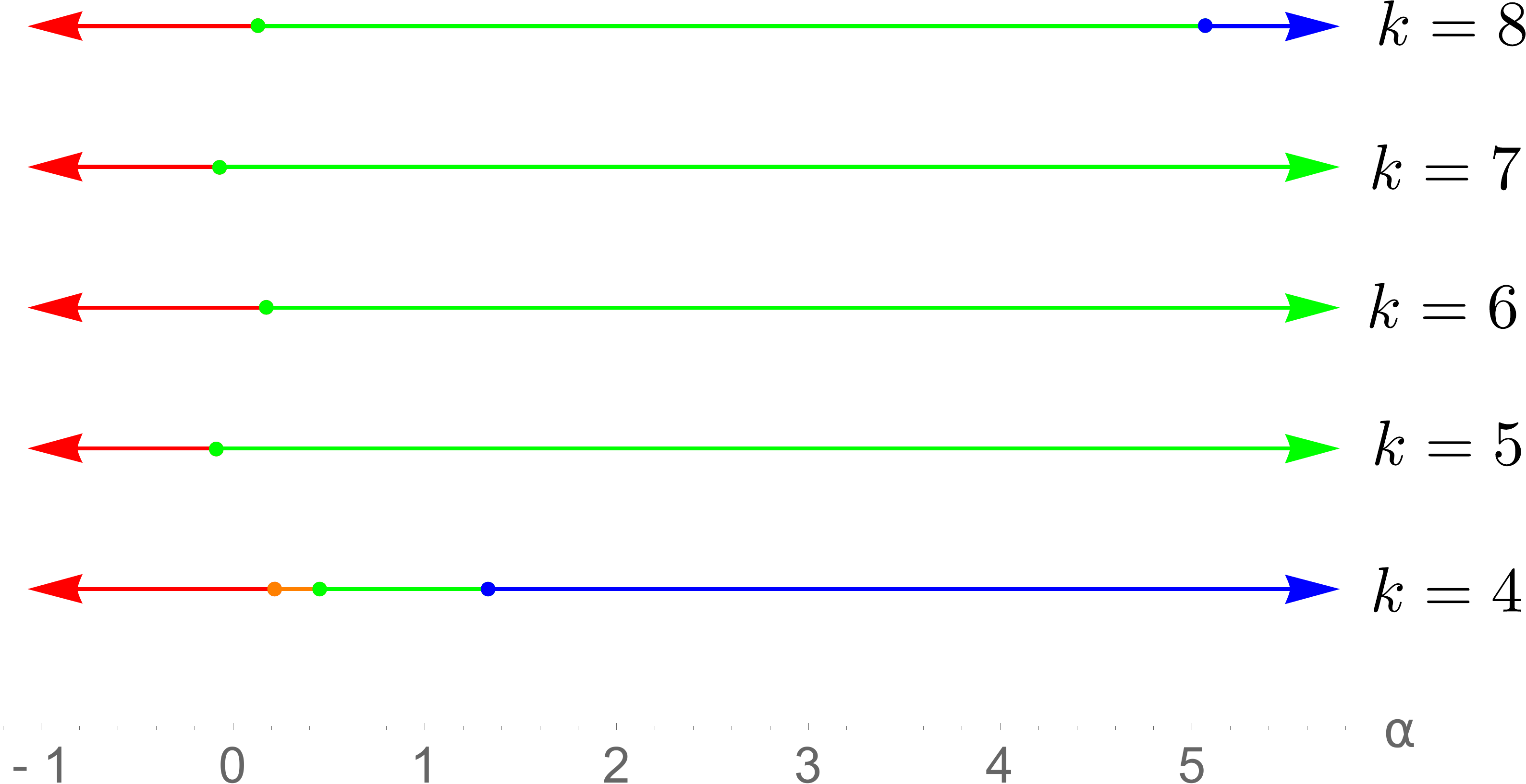}
\caption{\label{fig:phase-lines}
Phases of BC spin nets: Red indicates the Ashtekar-Lewandowski phase, green the factorising BC phase and blue the factorising phase with singular values alternating between 1 and 0. Only for $k=4$ we find a non-factorising phase, here orange, in between the Ashtekar-Lewandowski and the factorising BC phase. The arrows indicate that this phase continues up to $\pm \infty$. Across the different levels $k$, the transition between Ashtekar-Lewandowski and factorising BC is close to $\alpha=0$. It appears the transition from factorising BC to the factorising and alternating phase moves to larger $\alpha$ as $k$ is increased, however one would need to study even larger $k$ to confirm that.}
\end{figure}

Let us begin with the two phases found for any level $k$ of the quantum group, which are quite similar to the ones of the intertwiner model. Actually one of the fixed points is precisely a tensor product of two BC intertwiners:
\begin{itemize}
\item The first phase one finds for very small $\alpha$ is the phase in which only the trivial $(j^+,j^-) = (0,0)$ and the maximal representation $(j^+,j^-)= (j_{\text{max}},j_{\text{max}})$ (for even $k$) are allowed. As already discussed before, this is very similar to the Ashtekar-Lewandowski vacuum of LQG. This phase itself is not factorising as it actually requires that $J = J'$. The fixed point is summarized in the following matrix (for even $k$):
\begin{equation}
\left(
 \begin{matrix}
 1 & 0 & 0 & \dots & 0 \\
 0 & 0 & \dots & & 0 \\
 \vdots & \ddots & & & \vdots \\
 0 & \dots & & 0 & 1
 \end{matrix}
\right)
\end{equation}
\item The second phase we find appears for larger $\alpha$ and we call it `factorising BC phase'. It is characterised by allowing all representations $(J^+,J^-) = (J,J)$, where both copies of representations, i.e. $J^\pm$ and $(J^\pm)'$ are independent of one another. Summarized in a matrix of singular values this looks as follows (for $k=4$, $j_{\text{max}} = 2$):
\begin{equation}
\left(
 \begin{matrix}
  1 & 0  & 0 & 0 & 1 & 0 & 0 & 0 & 1 \\
  0 & 0  & 0 & 0 & 0 & 0 & 0 & 0 & 0 \\
  0 & \vdots  &   & \vdots  &   & \vdots   &   & \vdots & 0 \\
  0 &   &  &  & 0 &  &  &  & 0 \\
  1 & 0  & 0 & 0 & 1 & 0 & 0 & 0 & 1 \\
  0 &   &  &  & 0 &  &  &  & 0 \\
  0 &   &  &  & 0 &  &  &  & 0 \\
  0 &   &  &  & 0 &  &  &  & 0 \\
  1 & 0  & 0 & 0 & 1 & 0 & 0 & 0 & 1
 \end{matrix}
\right)
\end{equation}
As this model is factorising, this matrix can be written as tensor product of two vectors, one for $(J^+,J^-)$ and one for $((J^+)',(J^-)')$, which have entries $1$ if $J^+ = J^-$ and $0$ otherwise. These exactly characterize the original BC intertwiner fixed point.
\end{itemize}
For most levels $k$ of the quantum group we have studied we only find these two extended phases, that is both fixed points are attractive. In detail these were the levels $k=5$, $k=6$ and $k=7$. The parameter $\alpha$ at which the transition occurs varies for these three models, we summarize these values in fig. \ref{fig:phase-lines}. Moreover if we tune the system towards the phase transition we do not observe an (almost) scale invariance, that is an increase in the number of iterations that the models need to flow to a fixed point.  Instead the system rather quickly flows to one or the other fixed point  -- the non-vanishing singular values specifying the phase have converged after roughly ten to fifteen iterations. Thus it is unlikely that these phase transitions are of second order. 

\begin{figure}
\includegraphics[width=0.95\textwidth]{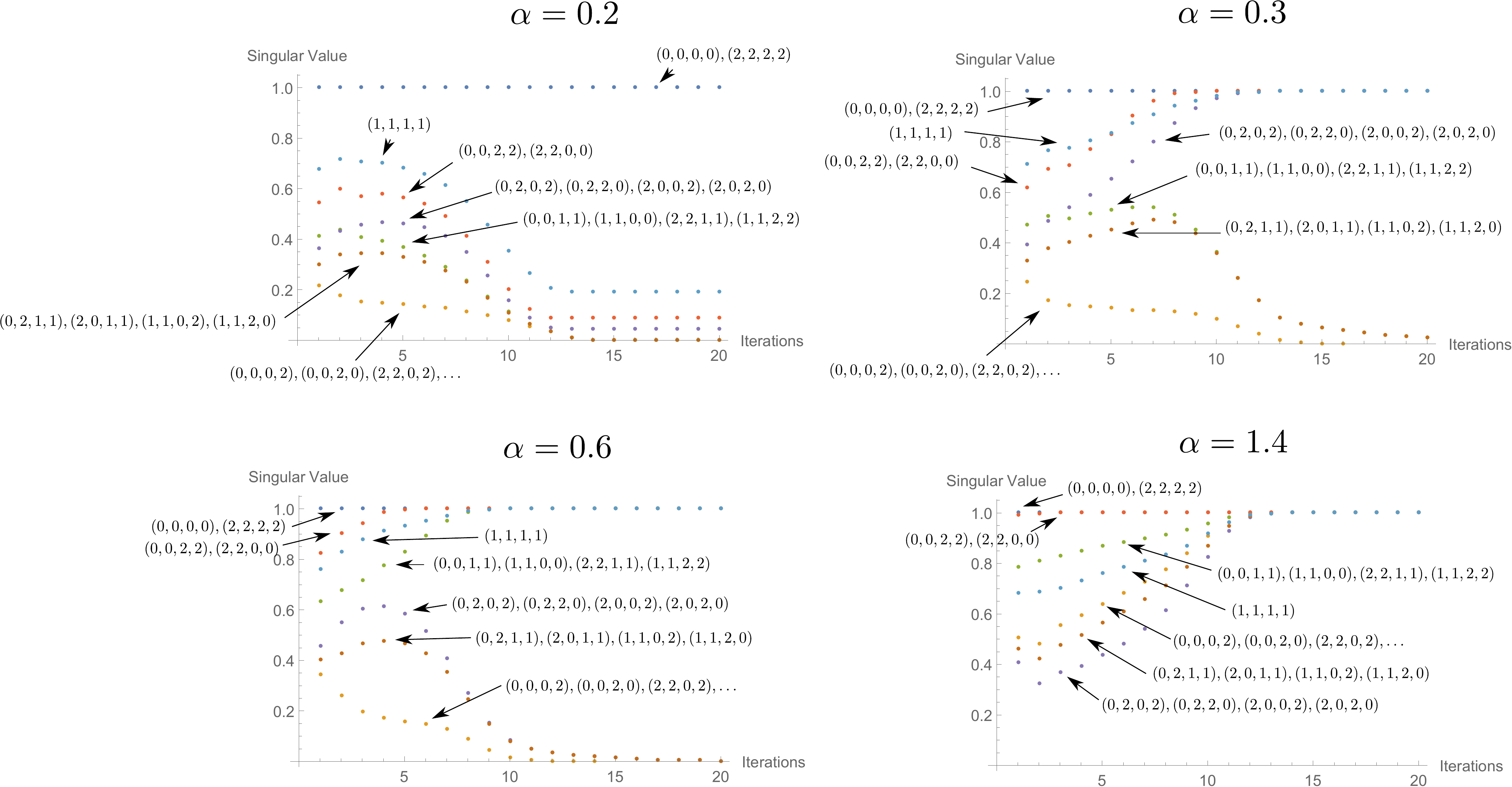}
\caption{\label{fig:sv-plots}
Plots of singular values (per channel $(j^+,j^-,(j^+)',(j^-)')$) for the $k=4$ BC model: The plots are for four different choices of $\alpha$ each flowing to a different fixed point characterizing one of the four phases. The colour coding of the plots is identical in each figure. Depending on the phase, the singular values of different channels converge to either zero or one. (The fact that some singular values do converge to a non-zero value smaller than one for $\alpha=0.2$ is a known, yet unphysical, feature of tensor network algorithms, which is overcome in the recent algorithm by Evenbly and Vidal \cite{vidal-evenbly}). Note that we do not plot those singular values that always converge to zero and thus do not help in differentiating the phases. Note that the convergence happens rather quickly after 10 to 15 iterations, even though some values of $\alpha$ are quite close to phase transitions. This already indicates that the transitions are not of second order.}
\end{figure}

There exist two models that possess a more interesting phase structure, namely for $k=4$ and $k=8$. There we find the following two phases:

\begin{itemize}
\item The first phase only occurs for the model $k=4$ and appears in between the Ashtekar-Lewandowski and the factorising BC phase. It is summarized best in a matrix:
\begin{equation}
 \left(
 \begin{matrix}
  1 & 0 & 0 & 0 & 0 & 0 & 0 & 0 & 1 \\
  0 & 0 & 0 & 0 & 0 & 0 & 0 & 0 & 0 \\
  0 & 0 & 1 & 0 & 0 & 0 & 1 & 0 & 0 \\
  0 & 0 & 0 & 0 & 0 & 0 & 0 & 0 & 0 \\
  0 & 0 & 0 & 0 & 1 & 0 & 0 & 0 & 0 \\
  0 & 0 & 0 & 0 & 0 & 0 & 0 & 0 & 0 \\
  0 & 0 & 1 & 0 & 0 & 0 & 1 & 0 & 0 \\
  0 & 0 & 0 & 0 & 0 & 0 & 0 & 0 & 0 \\
  1 & 0 & 0 & 0 & 0 & 0 & 0 & 0 & 1
 \end{matrix}
\right)
\end{equation}
As one can see only channels on the main diagonals are allowed, together with the condition that the sum of spins (individually for $(J^+,J^-)$ and $((J^+)',(J^-)')$) must be even. Thus the BC condition is broken on this fixed point. Clearly both copies of representations are not independent and the model is thus not factorising, without requiring that $J' = J^*$. Interestingly we have not found a similar phase in any other model. Due to the similarities of $\text{SU}(2)_k$ for $k=4$ restricted to integer representations to the finite group $S_3$, a similar phase / fixed point might exist for $S_3 \times S_3$ as well.
\item Another phase can be found for $k=4$ and $k=8$ for large $\alpha$, which is also violating the BC condition $J^+ = J^-$. Essentially all channels are allowed as long as the sum of spins for each copy of representations is even. Thus the matrix of singular values is alternating between $0$ and $1$ in each column and row (see for $k=4$):
\begin{equation}
 \left(
 \begin{matrix}
  1 & 0 & 1 & 0 & 1 & 0 & 1 & 0 & 1 \\
  0 & 0 & 0 & 0 & 0 & 0 & 0 & 0 & 0 \\
  1 & 0 & 1 & 0 & 1 & 0 & 1 & 0 & 1 \\
  0 & 0 & 0 & 0 & 0 & 0 & 0 & 0 & 0 \\
  1 & 0 & 1 & 0 & 1 & 0 & 1 & 0 & 1 \\
  0 & 0 & 0 & 0 & 0 & 0 & 0 & 0 & 0 \\
  1 & 0 & 1 & 0 & 1 & 0 & 1 & 0 & 1 \\
  0 & 0 & 0 & 0 & 0 & 0 & 0 & 0 & 0 \\
  1 & 0 & 1 & 0 & 1 & 0 & 1 & 0 & 1
 \end{matrix}
\right)
\end{equation}
This phase is factorising yet again, as we can write the matrix as a tensor product of two vectors, whose values alternate between $1$ and $0$. Thus we have found another fixed point of intertwiner models, to which the BC intertwiner model does not flow.
\end{itemize}
Due to these additional fixed points the phase structure for $k=8$ and in particular $k=4$ is more interesting than for the other models. Again these fixed points are attractive and come with extended phases. We have summarized the values of $\alpha$ at which the transitions occur in fig. \ref{fig:phase-lines}. Furthermore in fig. \ref{fig:sv-plots} we plot the flow of singular values for $k=4$ for four different values of $\alpha$, one for each phase.

Of course, more phases imply more phase transitions, which might possibly be of second order. Let us first focus on the case $k=4$, more precisely the transitions between the only non-factorising phase (orange in fig. \ref{fig:phase-lines}) we have found:

As we have discussed above in most models there exists a direct transition between the Ashtekar-Lewandowski and the factorising BC phase, which is not of second order. In the $k=4$ model, the non-factorising phase sits right in between the two before mentioned phases, thus splitting one phase transition into two. At the transition at lower $\alpha$, i.e. from Ashtekar-Lewandowski to the non-factorising phase, we observe only that the system requires a few more iterations (roughly 20) before converging to its fixed point. However no matter how close we tune $\alpha$ towards the transition, we do not find signs of an almost scale invariance. The situation is also very similar at the new transition to the factorising BC phase. Thus these two transitions are likely not of second order, too.

The last remaining transition, from factorising BC to the factorising and alternating model, is present in the models $k=4$ and $k=8$. However, as for the other phase transitions we do not find signs that the transition is of second order, in particular no (almost) scale invariance the further we tune the model towards the transition.

To sum up the initial BC spin net model, in contrast to the BC intertwiner, is not a fixed point of the renormalization group flow. Thus the model flows under coarse graining, where we find two to four different (extended) phases / fixed points depending on the level $k$ of the quantum group. It appears that we find more phases in case the level $k$ is a multiple of four, which might be related to the following fact: If it exists, the representation $\frac{j_{\text{max}}}{2} = \frac{k}{4}$ has the maximal quantum dimension (of the quantum group) and allows for largest number of coupling to other representations. As we restrict our models to integer representations this particular representation only exists for $k=4l$, $l \in \mathbb{N}$. If we increase $\alpha$ the respective weight of configurations containing $j=\frac{k}{4}$ grows faster than others while this representation furthermore couples to most other representations. We expect this to be the origin of the last factorising phase the model flows to for large $\alpha$, yet it seems that $\alpha$ has to be significantly larger for larger levels $k$ of the quantum group. 
 
 In fact, if one is looking for fixed points for the intertwiner models, leading to 2D topological models, the models with even levels $k$ and especially levels  $k=4l$, $l \in \mathbb{N}$ play a special role in featuring more fixed points or equivalently phases compared to the cases where $k$ is odd \cite{wojtek}. For these additional phases the condition of even spins also appears. We therefore believe that the  additional fixed points we found might appear for larger values of $k=4l$ as well.

A non-trivial phase structure with multiple phases also implies transitions between these phases. Second order phase transitions are particularly interesting for discrete models as they provide a non-trivial way of taking the continuum limit and obtaining propagating degrees of freedom. A typical sign for such a transition is an almost scale invariance close to the transition. In tensor networks this manifests itself as the tensors remain unchanged for a growing number iterations the closer the system is tuned towards the transition\footnote{Observing such a behaviour does not prove that the transition is of second order, but it is a strong indication, e.g. this is observed at the phase transition of the 2D Ising model.}. Unfortunately we observe no such behaviour for BC spin nets at any of the transitions we have found, no matter how close we tune towards the transition. Thus we conclude that the transitions of the BC spin net models are very likely not of second order and therefore do not allow for propagating degrees of freedom. 

These results allow us to draw a few tentative conclusions for the BC spin foam model. Of course, the non-existence of second order phase transitions in BC spin nets does not prove that the same holds for BC spin foams, but it can be taken as an indication.  (As mentioned spin nets can be interpreted as spin foams based on a particular 2--complex, which includes only two vertices, but a large number of edges.) 
The BC model has been criticized as implementing the simplicity constraints too strongly, which could suppress propagating degrees of freedom. The absence of a second order phase transition in the corresponding spin net model can be taken as an indicator that this is indeed the case. Of course this has to be confirmed by studying the BC spin foam models itself.

\section{Summary and discussion} \label{sec:discussion}

In this article we have employed tensor network renormalization to thoroughly study $\text{SU}(2)_k \times \text{SU}(2)_k$ intertwiner and spin net models, which are constructed analogously to 4D BC and EPRL/FK spin foam models. We have illustrated the numerical challenges in examining these models for larger levels $k$ of the quantum group and described how we have overcome these by a 3-valent version of familiar tensor network algorithms and further steps that reduced drastically the computational resources required. Let us briefly summarise and discuss the results.

The analogue BC models show an interesting structure. In the intertwiner case we find two fixed points, one is similar to the Asthekar-Lewandowski vacuum in LQG, the other is the initial BC intertwiner itself. Thus in a certain sense the BC model already implements a version of triangulation independence. For most levels $k$ of the quantum group BC spin nets similarly show two phases, again one of Ashtekar-Lewandowski type. The other is best described as a factorising implementation of the BC condition on the representations, that is $j^+ = j^-$. In the interpretation of spin nets as a melon spin foam \cite{q-spinnet}, i.e. two spin foam vertices connected by many edges, this implies a decoupling of the two spin foam vertices.  Since the simplicity constraints weaken the gluing or coupling of spin foam vertices, this can be interpreted as models in which the simplicity constraints are implemented too strongly.

 Notably the initial BC model is not a fixed point of spin net models. For particular levels $k$ we have found up to two {\it new} phases, one of which is not factorising.  It thus could represent an interesting new phase, where spin foam vertices are not decoupled, but nevertheless implement a version of the simplicity constraints.

None of these fixed points corresponds to BF theory, which implies that the simplicity constraints are implemented strong enough such that the BF symmetry is not recovered under coarse graining. However we have not found any signs indicating that the phase transitions are of second order, which can be taken as an indication that the simplicity constraints are implemented too strongly. This latter conclusion does however depend on how accurately the 2D spin net models mimic the 4D spin foam models:  a close relationship holds between 2D spin systems and 4D lattice gauge systems. Spin foams can be understood as generalized gauge systems \cite{holonomy1}, the question is therefore if this relationship survives the generalization. We hope that these questions can be resolved in the near feature by studying the coarse graining flow of 4D spin foam models, e.g. the BC model can readily studied via Monte-Carlo simulations \cite{positivity-bc,bc-monte,Khavkine}.

For the EPRL/FK model we have only been able to study intertwiner models, as the restrictions in imposing the simplicity constraints require large levels $k$ for non-trivial models. Generically we do not observe a convergence to a fixed point, only for parameters (in the one-parameter family of models parametrizing the measure) which disfavour representations with quantum dimension $d_j>1$, i.e. all but $j^\pm=0,j_{\text{max}}$,
, the model appears to flow to the Ashtekar-Lewandowski fixed point. This is rooted in the construction of the (Euclidean) model, where only very few entries in the initial tensor are actually non--zero. We expect this to be less severe for larger levels $k$ and in the Lie group case, however it could be a general flaw of the Euclidean theory. 
However these results must be taken with a grain of salt as the applied truncation scheme cannot account for all degrees of freedom relevant under coarse graining. 

We have however seen that the coarse graining flow, in particular in the EPRL/FK case, shows a surprising complexity. Of course to encounter such a complexity one needs a sufficiently large parameter space, in which the coarse graining flow takes place. Indeed tensor network algorithms provide such a large parameter space, in addition to a truncation scheme, adjusted to the dynamics of the system. As we have seen this leads of course to numerical challenges. We addressed these challenges with a range of techniques in this work, but a further improvement is needed.

In general one can always expect that the set of spin representations, allowed to appear in the initial models as prescribed by the simplicity constraints, is enlarged. This happens because under the coarse graining flow spins associated to finer building blocks are coupled to each other to give spins associated to coarse grained building blocks.  The question is then whether the flow leads to a complete washing out of simplicity constraints (if one flows to the BF model in which all spins are allowed) or whether the flow leads to some subset of allowed spins, that can be interpreted as stable form of the simplicity constraints. Another possibility is, as we have seen, a flow to the trivial phase, where only spin $j=0$ (or in the quantum group case also a $j_{\text{max}}$ with quantum dimension one) is allowed to appear. The tensor network algorithm employed here allows to track which spins are allowed and how relevant the spin configurations actually are. 

Interestingly we have not found that the simplicity constraints are completely washed out, that is a flow  to the BF models. This shows that the spin foam construction principle -- building models by imposing simplicity constraints on the BF model -- has the potential to lead to interesting models. 
For the BC spin net models we have found various fixed points, including a new fixed point, which does describe a non--factorizing spin net model. On the other hand the EPRL/FK intertwiner model showed a very intricate flow. We attribute that in part to the fact that in the Euclidean quantum group model only very few spin values are allowed initially.
 However, also in the  more general EPRL/FK models the set of allowed spins is indeed much more intricate than in the BC model. We thus expect a much more complex flow that will also depend on the Immirzi parameter. To study such aspects it is of course necessary to allow for a sufficiently large parameter space in which the flow can take space.

The results in this article are a first hint at the complex dynamics of spin foam models beyond a few building blocks. In order to explore this regime of many degrees of freedom, numerical algorithms and suitable truncations are required to efficiently identify and study the relevant dynamics. Once achieved this will allow us to contrast the models with observations and measurements, such that they can be eventually verified or falsified. To advance towards this goal works as the one presented here are necessary as they shed light on promising directions and technical issues that have to be overcome.

\begin{acknowledgments}
The authors wish to thank Florian Girelli for discussions at the start of this project. 
S.St. would like to thank Benjamin Bahr for discussions on the construction of the EPRL/FK model.

C.S. has been supported by an NSERC USRA fellowship and an NSERC Discovery grant awarded to Florian Girelli.
S.St. has been supported by the project BA 4966/1-1 of the German Research Foundation (DFG). 
This research was supported in part by Perimeter Institute for Theoretical Physics.  Research at Perimeter Institute is supported by the Government of Canada through the Department of Innovation, Science and Economic Development Canada and by the Province of Ontario through the Ministry of Research, Innovation and Science.
\end{acknowledgments}

\appendix 

\section{$\text{SU}(2)_k$ and graphical calculus} \label{app:q-basics}
In the following appendices we include several technical details necessary to thoroughly understand the calculations in this article, which are however not necessary to understand the main ideas. Moreover we will keep this brief as most of these topics have been more extensively addressed in \cite{q-spinnet,matter-toy}.

In this appendix we briefly include several basic facts about the quantum group $\text{SU}(2)_k$. The notation and conventions used are taken from \cite{biedenharn}, where one can also find a more detailed discussion of quantum groups. Interested readers should also consult \cite{yellowbook}.

By $\text{SU}(2)_k$ we actually mean the q-deformation ${\cal U}_q(\mathfrak{su}(2))$ of the universal enveloping algebra $\mathcal{U}(\mathfrak{su}(2))$ of the Lie algebra $\mathfrak{su}(2)$ as in \cite{biedenharn}. This algebra is generated by three operators $J_{\pm},J_z$ with commutation relations
\begin{eqnarray}
[J_z , J_{\pm} ]&=& \pm J_\pm    \quad ,   \nonumber \\
~ [J_+ , J_-] &=& \frac{q^{J_z}-q^{-J_z}}{q^{1/2}-q^{-1/2}} \quad .
\end{eqnarray}
Given the deformation parameter $q$ one defines quantum numbers of the quantum group:
\begin{equation}
[n]= \frac{q^{\frac{n}{2}}-q^{-\frac{n}{2}}  }{q^{\frac{1}{2}}-q^{-\frac{1}{2}}  } \quad .   
\end{equation}
For $\text{SU}(2)_k$, $q$ is a root of unity, $q=\exp(\frac{2\pi}{(k+2)}i)$, where $k \in \mathbb{N}$ is called the level of the quantum group. In this case quantum numbers are periodic 
\begin{eqnarray}
[n]&=&\frac{\sin(\frac{2\pi n}{2k+4})}{\sin(\frac{2\pi}{2k+4})}  \; ,
\end{eqnarray}
with zeros at $n=0$ and $n=k+2$.

As for $\text{SU}(2)$, the finite dimensional representations of $\text{SU}(2)_k$ are labelled by $j \in \frac{\mathbb{N}}{2}$ and can be defined on $2(j+1)$ dimensional representation spaces $V_j$. The quantum dimension $d_j$ of representation $j$ is defined as the quantum number of the classical dimension:
\begin{equation}
d_j:=[2j+1]  \quad .
\end{equation}
Given the periodicity of quantum numbers only representations $j \leq \frac{k}{2}$ have a strictly positive quantum dimension. Representations $j=0,\frac{1}{2},\ldots,\frac{k}{2}$ are called admissible, representations $j>\frac{k}{2}$ are of so--called quantum trace zero.

The tensor product of two representations $V_{j_1},V_{j_2}$ is defined via the co--product $\Delta$. The action of the $\text{SU}(2)_k$ algebra on $V_{j_1}\otimes V_{j_2}$ is defined as
\begin{eqnarray}
\Delta(J_\pm) \; &=& \; q^{-J_z/2} \otimes J_{\pm} \,+\, J_\pm  \otimes q^{J_z/2} \nonumber \\
\Delta(J_z) \; &=& \; \mathbb{I} \otimes J_z \,+\, J_z\otimes \mathbb{I} \quad .
\end{eqnarray}
The tensor product $V_{j_1}\otimes V_{j_2}$ can be decomposed into a direct sum of irreducible representations plus a part consisting of trace zero representations (which are modded out). With an orthogonal basis $|j,m\rangle$ in the representation spaces, the decomposition is given by Clebsch-Gordan coefficients
\begin{eqnarray}
|j,m\rangle &=&\sum_{m_1,m_2} {}_qC^{j_1j_2j}_{m_1m_2 m} \,|j_1m_1\rangle \otimes |j_2,m_2\rangle \quad .
\end{eqnarray}
If three admissible representations $j_I$, $j_K$ and $j_L$ are coupled in this way, the Clebsch--Gordan coefficients are non--vanishing if several conditions are satisfied:
\begin{eqnarray} \label{couplingcond}
j_I+j_K & \geq & j_L \; \text{for permutations} \; \{J,K,L\} \;\text{of}\; \{1,2,3\} \; ,\nonumber \\
j_1+j_2+j_3 &=& 0\!\mod 1 \; , \nonumber \\
j_1+j_2+j_3 &\leq & k \; .
\end{eqnarray}
The last condition in (\ref{couplingcond}) is special to the quantum deformed case at root of unity and indicates that  $V_{j_1}\otimes V_{j_2}$ can include trace zero parts, which can be modded out \cite{yellowbook}. However, some equations (for instance the definition of the $[6j]$ symbol) are only valid up to trace zero parts \cite{yellowbook}. 

In particular we have the completeness relation
\begin{eqnarray} \label{complete}
\sum_{m_3,\, j_3 \, \text{admiss.}} {}_qC^{j_1j_2j_3}_{m_1m_2m_3} \,\, {}_qC^{j_1j_2j_3}_{m'_1m'_2m_3}&=&\Pi^{j_1 j_2}_{m_1m_2\,,m'_1m'_2} \quad ,
\end{eqnarray}
where $\Pi^{j_1 j_2}_{m_1m_2\,,m'_1m'_2}$ projects out the trace zero part in $V_{j_1}\otimes V_{j_2}$. The orthogonality relation for the Clebsch-Gordan coefficients is given as
\begin{equation} \label{ortho}
\sum_{m_1,m_2} {}_qC^{j_1j_2j}_{m_1m_2m} \,{}_qC^{j_1j_2j'}_{m_1m_2m'}\,\,=\,\,\delta_{jj'}\delta_{mm'} \theta_{j_1j_2j} \; ,
\end{equation}
where $\theta_{j_1j_2j}=1$ if the coupling conditions (\ref{couplingcond}) are satisfied and vanishing otherwise.

\section{Diagrammatic Calculus} \label{app:graph}

When studying spin net models the notion of the dual representation is necessary. For quantum groups this is more complicated to define than in the classical case, but can be conveniently overcome with the graphical calculus invented in \cite{q-spinnet} (and also used in \cite{matter-toy}).

A special direction must be specified for the quantum group, which we will take to be the vertical direction. Then the drawings are interpreted as maps from a tensor product of representation spaces of $\text{SU}(2)_k$ (incoming lines from below) to a tensor product of representation spaces (outgoing lines on top). Each line carries a representation label $j$ and a magnetic index $m$. Clebsch-Gordan coefficients $_q\mathcal{C}_{m_1\,m_2\,m_3}^{j_1 \, j_2 \, j_3}$\footnote{This is not the standard Clebsch-Gordan coefficient defined in \cite{biedenharn}, but it is modified by the quantum dimension: ${}_q\mathcal{C}_{m_1\,m_2\,m_3}^{j_1 \, j_2 \, j_3} = {}_q C_{m_1\,m_2\,m_3}^{j_1 \, j_2 \, j_3 } \left(\sqrt{d_{j_3}}\right)^{-1}$.} are a basic example: They are interpreted as a map $V_{j_1} \otimes V_{j_2} \rightarrow V_{j_3}$, symbolizing how the spins $j_1$ and $j_2$ (with their respective magnetic indices) couple to $j_3$:
\begin{equation}\label{eq:clebsch}
\begin{tikzpicture} [baseline,scale=0.75]
\draw (0,-0.75) node {$j_1$}
      (0,-0.5) -- (0.5,0) -- (1,-0.5)
      (1,-0.75) node {$j_2$}
      (0.5,0) -- (0.5,0.5) 
      (0.5,0.75) node {$j_3$};
\end{tikzpicture}
:= {}_q\mathcal{C}_{m_1\,m_2\,m_3}^{j_1 \, j_2 \, j_3} \quad .
\end{equation}
A special case of this Clebsch-Gordan coefficient is given by $j_1 = j_2 = j$ and $j_3 = 0$, which we call `cap'. It represents a map $V_j\otimes V_j\rightarrow {\mathbb C}$:
\begin{equation} 
\begin{tikzpicture} [baseline,scale=0.75]
\draw (0,-0.75) node {$m$}
      (0,-0.5) -- (0,0) 
      (1,0) arc(-0:180:0.5)
      (0.5,0.75) node {$j$}
      (1,0) -- (1,-0.5)
      (1,-0.75) node {$m'$};
\end{tikzpicture}
:= {}_q \mathcal{C}^{j\,j\,0}_{m\,m'\,0} \sqrt{d_j} =(-1)^{j-m} q^{\frac{m}{2}} \delta_{m,-m'} \quad.
\end{equation}
From this `cap' we can similarly define a `cup' by requiring that they give the identity if we concatenate them:
\begin{equation}\label{identity}
\begin{tikzpicture}[baseline,scale=0.75]
\draw (0,-0.75) node {$m$}
      (0,-0.5) -- (0,0) 
      (1,0) arc(-0:180:0.5)
      (1,0) -- (1,-0.1)
      (1,-0.1) arc(-180:0:0.5)
      (2,-0.1) -- (2,0.6)
      (2,0.85) node {$m''$};
\end{tikzpicture}
= 
\begin{tikzpicture} [baseline,scale=0.75]
\draw (0,-0.75) node {$m$}
      (0,-0.5) -- (0,0.6)
      (0,0.85) node {$m''$};
\end{tikzpicture}
= \delta_m^{m''} \quad ,
\end{equation}
which gives:
\begin{equation}
\begin{tikzpicture}[baseline,scale=0.75]
\draw (0,0.75) node {$m$}
      (0,0.5) -- (0,0)
      (0,0) arc(-180:0:0.5)
      (1,0) -- (1,0.5)
        (0.5,-0.75) node {$j$}
      (1,0.75) node {$m'$};
\end{tikzpicture}
= (-1)^{j+m} q^{\frac{m}{2}} \delta_{m,-m'} \quad .
\end{equation}
Using `cups' and `caps' we obtain Clebsch-Gordan coefficients for the quantum group with inverse (here: complex conjugate) deformation parameter $\bar{q}$ as follows:
\begin{equation}
{}_{\bar{q}} \mathcal{C}_{m_1\, m_2 \, m_3}^{j_1 \, j_2 \, j_3} =
\begin{tikzpicture}[baseline,scale=0.75]
\draw (0,-0.55) -- (0,0) -- (-0.5,0.5)
      (0,0) -- (0.5,0.5)
      (0,-0.8) node {$j_3$}
      (-0.5,0.75) node {$j_2$}
      (0.5,0.75) node {$j_1$};
\end{tikzpicture}
\; = \;
\begin{tikzpicture}[baseline,scale=0.75]
\draw (-0.5,-1) -- (-0.5,-0.5) -- (0,0)
      (0,0) -- (0,0.7)
      (0,0) -- (0.5,-0.5)
      (0.5,-0.5) arc (-180:0:0.5)
      (1.5,-0.5) -- (1.5,0.7)
      (-0.5,-1.25) node {$j_3$}
      (0,0.95) node {$j_2$}
      (1.5,0.95) node {$j_1$};
\end{tikzpicture}
\; = \;
\begin{tikzpicture} [baseline,scale=0.75]
\draw (-1.5,0.7) -- (-1.5,-0.5)
      (-1.5,-0.5) arc (-180:0:0.5)
      (-0.5,-0.5) -- (0,0) -- (0,0.7)
      (0,0) -- (0.5,-0.5) -- (0.5,-1)
      (-1.5,0.95) node {$j_2$}
      (0,0.95) node {$j_1$}
      (0.5,-1.25) node {$j_3$};
\end{tikzpicture} \quad .
\end{equation}
This map is hence interpreted as mapping $V_{j_3} \rightarrow V_{j_1} \otimes V_{j_2}$, thus it is  dual to \eqref{eq:clebsch}. Of course one can analogously obtain \eqref{eq:clebsch} again:
\begin{equation}
\begin{tikzpicture} [baseline,scale=0.75]
\draw (-0.5,0.5) -- (0,0) -- (0,-0.7)
      (0,0) -- (0.5,0.5) 
      (1.5,0.5) arc(0:180:0.5)
      (1.5,0.5) -- (1.5,-0.7)
      (-0.5,0.75) node {$j_5$}
      (0,-0.95) node {$j_3$}
      (1.5,-0.95) node {$j_4$};
\end{tikzpicture}
\; = \;
\begin{tikzpicture} [baseline,scale=0.75]
\draw (-0.5,-0.5) -- (0,0) -- (0,0.7)
      (0,0) -- (0.5,-0.5)
      (-0.5,-0.75) node {$j_3$}
      (0.5,-0.75) node {$j_4$}
      (0,0.95) node {$j_5$};
\end{tikzpicture}
\; = \;
\begin{tikzpicture} [baseline,scale=0.75]
\draw (-1.5,-0.7) -- (-1.5,0.5) 
      (-0.5,0.5) arc(0:180:0.5)
      (-0.5,0.5) -- (0,0) -- (0,-0.7)
      (0,0) -- (0.5,0.5)
      (-1.5,-0.95) node {$j_3$}
      (0,-0.95) node {$j_4$}
      (0.5,0.75) node {$j_5$};
\end{tikzpicture} \quad .
\end{equation}
Concatenating these two maps gives a map $V_{j_3} \rightarrow V_{j_3}$ proportional to the identity.
\begin{equation}
\begin{tikzpicture} [baseline,scale=0.75]
\draw (0,-1) -- (0,-0.5) -- (-0.5,0) -- (0,0.5) -- (0,1)
      (0,-0.5) -- (0.5,0) -- (0,0.5)
      (0,-1.25) node {$j_3$}
      (-0.75,0) node {$j_1$}
      (0.75,0) node {$j_2$}
      (0,1.25) node {$j_3$};
\end{tikzpicture}
\; = \;
\begin{tikzpicture}[baseline,scale=0.75]
\draw (-1,-1) -- (-0.5,-0.5) -- (-0.5,0) -- (0,0.5) -- (0,1)
      (-0.5,-0.5) -- (0,-1) arc(-180:0:0.25) -- (0.5,0) -- (0,0.5)
      (-1,-1.25) node {$j_3$}
      (-0.75,0) node {$j_1$}
      (0.75,0) node {$j_2$}
      (0,1.25) node {$j_3$};
\end{tikzpicture}
\; = \; (-1)^{j_1 + j_2 - j_3} d_{j_3}^{-1} \delta_{m_3\,m'_3} \quad.
\end{equation}
With this graphical calculus already seen in the main body of the article, several important identities can be compactly written:

An important ingredient is the Haar projector $\mathcal{P}$, where we restrict ourselves here to the one for $\text{SU}(2)_k$. The version for $\text{SU}(2)_k \times \text{SU}(2)_k$ is obtained by tensoring the expression as seen in the main part of this article. A 4-valent intertwiner is given as follows:
\begin{equation} \label{eq:4-valent-basis}
\begin{tikzpicture}[baseline,scale=0.75]
\draw (-0.5,-1) -- (0,-0.5) -- (0,0.5) -- (-0.5,1)
      (0.5,-1) -- (0,-0.5)
      (0,0.5) -- (0.5,1)
      (-0.5,-1.25) node {$j_3$}
      (0.5,-1.25) node {$j_4$}
      (0.5,1.25) node {$j_1$}
      (-0.5,1.25) node {$j_2$}
      (0.25,0) node {$j_5$};
\end{tikzpicture}
\; = \; \sum_{m_5} {}_{\bar{q}} \mathcal{C}^{j_1\,j_2\,j_5}_{m_1\,m_2\,m_5} \; {}_q \mathcal{C}^{j_3\,j_4\,j_5}_{m_3\,m_4\,m_5} \quad .
\end{equation}
Its dual is defined by placing `cups' on its bottom legs and `caps' on its top ones avoiding crossing of legs. Graphically this is nicely expressed as:
\begin{equation}\label{dualize}
\begin{tikzpicture}[baseline,scale=0.75]
\draw (-0.25,-0.75) -- (0,-0.5) -- (0,0.5) -- (-0.25,0.75) arc(0:180:0.25)
      (0.25,-0.75) -- (0,-0.5)
      (0,0.5) -- (0.25,0.75) arc(0:180:0.75)
      (-0.25,-0.75) arc(0:-180:0.25)
      (0.25,-0.75) arc(0:-180:0.75)
      (-0.75,0.5) node {$j_1$}
      (-1.25,0.5) node {$j_2$}
      (-0.75,-0.5) node {$j_4$}
      (-1.25,-0.5) node {$j_3$}
      (0.25,0) node {$j_5$};
\end{tikzpicture}
\; = \;
\begin{tikzpicture}[baseline,scale=0.75]
\draw (-0.5,0.75) -- (-0.25,1) arc(180:0:0.5) -- (0.75,-1) arc(0:-180:0.5) -- (-0.5,-0.75)
      (-0.25,1) -- (0,0.75)
      (-0.25,-1) -- (0,-0.75)
      (-0,0.5) node {$j_1$}
      (-0.5,0.5) node {$j_2$}
      (-0,-0.5) node {$j_4$}
      (-0.5,-0.5) node {$j_3$}
      (1,0) node {$j_5$};
\end{tikzpicture}
\; = \; (-1)^{2 j_5} \sum_{m_5}\; q^{m_5} \; {}_{\bar{q}} \mathcal{C}^{j_1 \, j_2 \, j_5}_{m_1 \, m_2 \, m_5} \; {}_q \mathcal{C}^{j_3 \, j_4 \, j_5}_{m_3 \, m_4 \, m_5}
\quad .
\end{equation}
Diagrams \eqref{eq:4-valent-basis} and \eqref{dualize} determine $\mathcal{P}$ up to normalization. To compute $\mathcal{P} \cdot \mathcal{P}$ we have to evaluate the following diagram
\begin{equation}
\begin{tikzpicture}[baseline,scale=0.8]
\draw (-0.25,-0.75) -- (0,-0.5) -- (0,0.5) -- (-0.25,0.75) arc(0:180:0.25) -- (-1,0.5) -- (-1,-0.5)
      (0.25,-0.75) -- (0,-0.5)
      (0,0.5) -- (0.25,0.75) arc(0:180:0.75) -- (-1,0.5)
      (-0.25,-0.75) arc(0:-180:0.25) -- (-1,-0.5)
      (0.25,-0.75) arc(0:-180:0.75) -- (-1,-0.5)
      (-1.25,0) node {$j_5$}
      (0.25,0) node {$j'_5$}
      (-0.5,0.75) node {$j_1$}
      (-0.5,-0.75) node {$j_3$}
      (-0.5,1.75) node {$j_2$}
      (-0.5,-1.75) node {$j_4$};
\end{tikzpicture}
\; = \; (-1)^{j_1+j_2+j_3+j_4} \left(d_{j_5}\right)^{-1} \delta_{j_5 j'_5} \quad  .
\end{equation}
We obtain
\begin{equation}
 \mathcal{P}_{(\{m\},\{n\})}(j_1,j_2,j_3,j_4) := \sum_{j_5} (-1)^{j_1+j_2+j_3+j_4}\; d_{j_5} \;
\begin{tikzpicture}[baseline,scale=0.75]
\draw (-0.5,-1) -- (0,-0.5) -- (0,0.5) -- (-0.5,1)
      (0.5,-1) -- (0,-0.5)
      (0,0.5) -- (0.5,1)
      (-0.5,-1.25) node {$j_3$}
      (0.5,-1.25) node {$j_4$}
      (0.5,1.25) node {$j_1$}
      (-0.5,1.25) node {$j_2$}
      (0.25,0) node {$j_5$};
\end{tikzpicture}
\; \otimes \;
\begin{tikzpicture}[baseline,scale=0.75]
\draw (-0.5,0.75) -- (-0.25,1) arc(180:0:0.5) -- (0.75,-1) arc(0:-180:0.5) -- (-0.5,-0.75)
      (-0.25,1) -- (0,0.75)
      (-0.25,-1) -- (0,-0.75)
      (-0,0.5) node {$j_1$}
      (-0.5,0.5) node {$j_2$}
      (-0,-0.5) node {$j_4$}
      (-0.5,-0.5) node {$j_3$}
      (1,0) node {$j_5$};
\end{tikzpicture}
\quad ,
\end{equation}
where the magnetic indices $m$ are encoded in the first diagram, $n$ in the second.

The change of the recoupling scheme
\begin{equation}
\begin{tikzpicture}[baseline,scale=0.75]
\draw (-1,0.5) -- (-0.5,0) -- (0,0) -- (0.5,0.5)
      (-0.5,0) -- (-1,-0.5)
      (0,0) -- (0.5,-0.5)
      (-1,-0.75) node {$j_3$}
      (-1,0.75) node {$j_2$}
      (0.5,0.75) node {$j_1$}
      (0.5,-0.75) node {$j_4$}
      (-0.25,0.25) node {$j_6$};
\end{tikzpicture}
\;  = \; \sum_{j_5}  \sqrt{\frac{d_{j_5}}{d_{j_6}}} 
\left[
\begin{matrix}
\, j_1 \, & \, j_2 \, & \, j_5 \, \\
\, j_3 \, & \, j_4 \, & \, j_6 \,
\end{matrix}
\right] 
\begin{tikzpicture}[baseline,scale=0.75]
\draw (-0.5,-0.75) -- (0,-0.25) -- (0,0.25) -- (-0.5,0.75)
      (0.5,-0.75) -- (0,-0.25)
      (0,0.25) -- (0.5,0.75)
      (-0.75,-0.75) node {$j_3$}
      (0.75,-0.75) node {$j_4$}
      (0.75,0.75) node {$j_1$}
      (-0.75,0.75) node {$j_2$}
      (0.25,0) node {$j_5$};
\end{tikzpicture}
\end{equation}
is given by the $[6j]$ symbol, which is also defined by a graphical identity:
\begin{equation} \label{6j-def}
\begin{tikzpicture}[baseline,scale=0.6]
\draw (0,-2) -- (0,-1.5) -- (-0.5,-1) -- (-0.5,-0.5) -- (0.5,0.5) -- (0.5,1) -- (0,1.5) -- (0,2) arc(180:0:1) -- (2,-2) arc(0:-180:1)
      (0.,-1.5) -- (1,-0.5) -- (1,0) -- (0.5,0.5)
      (-0.5,-0.5) -- (-1,0) -- (-1,0.5) -- (0,1.5)
      (-0.75,-1) node {$j_1$}
      (1.3,-0.5) node {$j_2$}
      (-1.25,0) node {$j_3$}
      (0.8,1) node {$j_4$}
      (2.3,0) node{$j_5$}
      (-0.2,0.25) node {$j_6$};
\end{tikzpicture}
\; = \;
\begin{tikzpicture}[baseline,scale=0.6]
\draw (0,-2) -- (0,-1.5) -- (-1,-0.5) -- (-1,0) -- (-0.5,0.5) -- (-0.5,1) -- (0,1.5) -- (0,2) arc(180:0:1) -- (2,-2) arc(0:-180:1)
      (0.,-1.5) -- (0.5,-1) -- (0.5,-0.5) -- (1,0) -- (1,0.5) -- (0,1.5)
      (0.5,-0.5) -- (-0.5,0.5)
      (-1.25,-0.5) node {$j_1$}
      (0.8,-1) node {$j_2$}
      (1.3,0) node {$j_4$}
      (-0.75,1) node {$j_3$}
      (2.3,0) node {$j_5$}
      (0.2,0.25) node {$j_6$};
\end{tikzpicture}
\; = \;
\left \{
\begin{matrix}
\,j_1\, & \,j_2\, & \,j_5\, \\
\,j_4\, & \,j_3\, & \,j_6\,
\end{matrix}
\right \} =: \frac{(-1)^{j_1+j_2+j_3+j_4}}{\sqrt{d_{j_5}d_{j_6}}}\left[
\begin{matrix}
\, j_1 \, & \, j_2 \,  & \, j_5 \, \\
\, j_4 \, & \, j_3 \, & \, j_6 \,
\end{matrix}
\right] \quad .
\end{equation}
See \cite{q-spinnet} for a derivation.

A diagram worth mentioning is the following, as it appears in the 4-valent tensor network algorithm (see again \cite{q-spinnet} for a derivation):
\begin{equation} \label{eq:9j-graph}
\begin{tikzpicture}[baseline,scale=0.8]
\draw (-1,-1) -- (-0.5,-0.5) -- (-0.5,0.5) -- (-1,1) -- (0,1.5) arc(180:0:0.75) -- (1.5,-1.5) arc(0:-180:0.75) -- (-1,-1)
      (0,-1.5) -- (1,-1) -- (0.5,-0.5) -- (0.5,0.5) -- (1,1) -- (0,1.5)
      (-0.5,-0.5) -- (0.5,-0.5)
      (-0.5,0.5) -- (0.5,0.5)
      (0,-0.75) node {$j_3$}
      (-0.75,0) node {$j_2$}
      (0.75,0) node {$j_4$}
      (0,0.75) node {$j_1$}
      (-1.25,-1.) node {$j_7$}
      (1.25,-1.) node {$j_8$}
      (1.25,1.) node {$j_5$}
      (-1.25,1.) node {$j_6$}
      (1.75,0) node {$j_9$};
\end{tikzpicture} \quad .
\end{equation}
With the following two identities, it is possible to split this diagram into two $[6j]$ symbols. Fortunately this splitting is precisely the splitting necessary for the 3-valent algorithm such that the equation for the coarse grained tensor can be readily read off from the 4-valent algorithm (see \cite{matter-toy} for a more thorough derivation):
\begin{equation} \label{eq:identity1}
\begin{tikzpicture}[baseline,scale=0.75]
\draw (0,-1) -- (0,1) 
(1,-1) -- (1,1) 
         (-0.25,0) node {$j_2$}
      (1.25,0) node {$j_4$};
\end{tikzpicture}
\; = \;
\begin{tikzpicture}[baseline,scale=0.75]
\draw (-0.5,-1)-- (0,-0.5) -- (0,0.5) --(-0.5,1)
	(0,0.5) --(0.5,1)
	(0.5,-1)-- (0,-0.5)
      (0.3,0) node {$j_9$}
      (-0.75,1) node {$j_2$}
      (0.75,1) node {$j_4$}
       (-0.75,-1) node {$j_2$}
      (0.75,-1) node {$j_4$};
\end{tikzpicture}
\; (-1)^{j_2+j_4-j_9}{d_{j_9}} \quad  , \quad
\begin{tikzpicture}[baseline,scale=0.75]
\draw [pattern=north east lines] (0,0) circle[radius = 0.5];
\draw (0,0.5) -- (0,1.5) 
      (0,-0.5) -- (0,-1.5)
      (0.25,-1) node {$j_9$}
      (0.25,1) node {$j_9$};
\end{tikzpicture}
\; = \;
\begin{tikzpicture}[baseline,scale=0.75]
\draw [pattern=north east lines] (0,0) circle[radius = 0.5];
\draw (0,0.5) -- (0,1) arc(180:0:0.5) -- (1,-1) arc(0:-180:0.5)
      (0,-0.5) -- (0,-1)
      (1.25,0) node {$j_9$};
\end{tikzpicture}
\; \frac{(-1)^{2j_9}}{d_{j_9}} \;
\begin{tikzpicture}[baseline,scale=0.75]
\draw (0,-1) -- (0,1)
      (0.3,0) node {$j_9$};
\end{tikzpicture} \quad .
\end{equation}
Thus we obtain for \eqref{eq:9j-graph}:
\begin{align}
\begin{tikzpicture}[baseline,scale=0.8]
\draw (-1,-1) -- (-0.5,-0.5) -- (-0.5,0.5) -- (-1,1) -- (0,1.5) arc(180:0:0.75) -- (1.5,-1.5) arc(0:-180:0.75) -- (-1,-1)
      (0,-1.5) -- (1,-1) -- (0.5,-0.5) -- (0.5,0.5) -- (1,1) -- (0,1.5)
      (-0.5,-0.5) -- (0.5,-0.5)
      (-0.5,0.5) -- (0.5,0.5)
      (0,-0.75) node {$j_3$}
      (-0.75,0) node {$j_2$}
      (0.75,0) node {$j_4$}
      (0,0.75) node {$j_1$}
      (-1.25,-1.) node {$j_7$}
      (1.25,-1.) node {$j_8$}
      (1.25,1.) node {$j_5$}
      (-1.25,1.) node {$j_6$}
      (1.75,0) node {$j_9$};
\end{tikzpicture}
\; & = \;
(-1)^{j_2 + j_4 - j_9} d_{j_9}
\begin{tikzpicture}[baseline,scale=0.9]
\draw (-0.5,-0.75) -- (0,-0.25) -- (0,0.25) -- (-0.5,0.75) -- (0,1.25) arc(180:0:0.75) -- (1.5,-1.25) arc(0:-180:0.75) -- (0.5,-0.75)
      (0.5,-0.75) -- (0,-0.25)
      (0.5,0.75) -- (0,1.25)
      (-0.5,-0.75) -- (0,-1.25)
      (0,0.25) -- (0.5,0.75)
      (-0.5,0.75) -- (0.5,0.75)
      (-0.5,-0.75) -- (0.5,-0.75)
       (0,-0.5) node {$j_3$}
      (0,0.55) node {$j_1$}
      (-0.6,1) node {$j_6$}
      (0.6,1) node {$j_5$}
      (0.6,0.5) node {$j_4$}
      (-0.6,0.5) node {$j_2$}
      (0.6,-0.4) node {$j_4$}
      (-0.6,-0.4) node {$j_2$}
      (-0.6,-1) node {$j_7$}
      (0.6,-1) node {$j_8$}
      (0.25,0) node {$j_9$}
      (1.75,0) node {$j_9$};
\end{tikzpicture} = \;
(-1)^{j_2+j_4-j_9}
\begin{tikzpicture}[baseline,scale=0.9]
\draw (0,1.75) arc(180:0:0.5) -- (1.,0.75) arc(0:-180:0.5)
      (0,-0.75)  arc(180:0:0.5) -- (1,-1.75) arc(0:-180:0.5)
      (0,0.75)--(0.5,1.25)
        (0,1.75)--(0.5,1.25)
        (0,1.75)--(-0.5,1.25)
           (0,0.75)--(-0.5,1.25)
           (-0.5,1.25)--(0.5,1.25)
            (0,-0.75)--(0.5,-1.25)
        (0,-1.75)--(0.5,-1.25)
        (0,-1.75)--(-0.5,-1.25)
           (0,-0.75)--(-0.5,-1.25)
           (-0.5,-1.25)--(0.5,-1.25)
      (0.,-1.) node {$j_3$}
      (0.,1.1) node {$j_1$}
      (-0.6,1.6) node {$j_6$}
      (0.6,1.6) node {$j_5$}
      (0.6,0.75) node {$j_4$}
      (-0.6,0.75) node {$j_2$}
      (0.6,-0.75) node {$j_4$}
      (-0.6,-0.75) node {$j_2$}
      (-0.6,-1.6) node {$j_7$}
      (0.6,-1.6) node {$j_8$}
      (1.25,-1.25) node {$j_9$}
      (1.25,1.25) node {$j_9$};
\end{tikzpicture} = \; \nonumber \\
& = \; \frac{(-1)^{j_2 + j_4 + j_9} (-1)^{j_5+j_6+j_7+j_8} }{d_{j_9} \sqrt{d_{j_1} d_{j_3}}} 
 \left[
\begin{matrix}
\, j_2 \, & \, j_4 \,  & \, j_9 \, \\
\, j_5 \, & \, j_6 \, & \, j_1 \,
\end{matrix}
\right] 
\left[
\begin{matrix}
\, j_2 \, & \, j_4 \,  & \, j_9 \, \\
\, j_8 \, & \, j_7 \, & \, j_3 \,
\end{matrix}
\right] \; .
\end{align}

\section{Renormalization equation} \label{app:formula}
For the sake of completeness we provide the equations to compute the renormalized 3-valent tensor $\hat{S}$.

\begin{figure}
\includegraphics[scale=0.45]{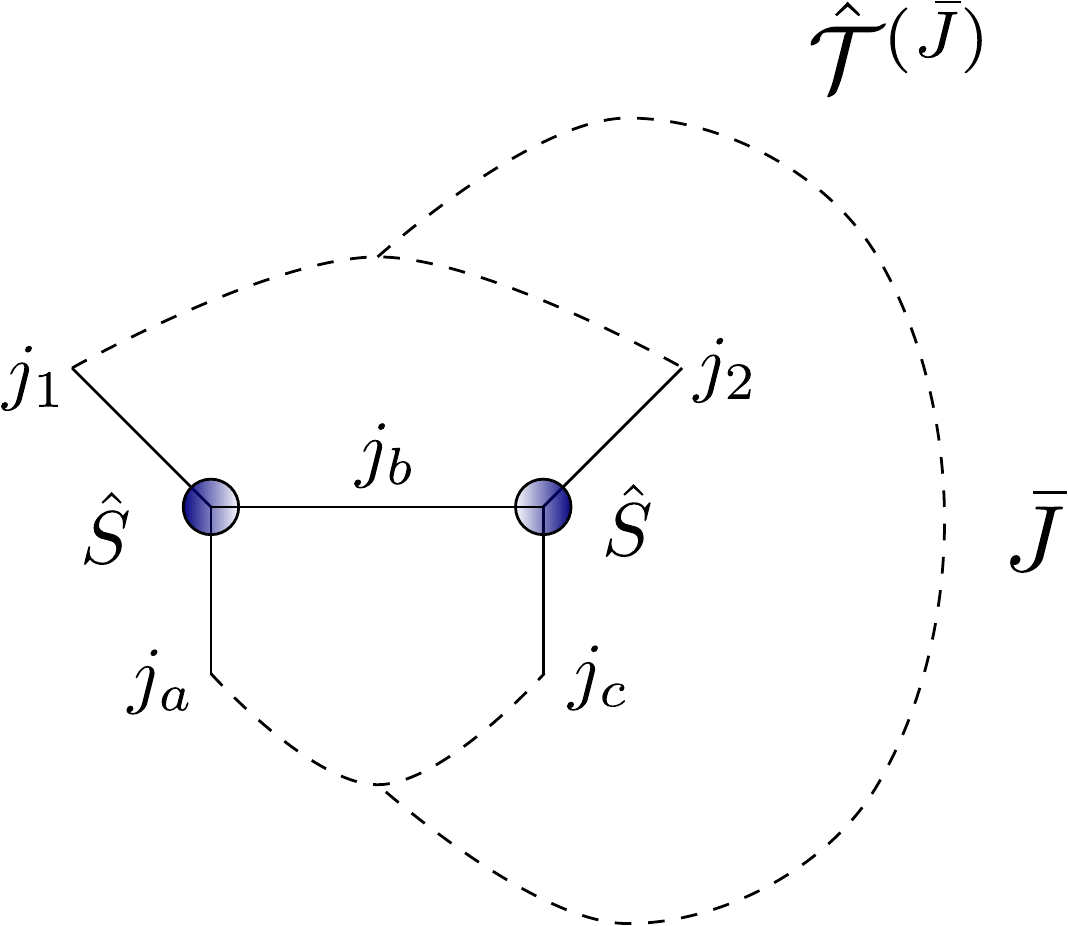}
\caption{\label{fig:int-tensor2}
Notation for intermediate tensor $\hat{\mathcal{T}}$}
\end{figure}

In the 3-valent algorithm, two 3-valent tensors $\hat{S}$ are contracted among a common edge\footnote{In principle, one has four different tensors $S_i$, $i=1,\dots,4$, but for the models under discussion here, they turn out to be all identical. Thus the algorithms is significantly simplified.} to an intermediate 4-valent tensor. For efficiency we directly compute the block-diagonal form of $\hat{\mathcal{T}}$:
\begin{align} \label{eq:intermediate-S1}
\hat{\mathcal{T}}^{(\{\bar{J}\})}(\{j_1\},\{j_2\};\{j_c\},\{j_a\})&=\sum_{\{j_b\}}
\frac{\sqrt{(-1)^{j^+_c + j^+_a + \bar{J}^+}}} {\sqrt{d_{\bar{J}^+}}\sqrt{d_{j^+_b}}}
\frac{\sqrt{(-1)^{j^-_c + j^-_a + \bar{J}^-}}} {\sqrt{d_{\bar{J}^-}}\sqrt{d_{j^-_b}}}
\frac{\sqrt{(-1)^{(j^+_c)' + (j^+_a)' + (\bar{J}^+)'}}} {\sqrt{d_{(\bar{J}^+)'}}\sqrt{d_{(j^+_b)'}}}
\frac{\sqrt{(-1)^{(j^-_c)' + (j^-_a)' + (\bar{J}^-)'}}} {\sqrt{d_{(\bar{J}^-)'}}\sqrt{d_{(j^-_b)'}}} \times \nonumber \\
& \times 
\sqrt{ d_{j^+_1}d_{j^+_2}}
\sqrt{ d_{j^-_1}d_{j^-_2}} 
\sqrt{ d_{(j^+_1)'}d_{(j^+_2)'}} 
\sqrt{ d_{(j^-_1)'}d_{(j^-_2)'}} \; \times \nonumber \\
& \times
\left[
\begin{matrix}
\, j^+_c \, & \, j^+_a \,  & \, \bar{J}^+ \, \\
\, j^+_1 \, & \, j^+_2 \, & \, j^+_b \,
\end{matrix}
\right]
\left[
\begin{matrix}
\, j^-_c \, & \, j^-_a \,  & \, \bar{J}^- \, \\
\, j^-_1 \, & \, j^-_2 \, & \, j^-_b \,
\end{matrix}
\right]
\left[
\begin{matrix}
\, (j^+_c)' \, & \, (j^+_a)' \,  & \, (\bar{J}^+)' \, \\
\, (j^+_1)' \, & \, (j^+_2)' \, & \, (j^+_b)' \,
\end{matrix}
\right]
\left[
\begin{matrix}
\, (j^-_c)' \, & \, (j^-_a)' \,  & \, (\bar{J}^-)' \, \\
\, (j^-_1)' \, & \, (j^-_2)' \, & \, (j^-_b)' \,
\end{matrix}
\right] \nonumber \\
& \times
(\hat{S})^{(\{j_1\})}(\{j_b\},\{j_a\})
(\hat{S})^{(\{j_2\})}(\{j_c\},\{j_b\})
\quad .
\end{align}
The $[6j]$ symbols stem from the treatment of magnetic indices to arrive at the block diagonal form. The notation is explained in fig. \ref{fig:int-tensor2}. Note that in the actual algorithm we work with superindices to only sum and store non--vanishing contributions. As before $\{\bar{J}\}$ denotes four $\text{SU}(2)_k$ representations $\bar{J}^+,\bar{J}^-,(\bar{J}^+)',(\bar{J}^-)'$.

To define a new effective 3-valent tensor $\hat{S}'$, we need to define a map mapping $\{j_a\}$, $\{j_c\}$ into a new coarser edge labelled by $\{\bar{J}\}$. As usual this is done by a singular value decomposition: From $\hat{\mathcal{T}}^{(\{\bar{J}\})}$ we define a matrix by grouping together the coarse edges $\{j_1\}$, $\{j_2\}$ and the finer edges $\{j_a\}$, $\{j_c\}$:
\begin{equation}
\tilde{\hat{S}}^{(\{\bar{J}\})}_{(\{j_1\},\{j_2\});(\{j_a\},\{j_c\})} = \sum_i \; U^{(\{\bar{J}\})}_{(\{j_1\},\{j_2\});i} \; \lambda^{(\{\bar{J}\})}_i \; \left( V^{(\{\bar{J}\})}_{(\{j_a\},\{j_c\});i} \right)^\dagger \quad .
\end{equation}
In our truncation scheme, we take over one singular value per block $\{\bar{J}\}$. Thus we obtain $(\hat{S}')^{(\{\bar{J}\})}$ by contracting the legs $\{j_a\}$, $\{j_c\}$ with the map $V^{(\{\bar{J}\})}_{(\{j_a\},\{j_c\});1}$\footnote{To not alter the partition function one actually inserts $V V^\dagger$. $V^\dagger$ gets contracted with the `opposite' tensor.}. However as $V$ is a unitary matrix it immediately follows that $(\hat{S}')^{(\{\bar{J}\})}$ is given by:
\begin{equation}
\left(\hat{S}'\right)^{(\{\bar{J}\})}\,(\{j_1\},\{j_2\}) = U^{(\{\bar{J}\})}_{(\{j_1\},\{j_2\});1} \; \lambda^{(\{\bar{J}\})}_1 \quad .
\end{equation}

\section{Normalisation BC model} \label{app:BC-norm}

In this section we will briefly derive the normalisation of the BC model. We fix it by requiring that the BC tensor $T_{\text{BC}}$ \eqref{eq:BC-tensor} contracted with itself gives $T_{\text{BC}}$ again:
\begin{equation}
T_{\text{BC}} \circ T_{\text{BC}} \overset{!}{=} T_{\text{BC}} \quad .
\end{equation}
$T_{\text{BC}} \circ T_{\text{BC}}$ is of the following form:
\begin{equation}
\left( T_{\text{BC}} \circ T_{\text{BC}} \right) (\{j\},\{m^\pm\},\{\tilde{n}^\pm\}) = c_{\{j_i\}}^2 \; \;
\begin{tikzpicture}[baseline,scale=0.85]
\draw (-0.25,-1.5) arc(180:0:0.5);
\draw (-0.75,-1.5) arc(180:0:1);
\draw (0.75,1.5) arc(0:-180:0.5);
\draw (1.25,1.5) arc(0:-180:1);
\draw (0.25,0.25) node {$j_1$}
(0.25,1.5) node {$j_2$}
(0.25,-0.25) node {$j_4$}
(0.25,-1.5) node {$j_3$}
(-0.5,0) node {$m^-$}
(1.25,0) node {$m^+$};
\end{tikzpicture}
\; \otimes \;
\begin{tikzpicture}[baseline,scale=0.85]
\draw (-0.25,0.5) arc(180:0:0.5);
\draw (-0.75,0.5) arc(180:0:1);
\draw (0.75,-0.5) arc(0:-180:0.5);
\draw (1.25,-0.5) arc(0:-180:1);
\draw (0.25,0.5) node {$j_1$}
(0.25,1.75) node {$j_2$}
(0.25,-0.5) node {$j_4$}
(0.25,-1.75) node {$j_3$}
(-0.5,0) node {$\tilde{n}^+$}
(1.25,0) node {$\tilde{n}^-$};
\end{tikzpicture} \times
\underbrace{
\begin{tikzpicture}[baseline,scale=0.85]
\draw (-0.25,-1.25) arc(180:-180:0.5);
\draw (-0.75,-1.25) arc(180:-180:1);
\draw (0.75,1.25) arc(0:-360:0.5);
\draw (1.25,1.25) arc(0:-360:1);
\draw (0.25,0.5) node {$j_1$}
      (0.25,1.25) node {$j_2$}
      (0.25,-0.5) node {$j_4$}
      (0.25,-1.25) node {$j_3$};
\end{tikzpicture}}_{(-1)^{2(j_1 + j_2 + j_3 + j_4)} d_{j_1} d_{j_2} d_{j_3} d_{j_4}} \quad .
\end{equation}
For the normalisation constant $c_{\{j_i\}}$ we thus obtain the following condition:
\begin{equation}
c_{\{j_i\}}^2 \underbrace{(-1)^{2(j_1 + j_2 + j_3 + j_4)}}_{=1} d_{j_1} d_{j_2} d_{j_3} d_{j_4} \overset{!}{=} c_{\{j_i\}} \quad \implies \quad c_{\{j_i\}} = \left( d_{j_1} d_{j_2} d_{j_3} d_{j_4} \right)^{-1} \quad .
\end{equation}
As explained in the main body of the article the normalisation in spin foam models is not uniquely fixed. Thus we choose $c_{\{j_i\}} = \left( d_{j_1} d_{j_2} d_{j_3} d_{j_4} \right)^{\alpha}$ to accommodate for this uncertainty.

\section{Normalisation of EPRL model} \label{app:EPRL-norm}

The derivation for the normalisation of the EPRL spin net model is analogous to the BC case. Again we study $T_{\text{EPRL}}$ \eqref{eq:EPRL-tensor} contracted with itself, which should give $T_{\text{EPRL}}$ again:
\begin{equation}
T_{\text{EPRL}} \circ T_{\text{EPRL}} \overset{!}{=} T_{\text{EPRL}} \quad .
\end{equation}
For $T_{\text{EPRL}} \circ T_{\text{EPRL}}$ we obtain:
\begin{equation}
\left(T_{\text{EPRL}} \circ T_{\text{EPRL}} \right) (\{j\},\{m^\pm\},\{\tilde{n}^\pm\}) := \sum_{l,l'} c_{\{l\}} c_{\{l'\}}
\begin{tikzpicture}[baseline,scale=0.85]
\draw (-0.5,-1) -- (0,-0.5) -- (0,0.5) -- (-0.5,1)
      (0.5,-1) -- (0,-0.5)
      (0,0.5) -- (0.5,1)
      (-0.45,-0.55) node {$l_3$}
      (0.45,-0.55) node {$l_4$}
      (0.45,0.55) node {$l_1$}
      (-0.45,0.55) node {$l_2$}
      (0.2,0) node {$l$};
\draw (-0.5,1) -- (-0.75,1.25);
\draw (-0.5,1) -- (-0.25,1.25);
\draw (-0.5,-1) -- (-0.75,-1.25);
\draw (-0.5,-1) -- (-0.25,-1.25);
\draw (0.5,1) -- (0.75,1.25);
\draw (0.5,1) -- (0.25,1.25);
\draw (0.5,-1) -- (0.25,-1.25);
\draw (0.5,-1) -- (0.75,-1.25);
\draw (0.25,-1.5) node {$j^-_4$};
\draw (0.75,-1.5) node {$j^+_4$};
\draw (-0.25,-1.5) node {$j^+_3$};
\draw (-0.75,-1.5) node {$j^-_3$};
\draw (0.25,1.5) node {$j^-_1$};
\draw (0.75,1.5) node {$j^+_1$};
\draw (-0.25,1.5) node {$j^+_2$};
\draw (-0.75,1.5) node {$j^-_2$};
\end{tikzpicture}
\; \otimes \;
\begin{tikzpicture}[baseline,scale=0.85]
\draw (-0.75,1.0) -- (-0.25,1.5) arc(180:0:0.75) -- (1.25,-1.5) arc(0:-180:0.75) -- (-0.75,-1.0)
      (-0.25,1.5) -- (0.25,1.)
      (-0.25,-1.5) -- (0.25,-1.0)
      (0.4,1.45) node {$l_1$}
      (-0.9,1.45) node {$l_2$}
      (0.4,-1.45) node {$l_4$}
      (-0.9,-1.45) node {$l_3$}
      (1.5,0.) node {$l'$};
\draw (0.25,1.) -- (0.5,0.75);
\draw (0.25,1.0) -- (0.,0.75);
\draw (-0.75,1.0) -- (-1.0,0.75);
\draw (-0.75,1.0) -- (-0.5,0.75);
\draw (-0.75,-1.0) -- (-0.5,-0.75);
\draw (-0.75,-1.0) -- (-1.0,-0.75);
\draw (0.25,-1.0) -- (0.5,-0.75);
\draw (0.25,-1.0) -- (0.,-0.75);
\draw (-1.0,0.5) node {$j^-_2$};
\draw (-0.5,0.5) node {$j^+_2$};
\draw (0.5,0.5) node {$j^+_1$};
\draw (0,0.5) node {$j^-_1$};
\draw (-1.0,-0.5) node {$j^-_3$};
\draw (-0.5,-0.5) node {$j^+_3$};
\draw (0.5,-0.5) node {$j^+_4$};
\draw (0.,-0.5) node {$j^-_4$};
\end{tikzpicture} \times
\begin{tikzpicture}[baseline,scale=1.0]
\draw (-0.5,-1) -- (0,-0.5) -- (0,0.5) -- (-0.5,1)
      (0.5,-1) -- (0,-0.5)
      (0,0.5) -- (0.5,1)
      (-0.45,-0.55) node {$l_3$}
      (0.45,-0.55) node {$l_4$}
      (0.45,0.55) node {$l_1$}
      (-0.45,0.55) node {$l_2$}
      (0.2,0) node {$l'$};
\draw (-0.5,1) -- (-1.0,1.5);
\draw (-0.5,1) -- (-0.25,1.25);
\draw (-0.5,-1) -- (-1.0,-1.5);
\draw (-0.5,-1) -- (-0.25,-1.25);
\draw (0.5,1) -- (1.,1.5);
\draw (0.5,1) -- (0.25,1.25);
\draw (-0.25,-1.25) -- (-0.25,-1.75);
\draw (-0.25,1.25) -- (-0.25,1.75);
\draw (0.25,-1.25) -- (0.25,-1.75);
\draw (0.25,1.25) -- (0.25,1.75);
\draw (0.5,-1) -- (0.25,-1.25);
\draw (0.5,-1) -- (1.0,-1.5);
\draw (-1.,1.5) -- (-0.5,2.0) -- (-0.25,1.75);
\draw (1.0,1.5) -- (0.5,2.0) -- (0.25,1.75);
\draw (-1.0,-1.5) -- (-0.5,-2.0) -- (-0.25,-1.75);
\draw (1.0,-1.5) -- (0.5,-2.0) -- (0.25,-1.75);
\draw (-0.5,2.0) -- (0,2.5) -- (0.5,2.0);
\draw (-0.5,-2.0) -- (0,-2.5) -- (0.5,-2.0);
\draw (0.,2.5) arc(180:0:1.0) -- (2.0,-2.5) arc(0:-180:1.0);
\draw (0.5,-1.5) node {$j^-_4$};
\draw (1.25,-1.5) node {$j^+_4$};
\draw (-0.5,-1.5) node {$j^+_3$};
\draw (-1.25,-1.5) node {$j^-_3$};
\draw (0.5,1.5) node {$j^-_1$};
\draw (1.25,1.5) node {$j^+_1$};
\draw (-0.5,1.5) node {$j^+_2$};
\draw (-1.25,1.5) node {$j^-_2$};
\draw (-0.55,2.25) node {$l_2$};
\draw (0.55,2.25) node {$l_1$};
\draw (-0.55,-2.35) node {$l_3$};
\draw (0.55,-2.35) node {$l_4$};
\draw (1.75,0) node {$l$};
\end{tikzpicture}
 \quad .
\end{equation}
The last diagram is straightforward to compute given the graphical identities in appendix \ref{app:graph}:
\begin{align}
\begin{tikzpicture}[baseline,scale=1.0]
\draw (-0.5,-1) -- (0,-0.5) -- (0,0.5) -- (-0.5,1)
      (0.5,-1) -- (0,-0.5)
      (0,0.5) -- (0.5,1)
      (-0.45,-0.55) node {$l_3$}
      (0.45,-0.55) node {$l_4$}
      (0.45,0.55) node {$l_1$}
      (-0.45,0.55) node {$l_2$}
      (0.2,0) node {$l'$};
\draw (-0.5,1) -- (-1.0,1.5);
\draw (-0.5,1) -- (-0.25,1.25);
\draw (-0.5,-1) -- (-1.0,-1.5);
\draw (-0.5,-1) -- (-0.25,-1.25);
\draw (0.5,1) -- (1.,1.5);
\draw (0.5,1) -- (0.25,1.25);
\draw (-0.25,-1.25) -- (-0.25,-1.75);
\draw (-0.25,1.25) -- (-0.25,1.75);
\draw (0.25,-1.25) -- (0.25,-1.75);
\draw (0.25,1.25) -- (0.25,1.75);
\draw (0.5,-1) -- (0.25,-1.25);
\draw (0.5,-1) -- (1.0,-1.5);
\draw (-1.,1.5) -- (-0.5,2.0) -- (-0.25,1.75);
\draw (1.0,1.5) -- (0.5,2.0) -- (0.25,1.75);
\draw (-1.0,-1.5) -- (-0.5,-2.0) -- (-0.25,-1.75);
\draw (1.0,-1.5) -- (0.5,-2.0) -- (0.25,-1.75);
\draw (-0.5,2.0) -- (0,2.5) -- (0.5,2.0);
\draw (-0.5,-2.0) -- (0,-2.5) -- (0.5,-2.0);
\draw (0.,2.5) arc(180:0:1.0) -- (2.0,-2.5) arc(0:-180:1.0);
\draw (0.5,-1.5) node {$j^-_4$};
\draw (1.25,-1.5) node {$j^+_4$};
\draw (-0.5,-1.5) node {$j^+_3$};
\draw (-1.25,-1.5) node {$j^-_3$};
\draw (0.5,1.5) node {$j^-_1$};
\draw (1.25,1.5) node {$j^+_1$};
\draw (-0.5,1.5) node {$j^+_2$};
\draw (-1.25,1.5) node {$j^-_2$};
\draw (-0.55,2.25) node {$l_2$};
\draw (0.55,2.25) node {$l_1$};
\draw (-0.55,-2.35) node {$l_3$};
\draw (0.55,-2.35) node {$l_4$};
\draw (1.75,0) node {$l$};
\end{tikzpicture}
\; & = \; (-1)^{\sum_{i=1}^4 j_i^+ + j_i^- -l_i } \left(d_{l_1} d_{l_2} d_{l_3} d_{l_4} \right)^{-1} \times
\begin{tikzpicture}[baseline,scale=1.0]
\draw (-0.5,-1) -- (0,-0.5) -- (0,0.5) -- (-0.5,1)
      (0.5,-1) -- (0,-0.5)
      (0,0.5) -- (0.5,1)
      (-0.75,-1.0) node {$l_3$}
      (0.75,-1.0) node {$l_4$}
      (0.75,1.0) node {$l_1$}
      (-0.75,1.0) node {$l_2$}
      (0.2,0) node {$l'$};
\draw (-0.5,-1) -- (0,-1.5) -- (0.5,-1);
\draw (-0.5,1) -- (0,1.5) -- (0.5,1);
\draw (0,1.5) arc(180:0:0.75) -- (1.5,-1.5) arc(0:-180:0.75);
\draw (1.25,0) node {$l$};
\end{tikzpicture} \nonumber \\
& = \; (-1)^{\sum_{i=1}^4 j_i^+ + j_i^- - 2 l} \left(d_{l_1} d_{l_2} d_{l_3} d_{l_4} \right)^{-1} d_l^{-2} \delta_{l l'} \quad .
\end{align}
This implies for the normalisation constant $c_{\{l\}}$:
\begin{equation}
c_{\{l\}}^2 (-1)^{\sum_{i=1}^4 j_i^+ + j_i^- - 2 l} \left(d_{l_1} d_{l_2} d_{l_3} d_{l_4} \right)^{-1} \left( d_l \right)^{-2} \overset{!}{=} c_{\{l\}} \quad \implies \quad c_{\{l\}} = (-1)^{\sum_{i=1}^4 j_i^+ + j_i^- - 2 l} \; d_{l_1} d_{l_2} d_{l_3} d_{l_4} d_l^{2} \quad .
\end{equation}
Again as the normalisation is not uniquely determined in spin foams we keep this arbitrary we choose $c_{\{l\}} = (-1)^{\sum_{i=1}^4 j_i^+ + j_i^- - 2 l} \; \left( d_{l_1} d_{l_2} d_{l_3} d_{l_4} \right)^\alpha d_l^{2}$.

\section{Derivation of EPRL amplitude} \label{app:EPRL-diagram}

In this appendix we will quickly outline how to derive the diagrams in the 3-valent EPRL model from the 4-valent one. Essentially one applies the identities \eqref{eq:identity1}. We will demonstrate this only for one diagram in \eqref{eq:EPRL-formula}, as it follows for the other diagram analogously:

\begin{align}
\begin{tikzpicture}[baseline,scale=0.85]
\draw (-0.75,1.0) -- (-0.25,1.5) arc(180:0:0.75) -- (1.25,-1.5) arc(0:-180:0.75) -- (-0.75,-1.0)
      (-0.25,1.5) -- (0.25,1.)
      (-0.25,-1.5) -- (0.25,-1.0)
      (0.4,1.45) node {$l_1$}
      (-0.9,1.45) node {$l_2$}
      (0.4,-1.45) node {$l_4$}
      (-0.9,-1.45) node {$l_3$}
      (1.5,0.) node {$l$};
\draw (0.25,1.) -- (0.5,0.75);
\draw (0.25,1.0) -- (0.,0.75);
\draw (-0.75,1.0) -- (-1.0,0.75);
\draw (-0.75,1.0) -- (-0.5,0.75);
\draw (-0.75,-1.0) -- (-0.5,-0.75);
\draw (-0.75,-1.0) -- (-1.0,-0.75);
\draw (0.25,-1.0) -- (0.5,-0.75);
\draw (0.25,-1.0) -- (0.,-0.75);
\draw (-1.25,0.75) node {$j^-_2$};
\draw (0.75,0.75) node {$j^+_1$};
\draw (-1.25,-0.75) node {$j^-_3$};
\draw (0.75,-0.75) node {$j^+_4$};
\draw (-1.,0.) node {$(J^-)'$};
\draw (0.65,0.) node {$(J^+)'$};
\draw (-1.0,-0.75) -- (-0.5,-0.25)  -- (-0.35,-0.4);
\draw (-0.15,-0.6) -- (0.,-0.75);
\draw (-0.5,-0.75) -- (-0.,-0.25) -- (0.5,-0.75);
\draw (-1.0,0.75) -- (-0.5,0.25) -- (0.,0.75);
\draw (-0.5,0.75) -- (-0.4,0.65);
\draw (-0.15,0.4) -- (0.,0.25) -- (0.5,0.75);
\draw (0,0.25) -- (0,-0.25);
\draw (-0.5,0.25) -- (-0.5,-0.25);
\end{tikzpicture}
\; & = \; (-1)^{(J^-)' + (J^+)' - l} d_l \;
\begin{tikzpicture}[baseline,scale=1.0]
\draw (-0.75,1.5) -- (-0.25,2.0) arc(180:0:0.75) -- (1.25,-2.0) arc(0:-180:0.75) -- (-0.75,-1.5)
      (-0.25,2.0) -- (0.25,1.5)
      (-0.25,-2.0) -- (0.25,-1.5)
      (0.4,1.95) node {$l_1$}
      (-0.9,1.95) node {$l_2$}
      (0.4,-1.95) node {$l_4$}
      (-0.9,-1.95) node {$l_3$}
      (1.5,0.) node {$l$};
\draw (0.25,1.5) -- (0.5,1.25);
\draw (0.25,1.5) -- (0.,1.25);
\draw (-0.75,1.5) -- (-1.0,1.25);
\draw (-0.75,1.5) -- (-0.5,1.25);
\draw (-0.75,-1.5) -- (-0.5,-1.25);
\draw (-0.75,-1.5) -- (-1.0,-1.25);
\draw (0.25,-1.5) -- (0.5,-1.25);
\draw (0.25,-1.5) -- (0.,-1.25);
\draw (-1.25,1.25) node {$j^-_2$};
\draw (0.75,1.25) node {$j^+_1$};
\draw (-1.25,-1.25) node {$j^-_3$};
\draw (0.75,-1.25) node {$j^+_4$};
\draw (-1.,0.5) node {$(J^-)'$};
\draw (0.65,0.5) node {$(J^+)'$};
\draw (-1.,-0.5) node {$(J^-)'$};
\draw (0.65,-0.5) node {$(J^+)'$};
\draw (0,0) node {$l$};
\draw (-1.0,-1.25) -- (-0.5,-0.75)  -- (-0.35,-0.9);
\draw (-0.15,-1.1) -- (0.,-1.25);
\draw (-0.5,-1.25) -- (-0.,-0.75) -- (0.5,-1.25);
\draw (-1.0,1.25) -- (-0.5,0.75) -- (0.,1.25);
\draw (-0.5,1.25) -- (-0.4,1.15);
\draw (-0.15,0.9) -- (0.,0.75) -- (0.5,1.25);
\draw (0,0.75) -- (0,0.5) -- (-0.25,0.25) -- (-0.25,-0.25) -- (0,-0.5) -- (0.,-0.75);
\draw (-0.5,0.75) -- (-0.5,0.5) -- (-0.25,0.25);
\draw (-0.5,-0.75) -- (-0.5,-0.5) -- (-0.25,-0.25);
\end{tikzpicture} \nonumber \\
& = \; \underbrace{(-1)^{(J^-)' + (J^+)' - l} d_l \times \frac{(-1)^{4l}}{d_l} }_{(-1)^{(J^-)' + (J^+)' - l}} \;
\begin{tikzpicture}[baseline,scale=1.0]
\draw (-0.75,2.5) -- (-0.25,3.0) arc(180:0:0.75) -- (1.25,1.0); 
\draw (1.25,-1.) -- (1.25,-3.) arc(0:-180:0.75) -- (-0.75,-2.5)
      (-0.25,3.0) -- (0.25,2.5)
      (-0.25,-3.0) -- (0.25,-2.5)
      (0.4,2.95) node {$l_1$}
      (-0.9,2.95) node {$l_2$}
      (0.4,-2.95) node {$l_4$}
      (-0.9,-2.95) node {$l_3$}
      (1.5,1.875) node {$l$};
\draw (0.25,2.5) -- (0.5,2.25);
\draw (0.25,2.5) -- (0.,2.25);
\draw (-0.75,2.5) -- (-1.0,2.25);
\draw (-0.75,2.5) -- (-0.5,2.25);
\draw (-0.75,-2.5) -- (-0.5,-2.25);
\draw (-0.75,-2.5) -- (-1.0,-2.25);
\draw (0.25,-2.5) -- (0.5,-2.25);
\draw (0.25,-2.5) -- (0.,-2.25);
\draw (-1.25,2.25) node {$j^-_2$};
\draw (0.75,2.25) node {$j^+_1$};
\draw (-1.25,-2.25) node {$j^-_3$};
\draw (0.75,-2.25) node {$j^+_4$};
\draw (-1.,1.5) node {$(J^-)'$};
\draw (0.65,1.5) node {$(J^+)'$};
\draw (-1.,-1.5) node {$(J^-)'$};
\draw (0.65,-1.5) node {$(J^+)'$};
\draw (1.5,-1.875) node {$l$};
\draw (-1.0,-2.25) -- (-0.5,-1.75)  -- (-0.35,-1.9);
\draw (-0.15,-2.1) -- (0.,-2.25);
\draw (-0.5,-2.25) -- (-0.,-1.75) -- (0.5,-2.25);
\draw (-1.0,2.25) -- (-0.5,1.75) -- (0.,2.25);
\draw (-0.5,2.25) -- (-0.4,2.15);
\draw (-0.15,1.9) -- (0.,1.75) -- (0.5,2.25);
\draw (0,1.75) -- (0,1.5) -- (-0.25,1.25) -- (-0.25,1.0) arc(-180:0:0.75);
\draw (1.25,-1) arc(0:180:0.75) -- (-0.25,-1.25) -- (0,-1.5) -- (0.,-1.75);
\draw (-0.5,1.75) -- (-0.5,1.5) -- (-0.25,1.25);
\draw (-0.5,-1.75) -- (-0.5,-1.5) -- (-0.25,-1.25);
\end{tikzpicture} \quad .
\end{align}
To arrive at the final expression it remains to include the identity for the $\mathcal{R}$ matrices, which is explained in the main part of the paper.

\bibliographystyle{utphys}
\bibliography{biblio}

\end{document}